\pdfoutput=1
\documentclass[a4paper]{article}

\usepackage{hyperref}
\usepackage{amsmath,amsfonts,amssymb}
\usepackage{amsthm}
\usepackage{color}
\usepackage{tikz}
\usepackage{graphicx}
\usepackage{cite}

\def\ds{\displaystyle}
\def\bea{\begin{array}{c}}
\def\ea{\end{array}}
\def\be{\begin{equation}\bea\ds}
\def\ee{\ea\end{equation}}
\def\bee{\begin{equation}\begin{array}{rcl}\ds}
\def\eee{\end{array}\end{equation}}

\def\tr{{\rm Tr}\,}
\def\nn{\nonumber}

\def\Hc{{\mathcal{H}}}
\def\Mc{{\mathcal{M}}}

\def\tr{{\rm tr}\,}
\def\Tr{{\rm Tr}\,}

\def\nn{\nonumber}

\title{Entanglement Classification from a Topological Perspective}

\author{Dmitry Melnikov}
\date{}

\begin{document}

\thispagestyle{empty}

\maketitle

\begin{center}
\textit{\small International Institute of Physics, Federal University of 
Rio Grande do Norte, \\ Campus Universit\'ario, Lagoa Nova, Natal-RN  
59078-970, Brazil}
\\ \vspace{0.6cm}
\textit{\small Institute for Theoretical and Experimental Physics, NRC Kurchatov Institute, \\ B.~Cheremushkinskaya 25, 117218 Moscow, Russia}
\\ \vspace{2cm}

\end{center}

\vspace{-1cm}

\begin{abstract}
Classification of entanglement is an important problem in Quantum Resource Theory. In this paper we discuss an embedding of this problem in the context of Topological Quantum Field Theories (TQFT). This approach allows classifying entanglement patterns in terms of topological equivalence classes. In the bipartite case a classification equivalent to the one by Stochastic Local Operations and Classical Communication (SLOCC) is constructed by restricting to a simple class of connectivity diagrams. Such diagrams characterize quantum states of TQFT up to braiding and tangling of the ``connectome.'' In the multipartite case the same restricted topological classification only captures a part of the SLOCC classes, in particular, it does not see the W entanglement of three qubits. Nonlocal braiding of connections may solve the problem, but no finite classification is attempted in this case. Despite incompleteness, the connectome classification has a straightforward generalization to any number and dimension of parties and has a very intuitive interpretation, which might be useful for understanding specific properties of entanglement and for design of new quantum resources.  
\end{abstract} 

\section{Introduction}

The term entanglement, in reference to intricate correlations existing between parts of a quantum system, raises natural analogies with topology. An early discussion of this observation appears in~\cite{Aravind:1997}, for example. Specifically, knot theory is a branch of topology that studies classical entanglement of one-dimensional objects, such as curves, ropes or ribbons, and the question is whether analogy between this classical entanglement and quantum correlations has any strict mathematical meaning, or whether there is any practical use of it for quantum physics.

Topological quantum field theories (TQFTs)~\cite{Atiyah:1989vu,Witten:1988ze} offer an obvious setup to address such questions. These theories appeared as a link connecting knot theory and quantum mechanics (quantum field theory), endowing knots and links with a notion of quantum correlators~\cite{Witten:1988hf}. In the instances of quantum mechanics appearing in the context of TQFT all observables are expressed in terms of the topological invariants of knots, and so must be the entanglement and its measures. The question asked in~\cite{Aravind:1997} is more specific: whether the character of quantum entanglement is encoded in the linking patterns of knots. This question can be motivated by the classification problems in knots and in quantum information theory. On one hand, classification and counting of knots is a fundamental problem of knot theory. On the other, one is interested in understanding possible patterns of entanglement. The latter is an important problem of Quantum Resource Theory: the structure of the state's entanglement defines quantum tasks, for which this state is a suitable resource.

Further motivation stems from the fact that physical systems governed by TQFTs, such as two-dimensional electrons in strong magnetic fields, or concept condensed matter systems with Majorana fermions, can serve a basis for a quantum computer, whose states are naturally robust against decoherence~\cite{Kitaev:1997wr,Freedman:2000rc,Kitaev:2000nmw}. From the theoretical point of view, the most natural set of basic operations (gates) in such a computer is provided by braiding of the quasiparticles in the above systems -- a procedure of creating knots (links) from the particles' worldlines. Then the question of~\cite{Aravind:1997} becomes a question of vocabulary, translation of known quantum algorithms into the language of mathematics of knot theory.

Different aspects of topological quantum computation and physical realizations of topological phases of matter have been by now extensively addressed in the literature. The canonical reference for this subject is the review~\cite{Nayak:2008zza}. In the present note we would like to take a step back towards the original question and discuss the problem of quantum entanglement classification from the point of view of formal TQFT. A systematic early effort to discuss the entanglement features of knots and to build the knot theory version of the quantum computer based on the ideas similar to the ones in~\cite{Aravind:1997} was taken in the following (likely incomplete) set of references: \cite{Kauffman:2002qua,Kauffman:2003cri,Kauffman:2004bra,Kauffman:2004qua,Lomonaco:2008qua,Melnikov:2017bjb,Kauffman:2019top,Padmanabhan:2019qed}. (See a review of these results and further references in~\cite{Kauffman:2013bh}.)

The purpose of this paper is to discuss what knots, or rather TQFTs, teach us about quantum entanglement, or how they encode the entanglement and how they classify its different types. Besides~\cite{Aravind:1997}, this expands and complements similar recent studies, e.g.~\cite{Dong:2008ft,Salton:2016qpp,Balasubramanian:2016sro,Chun:2017hja,Dwivedi:2017rnj,Balasubramanian:2018por,Melnikov:2018zfn,Quinta:2018scm,Dwivedi:2019bzh,Dwivedi:2020jyx,Bao:2021gzu,Buican:2021axn}. Here we will start from the axiomatic approach to TQFT, as in~\cite{Melnikov:2018zfn}, and compare different forms of entanglement suggested by topology to the Quantum Resource Theory classification based on Stochastic Local Operations and Classical Communication (SLOCC). The topologies surging in the present approach will be different from that of~\cite{Balasubramanian:2016sro}, for example, in that they will be based on tangles (open lines), rather than links (closed lines).

Axiomatic TQFT associates a quantum state to a topological space with a boundary (we will review the axiomatic definition in one of the following sections). The boundary defines the kind of Hilbert space the state belongs to, while the bulk -- the details of the state in this Hilbert space. Different spaces can be glued along equivalent boundaries to realize scalar products, matrix multiplication, or more general tensor contractions. An important realization of the TQFT axioms is provided by functional integrals of metric independent action functionals (partition functions).

Topological spaces are equivalent up to homeomorphisms. In comparison to metric spaces (nontopological partition functions) they already define classes of equivalence of quantum states.  Bulk-disconnected boundaries correspond to product states. Disconnected boundary components introduce the notion of locality. All this sets a stage for the discussion of entanglement and its classification.

The naive classification by topology in a generic TQFT gives an infinite number of discrete topological classes. This is problematic for the following reasons. First, in Quantum Resource Theory one is interested in a finite classification of quantum tasks and hence needs a finite number of entanglement classes, so a coarse graining of the infinity of possibilities is desired. This problem is solved for specific TQFTs, with finite-dimensional Hilbert spaces, like Chern-Simons theory: there are linear relations between states of different topology, and the total number of independent topologies is finite. However, the relations make the classification problem more complicated. Second, the fact that the topologies form a discrete set does not guarantee that all possible entanglement patterns are reproduced by the topologies. In such a scenario, one may hope that representatives of all the entanglement classes can be constructed with the required precision.

In this work a coarse-grained classification is proposed in a class of TQFT, based on coarse graining of the connectivity classes of the topological spaces. Specifically, we will consider $SU(2)_k$ Chern-Simons theories in a 3-sphere with multiple 2-sphere boundaries. The 2-spheres will be allowed to have punctures, and the 3-sphere will have 1D defects (Wilson lines).\footnote{In order to avoid the complication of additional linear relations mentioned above and to make the discussion generic, $k$ will always be assumed large, typically much larger than the number of punctures. Alternatively, one can assume analytic continuation to noninteger $k$.} Different topologies will be characterized by different wiring of 1D defects and the connectivity classes will be defined by the adjacency matrices of the corresponding graphs. To address the completeness of such a classification, it will be compared to the SLOCC one.

The connectivity classification can be illustrated in the simplest case of a pair of qubits as follows:
\begin{eqnarray}
\begin{array}{c}
     \includegraphics[scale=0.15]{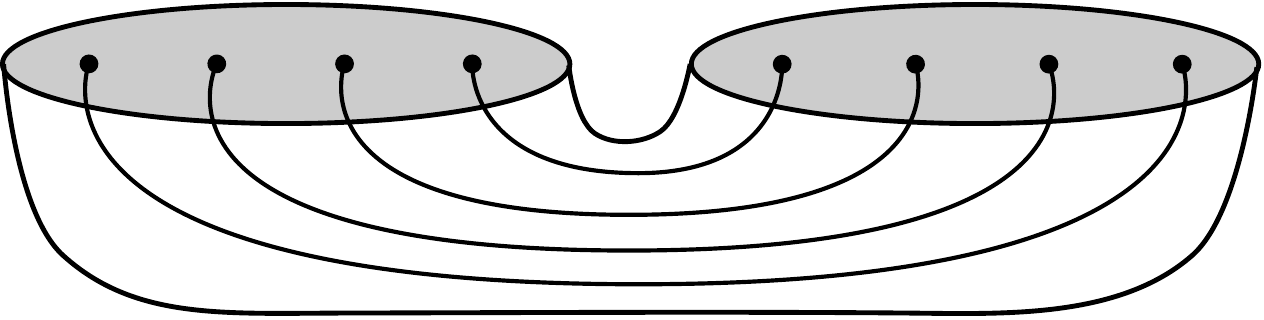}
\end{array} & \longleftrightarrow & \text{entangled} \nn\\
&& \label{entanglement}\\
\begin{array}{c}
\includegraphics[scale=0.15]{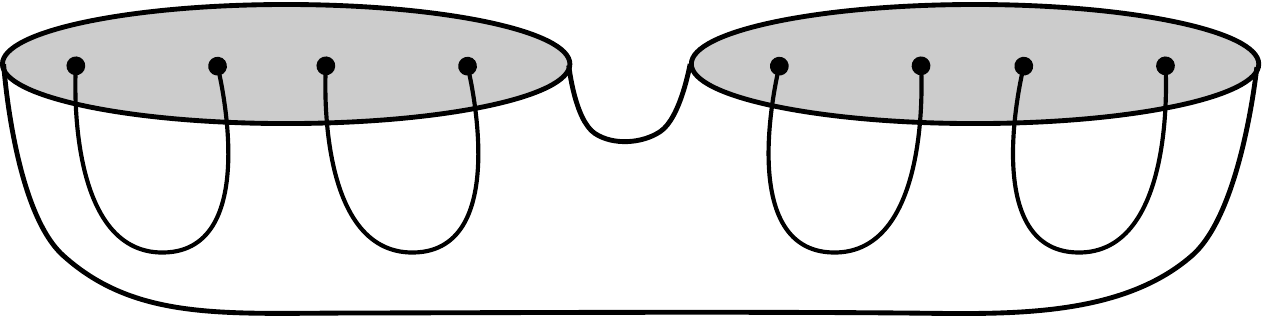}
\end{array} & \longleftrightarrow & \text{unentangled} \nn
\end{eqnarray}
Shaded discs in these heuristic diagrams represent boundary spheres $S^2$ with punctures (qubits) and lines -- evolution of the punctures in the 3D bulk (Wilson lines, correlations). It will be argued that these are the only relevant ways of connecting the punctures up to local braiding and tangling. Further details of the identification are provided in the body of the paper. 

Simple connectivity classification does not generalize to the case of generic qudits in a straightforward way. The main problem is that the dimension of the Hilbert space grows too fast with the number of additional Wilson lines. One needs to project these large Hilbert spaces onto appropriate subspaces. Necessary projectors exist in TQFT, known as Jones-Wenzl projectors~\cite{Jones:1985dw,Wenzl:1985seq}. They allow building a connectivity classification matching the SLOCC one in the bipartite case. This result is a consequence of the fact that the bipartite SLOCC classification is related to a simple property of reduced density matrices and can be reproduced in terms of highest weight vectors of irreducible representation of $SU(2)$, to which the Jones-Wenzl projectors have a direct relation.

With the growth of the number of parties the generalization of the SLOCC classification quickly becomes inefficient. It is understood that the tripartite case of qubits has four nonequivalent classes~\cite{Dur:2000zz}, while for four~\cite{Verstraete:2002four} and more parties the SLOCC classification is only partially successful~\cite{Horodecki:2009zz}. The multipartite entanglement classes are determined up to a number of parameters not fixed by the SLOCC. In other words, the precise classification gives an infinite number of entanglement classes, making the full version impractical and requiring some further coarsening.

The situation with the topological classification developed here is slightly different. On one hand, it remains finite for any number of parties, but on the other, it becomes more coarse than needed. The extension of the above approach to three and more qubits is not able to capture all the SLOCC classes, including the case of the tripartite entanglement. In the latter the connectivity diagrams reproduce only three out of four classes. The most intricate $W$ class is not captured by the connectivity, but can rather be achieved through additional nonlocal braiding. In other words, there seem to be no simple topological representation for the $W$ states, though it can be constructed with needed approximation by a sufficient number of nonlocal unitaries or local nonunitary invertible operations. 

In summary, the topological classification provides a very intuitive interpretation of entanglement as a resource (Wilson lines) shared between the parties, and a specific form entanglement is defined by a specific distribution of this resource (wiring of the Wilson lines). By discrete nature of TQFT, they cannot capture all the possible states in the Hilbert space, but rather approximate any given state with a desired accuracy. As a result, some of the multipartite SLOCC entanglement classes seem inaccessible for the topological classification, making the latter more coarse than the former. It is, nevertheless, generalizable to any number of parties and we will show that in general, the topological approach is an interesting tool, which can potentially lead to new understanding of the properties of entanglement.  

Finally, it is interesting to compare the topological classification with other examples of coarse classifications, such as~\cite{Sawicki:2012cri,Walter:2013ent,Sawicki:2014con,Maciazek:2018asy}. The latter works associate different forms of entanglement with different geometrical (rather than topological) forms -- the polytopes, which appear as spaces characterizing irreducible action of SLOCC operators on local density matrices. The polytope classification detects the W class in the 3-qubit case and considerably more classes of genuine 4-qubit entanglement (seven versus two) as compared to the connectivity classification. However, the topological one has an advantage of the entanglement information being represented by the state itself, rather than by its complex image in an auxiliary space. The latter difference should be relevant for entanglement engineering.

The remainder of this paper is organized as follows. In section~\ref{sec:reptheory} we will review a connection of the classification of the bipartite entanglement with the representation theory and highest weight vectors of irreducible representations. Section~\ref{sec:tqft} gives a brief review of the axioms of TQFT. Necessary technical tools will be introduced in section~\ref{sec:entanglement}, in which the topological entanglement is discussed as the avatar of the quantum one. We will review how entanglement can be classified in one, two and three-dimensional TQFT. The one-dimensional case will contain the simplest realization of the main idea that will subsequently be embedded in higher dimensions. In the following part of the paper we will focus on the 3D case with boundaries being 2-spheres with punctures. Since the simple connectivity classification does not properly capture the classes of bipartite entanglement for generic qudits, in section~\ref{sec:Jones-Wenzl} we will elaborate it by introducing necessary projectors. In section~\ref{sec:3partite} we will discuss the classification of the 3-partite entanglement and the fate of the W-type entanglement in the topological classification. We will summarize the results of the paper in section~\ref{sec:conclusions}.

\section{Entanglement and representation theory}
\label{sec:reptheory}

Before discussing the topological interpretation of entanglement let us first review the connection of the latter with representations of unitary groups. We will illustrate it by the example of $SU(2)$ and focus on the bipartite entanglement. Further details and generalizations can be found elsewhere. See~\cite{Walter:2013ent} for alternative ways of classifying entanglement via irreducible representations.

For $SU(2)$, a qubit is realized by the fundamental representation of the group. In the case of a pair of qubits the states are decomposed into a pair of irreducible representations ${\bf 1}$ and ${\bf 3}$:
\be
\begin{array}{r}
{\bf 1}:\\
\\
{\bf 3}: 
\end{array} \qquad
\begin{array}{ccc}
& \frac{1}{\sqrt{2}}\left(|01\rangle - |10\rangle\right) & \\
\\
|00\rangle & \frac{1}{\sqrt{2}}\left(|01\rangle + |10\rangle\right) & |11\rangle
\end{array}
\ee
The highest weight vectors of the two representations correspond to two classes of SLOCC entanglement (Bell and separable). 

In the case of a pair of qutrits (here we label spin projections as $+1\to 0$, $0\to 1$, and $-1\to 2$), the basis of irreducible representations can be chosen as
\be
\begin{array}{r}
{\bf 1}:\\
\\
{\bf 3}: \\
\\
{\bf 5}:
\end{array} \qquad
\begin{array}{ccccc}
&  & \frac{1}{\sqrt{3}}\left(|0,2\rangle -|1,1\rangle + |2,0\rangle\right) &  &\\
\\
& \frac{1}{\sqrt{2}}\left(|01\rangle - |10\rangle\right) & \frac{1}{\sqrt{2}}\left(|02\rangle - |20\rangle\right) &  \frac{1}{\sqrt{2}}\left(|21\rangle - |12\rangle\right) & \\
\\
|00\rangle & \frac{1}{\sqrt{2}}\left(|01\rangle + |10\rangle\right) &  \frac{1}{\sqrt{6}}\left(|0,2\rangle +2|1,1\rangle + |2,0\rangle\right) & \frac{1}{\sqrt{2}}\left(|21\rangle + |12\rangle\right) & |22\rangle 
\end{array}
\ee
Computing the reduced density matrices of the highest weight vectors, one finds that for ${\bf 1}$, ${\bf 3}$, and ${\bf 5}$ representations they have ranks three, two, and one respectively. In other words the highest weights belong to the three possible SLOCC classes of qutrits.

For a general qudit pair characterized by spins $(J_1,J_2)$ one expects that there are $2\min(J_1,J_2)+1$ classes of entanglement. This number matches the number of irreducible representations of $SU(2)$ in the expansion of the tensor product $J_1\otimes J_2$. What remains to be checked is whether the reduced density matrices of the highest weight vectors of the irreps span ranks ranging from $1$ to $2\min(J_1,J_2)+1$. Indeed, it is not hard to see that the reduced matrices of the highest weight vectors are of the Schmidt form, with the number of the (Clebsch-Gordan) coefficients equal to $2\min(J_1,J_2)+1$. Moreover, each of these coefficients is nonvanishing, as can be checked from their explicit form. Therefore, all the SLOCC classes of bipartite entanglement are labeled by the highest weight vectors of $SU(2)$.

Let us now check the case of three qubits. The basis of the expansion of the tensor product ${\bf 2}\otimes {\bf 2}\otimes {\bf 2}$ can be chosen in the following form:
\be
\begin{array}{r}
{\bf 2}:\\
\\
{\bf 4}: \\
\\
{\bf 2}:
\end{array} \qquad
\begin{array}{cccc}
 &  \frac{1}{\sqrt{2}}\left(|001\rangle + |100\rangle\right)\,, & \frac{1}{\sqrt{2}}\left(|110\rangle + |011\rangle\right)\,,  & \\ \\
 |000\rangle\,, & \frac{1}{\sqrt{3}}\left(|001\rangle + |010\rangle + |100\rangle\right) & \frac{1}{\sqrt{3}}\left(|110\rangle + |101\rangle + |011\rangle\right) \,, & |111\rangle\,, \\ \\
  & \frac{1}{\sqrt{6}}\left(|001\rangle - 2|010\rangle + |100\rangle\right)\,, & \frac{1}{\sqrt{6}}\left(|110\rangle - 2|101\rangle + |011\rangle\right)\,. &  
\end{array}
\ee
One sees that the highest weight vectors span all but Greenberger-Horne-Zeilinger (GHZ) classes of the SLOCC entanglement~\cite{Dur:2000zz}. In particular, states like $|001\rangle-2|010\rangle+|100\rangle$ are of W type. The fact that W class and not GHZ appears in the highest weights is not obvious a priori since W entanglement is more special: unlike the GHZ one, it corresponds to a subset of measure zero in the full Hilbert space. 


\section{Topological quantum field theories}
\label{sec:tqft}

In this section we briefly introduce the axiomatic definition of TQFT~\cite{Atiyah:1989vu,Witten:1988ze}, given in terms of category theory. This definition is the main framework that will be used in this paper. The reader can also check~\cite{Kauffman:2013bh,Melnikov:2018zfn,Melnikov:2020mno} for further details.

A TQFT is a map (functor) from (the category of) topological spaces (or the category of cobordisms) to (the category of) linear spaces. Linear maps are morphisms in the category of the linear spaces and morphisms in the topological space category are cobordisms -- topological spaces whose boundaries are the objects of the same category.

Practically this means the following. Let us consider a set of hypersurfaces $\Sigma$. To each of these $\Sigma$ the TQFT map $F$ associates a Hilbert space $\Hc_\Sigma$,
\be
\begin{array}{c}
     \includegraphics[scale=0.3]{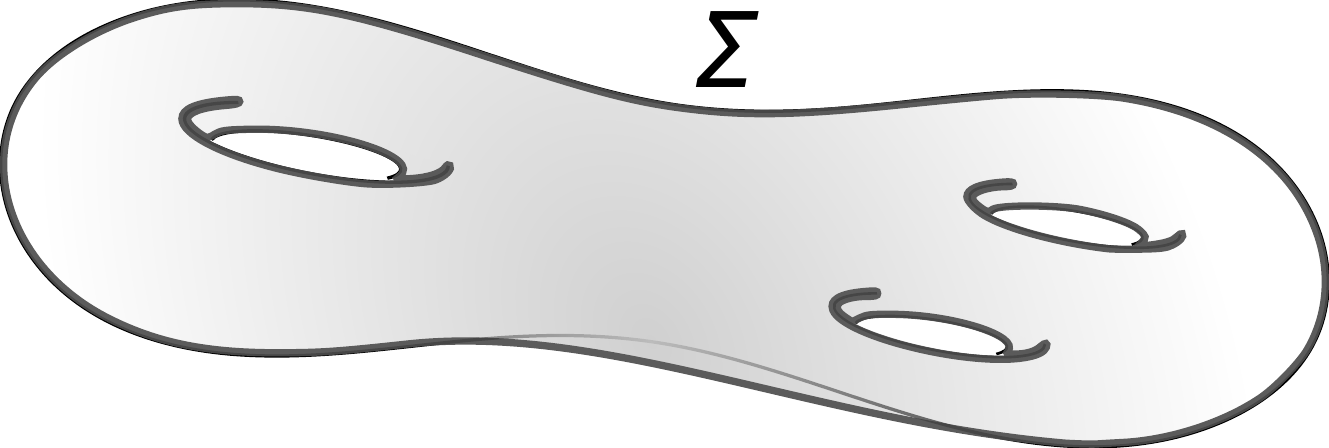} 
\end{array}
\quad \to \quad \Hc_\Sigma \ = \ F(\Sigma)\,.
\ee
If the dimension of $\Sigma$ is $D$, then any $(D+1)$-dimensional manifold $\Mc$, whose boundary is $\Sigma$, $\Sigma=\partial\Mc$, is mapped by TQFT to a vector in $\Hc$,
\be
F:\ 
\begin{array}{c}
\includegraphics[scale=0.15]{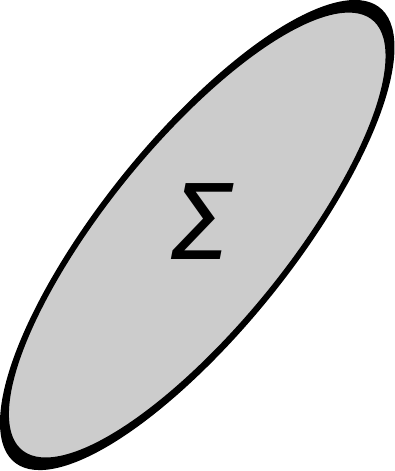}
\end{array} \ \to \  \Hc_\Sigma\,, \qquad  \begin{array}{c}
\includegraphics[scale=0.12]{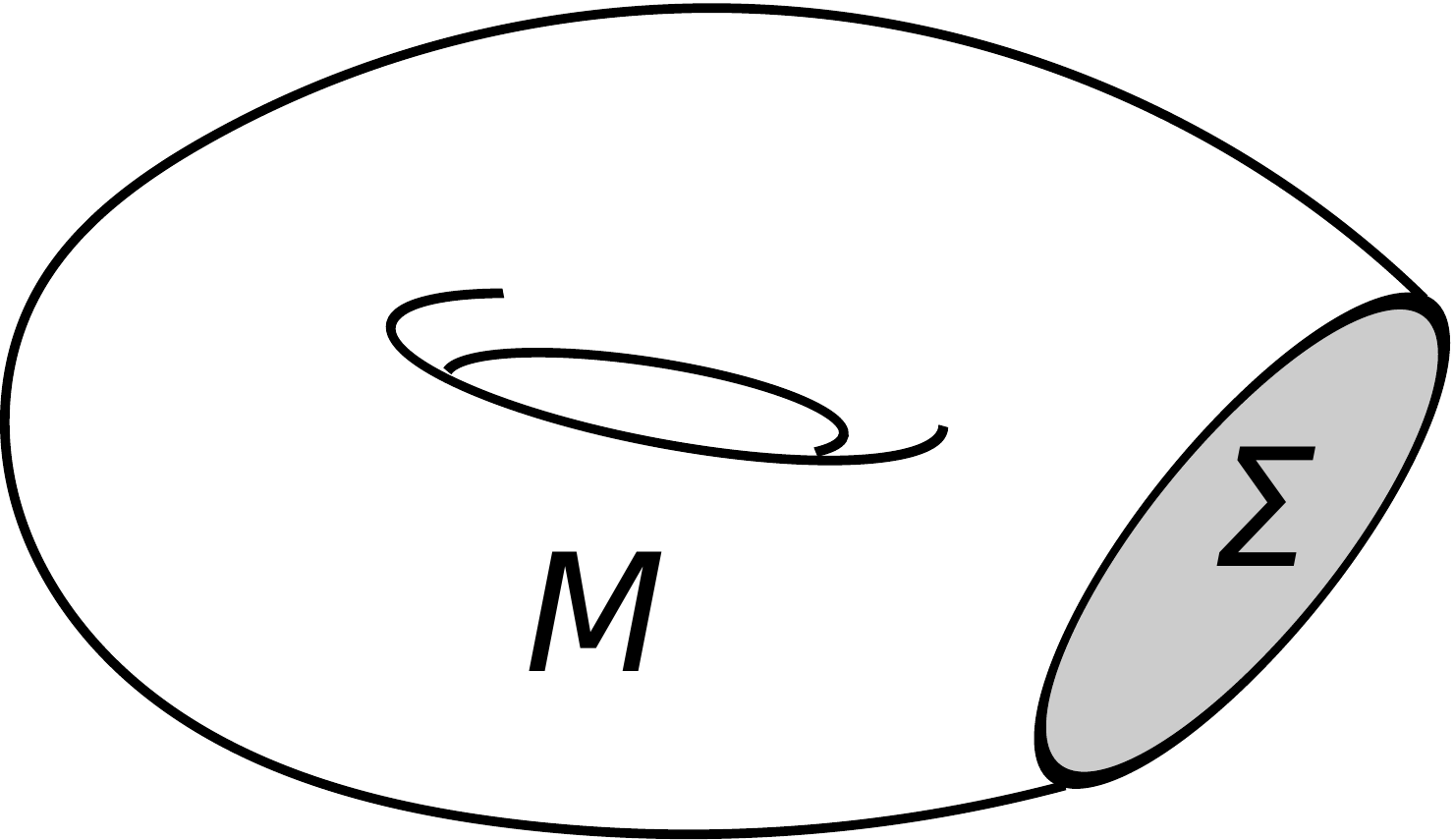}
\end{array} \quad \to \quad  v = F(\Mc)\in \Hc_\Sigma\,.
\ee
In other words, if we have a $D$-dimensional hypersurface representing a Hilbert space, then all possible $(D+1)$-dimensional spaces that can be glued to this hypersurface are vectors in the Hilbert space.

Cobordisms of hypersurfaces, that is $(D+1)$-dimensional spaces $\Mc$ with boundaries $\partial\Mc=\Sigma_1\cup \Sigma_2$, are mapped to linear maps between spaces $\Hc_{\Sigma_1}$ and $\Hc_{\Sigma_2}$. In particular, $\Mc$ with $\partial\Mc = \Sigma\cup\overline{\Sigma}$, where by $\overline{\Sigma}$ we understand a space with an opposite orientation to $\Sigma$, such that $F(\overline{\Sigma})=\Hc_{\overline{\Sigma}}=\Hc_\Sigma^\ast$, the dual space, are mapped to linear operators (matrices) on $\Hc_\Sigma$. The natural composition of cobordisms, which is gluing along the common boundary, is mapped to the composition of linear operators. For example, from $\Mc_1$ and $\Mc_2$, such that $\partial\Mc_1=\Sigma_1\cup\overline{\Sigma_2}$ and $\partial\Mc_2=\Sigma_2\cup\overline{\Sigma_3}$, one can form a cobordism $\Mc_3$ with $\partial\Mc_3=\Sigma_1\cup\overline{\Sigma_3}$,
\be
\begin{array}{c}\includegraphics[scale=0.3]{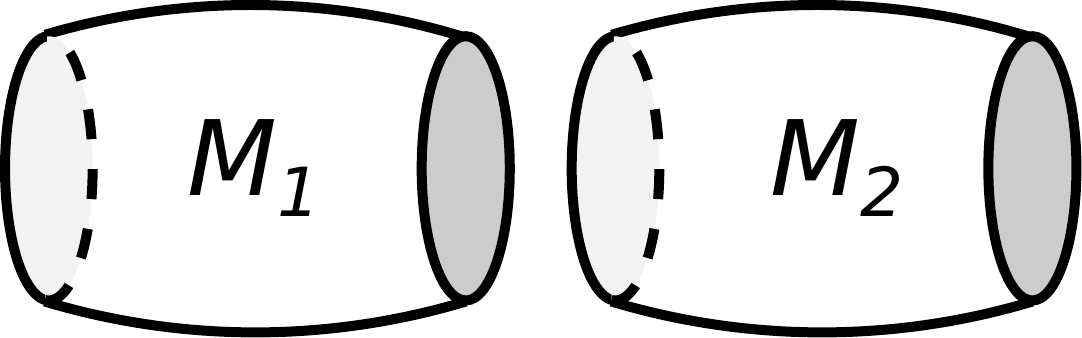}\end{array} \ = \ 
\begin{array}{c}\includegraphics[scale=0.3]{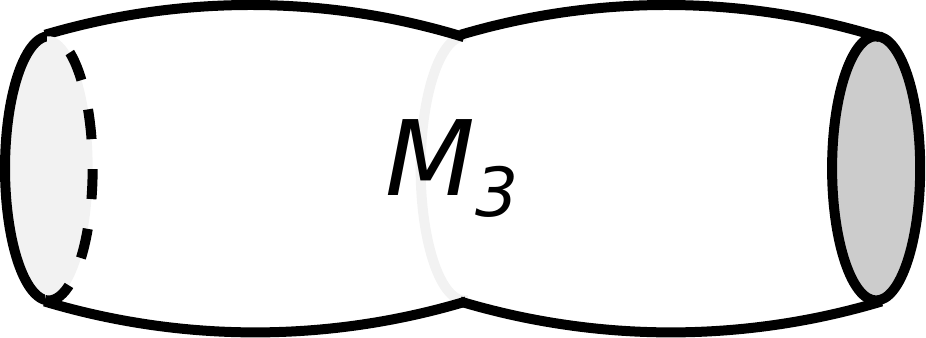} \end{array} \qquad \to \qquad F(\Mc_2)\circ F(\Mc_1) \ = \ F(\Mc_3)\,.
\ee

These definitions are supplemented by the following set of axioms.
\begin{itemize}
    \item A closed $(D+1)$-dimensional space, that is $\Mc$ with $\partial\Mc=\emptyset$ is mapped to a zero-dimensional Hilbert space, that is $\mathbb{C}$. This allows to understand states as images of cobordisms from $\emptyset$ to some $\Sigma$, in the sense of general equivalence of states and linear operators. Hence, one can introduce the internal product of states through cobordisms $\emptyset\to\emptyset$ as the following diagram suggests:
    \be
    \begin{array}{c}
 \includegraphics[scale=0.15]{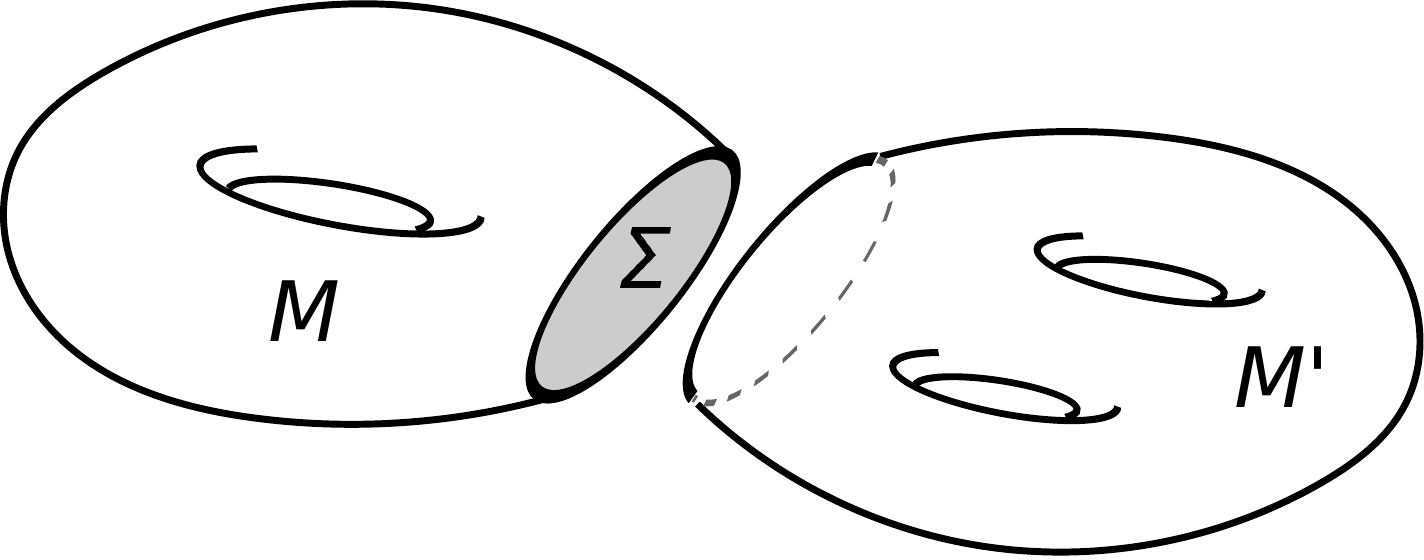}
 \end{array}
 \ = \ 
 \begin{array}{c}
 \includegraphics[scale=0.15]{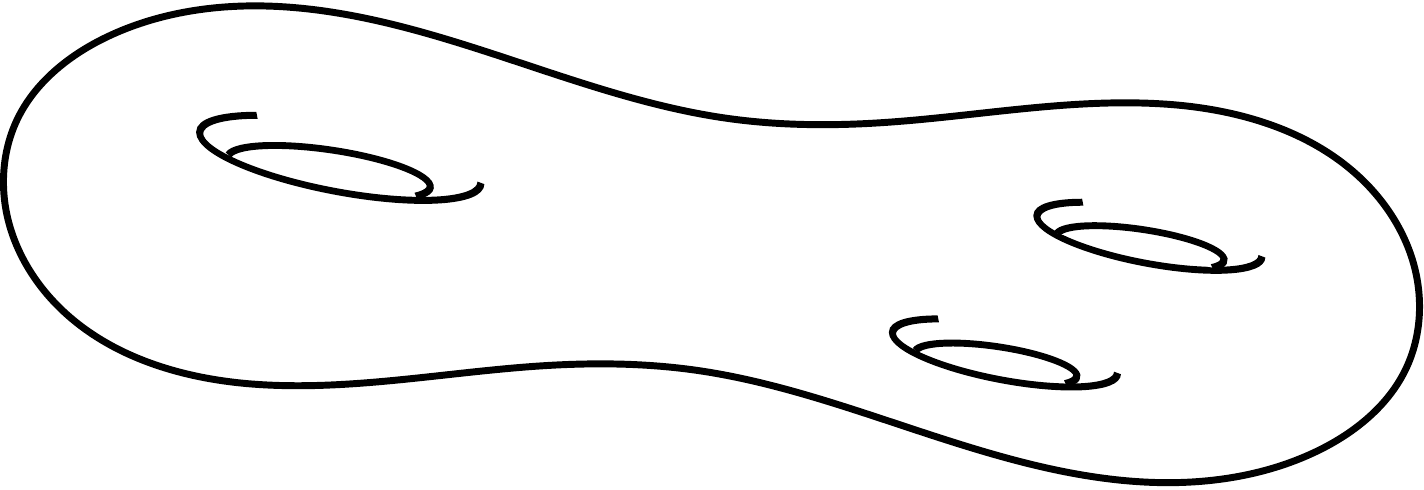}
 \end{array}
  \quad \to\quad  \ u^\ast v\,.
    \ee
    \item Featureless cylinder $I=\Sigma\times[0,1]$ (a trivial evolution of $\Sigma$) is mapped to the identity operator on $\Hc_\Sigma$.
    \item A hypersurface $\Sigma$ consisting of several disconnected components $\Sigma=\Sigma_1\cup\Sigma_2\cup\cdots\cup\Sigma_k$ is mapped to the tensor product of the corresponding vector spaces,
    \be
    \label{separability}
    F(\Sigma_1\cup\Sigma_2\cup\cdots\cup\Sigma_k) \ = \ \Hc_{\Sigma_1}\otimes \Hc_{\Sigma_2}\otimes \cdots\otimes \Hc_{\Sigma_k}\,.
    \ee
\end{itemize}

The above axioms are intended to realized analogs of matrix multiplication and more general operations with tensors in terms of topological spaces. An explicit realization of such a functor can be realized in terms of the functional integral (partition function) of metric independent action functionals. In most of this work, we will assume that the corresponding action is that of a specific Chern-Simons theory.


\section{Entanglement and TQFT}
\label{sec:entanglement}

The axiomatic definition of TQFT is a convenient starting point to discuss entanglement. From the property~(\ref{separability}) of TQFT, there is a  natural presentation for separable,
\be
\label{separable}
|\Psi_{\rm sep}\rangle \ = \ \begin{array}{c}\includegraphics[scale=0.25]{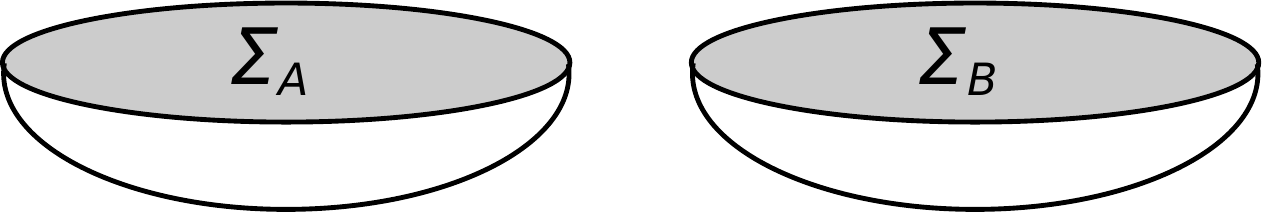}\end{array},
\ee
and entangled
\be
\label{EntState}
 |\Psi_{\rm ent}\rangle \ = \ \begin{array}{c}\includegraphics[scale=0.25]{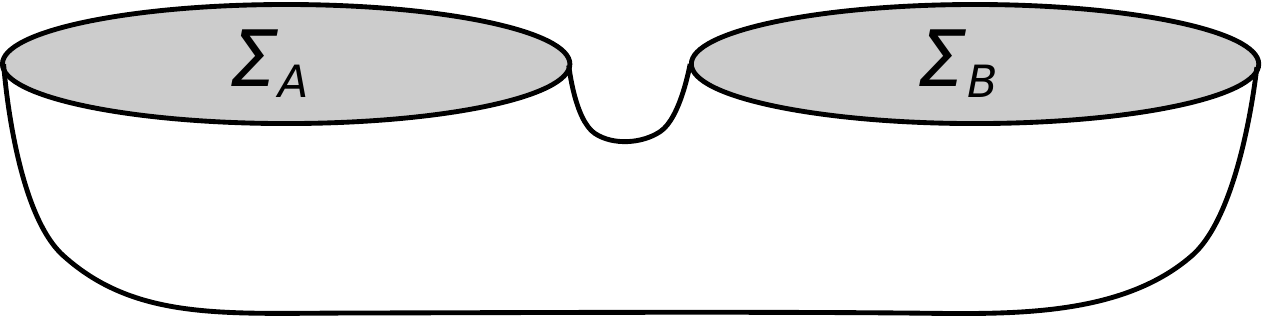}\end{array}
\ee
states~\cite{Melnikov:2018zfn}. Here, the two heuristic diagrams represent two distinct classes of cobordisms connecting two boundaries $\Sigma_A$ and $\Sigma_B$, disconnected and connected. One way to check that the first state is separable and the second is entangled is to compute the von Neumann entropy. The reduced density matrices of the subsystem $A$ for the two states are 
\label{rhoreduced}
 \be
 {\rho}_{\rm sep}(A) \ = \ \left[\begin{array}{c}\includegraphics[scale=0.2]{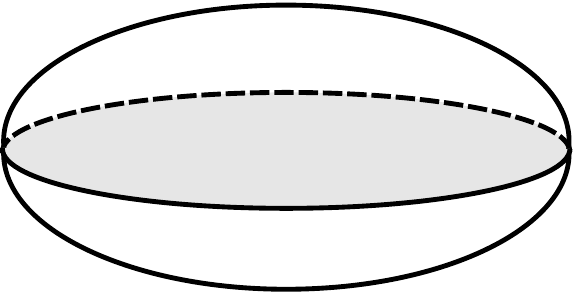}\end{array}\right]^{-1}  \begin{array}{c}\includegraphics[scale=0.25]{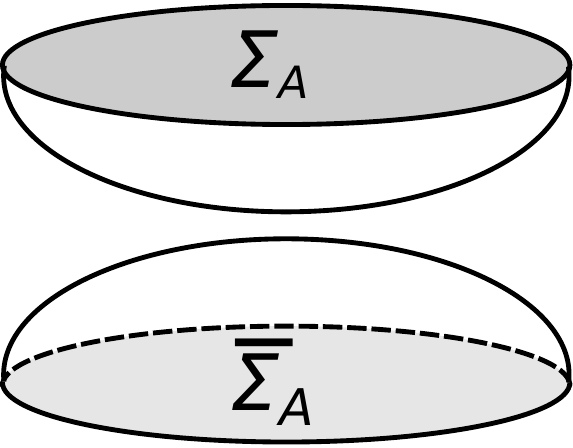}
 \end{array}  
 \quad \text{and} \quad 
 {\rho}_{\rm ent}(A) \ = \ \left[\begin{array}{c}\includegraphics[scale=0.1]{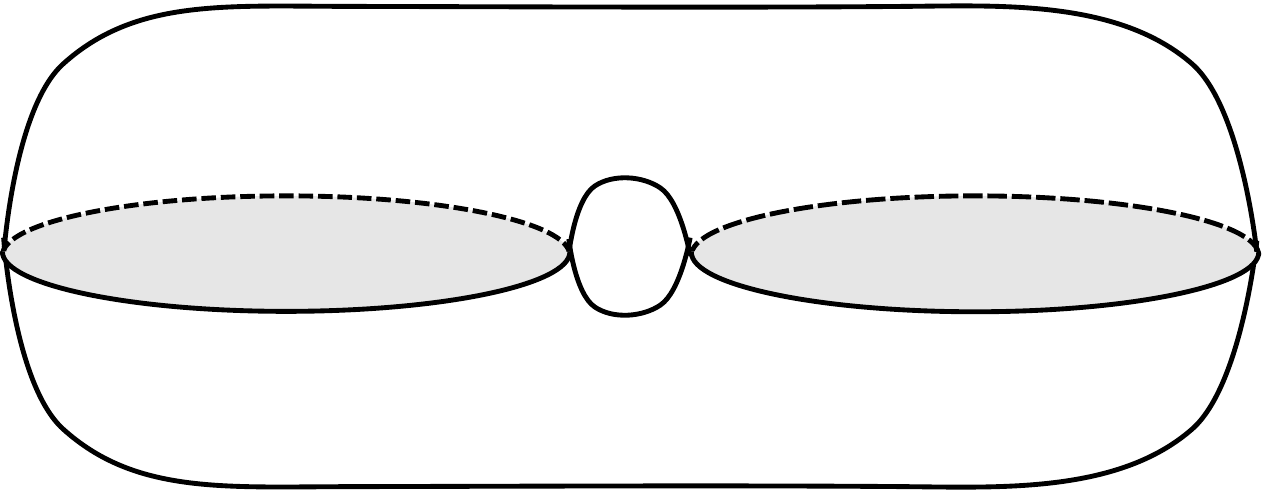}\end{array}\right]^{-1}  \begin{array}{c}\includegraphics[scale=0.25]{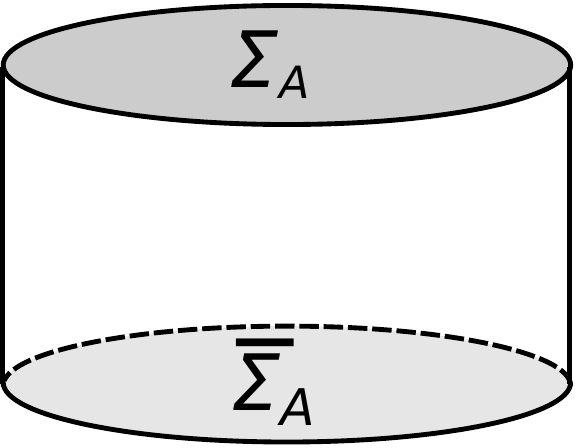}\end{array},
 \ee
 where partial tracing is obtained by gluing together two copies of the same state along $\Sigma_B$. The diagrams in the square brackets are the normalization factors encoded by closed topological spaces -- full traces.
 
 The von Neumann entropy, $S_{\rm E}=-\tr\rho(A)\log\rho(A)$, can be computed using the replica trick. It is not hard to see~\cite{Melnikov:2018zfn,Melnikov:2020mno} that for the first state $\tr\rho_{\rm ent}^n=1$, for any $n$, and $S_{\rm E}=0$, while for the ``entangled'' one it is
\be
\label{SE}
 {S}_{{\rm E}}(A) \ = \ \log \left[\begin{array}{c}\includegraphics[scale=0.25]{donut.pdf}\end{array}\right].
\ee

In order to explicitly compute this number one needs to be more precise about the structure of the corresponding manifold. If in state~(\ref{EntState}) the two boundaries $\Sigma_A$ and $\Sigma_B$ are homeomorphic $\Sigma_A\simeq\Sigma_B=\Sigma$ and connected by a cobordism of the simplest topology $\Sigma\times[0,1]$ then one can understand the reduced density matrix, the cylinder in~(\ref{rhoreduced}), as an identity operator, so the von Neumann entropy is just the logarithm of the trace of the identity over the Hilbert subspace ${\cal H}_A$. The latter equals $\dim{\cal H}_A$, which means that states of type~(\ref{EntState}) are maximally entangled.

One can also consider cobordisms of different topology and multipartite systems, as in the following example: 
\be
\includegraphics[scale=0.2]{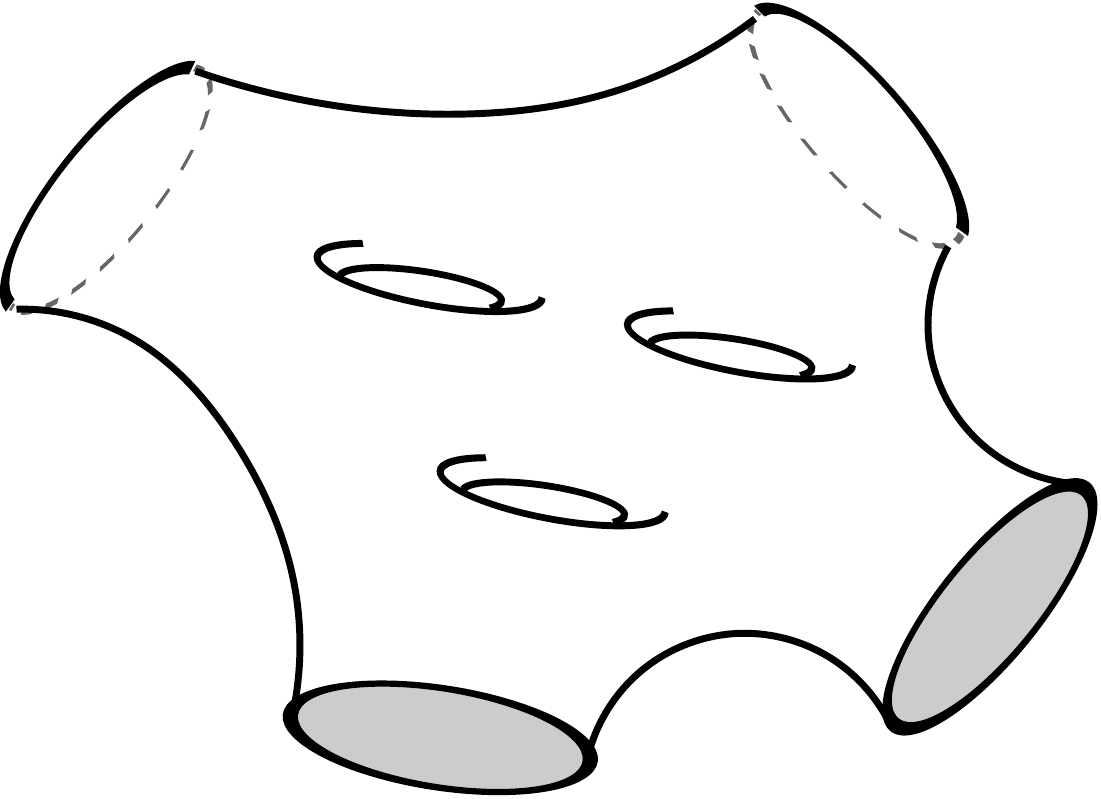}
\ee
A natural question in this case is about the relation between the variation of topology associated to the states and the variation of the properties of such states from the point of view of Quantum Resource Theory. In other words, whether different topologies correspond to qualitatively different entanglement properties and whether the topology makes a difference for different quantum tasks.  

One very well-known classification of states' entanglement and their relevance for different quantum tasks is that by action of Stochastic Local Operations and Classical Communication (SLOCC)~\cite{Dur:2000zz}. In that classification two quantum states are equivalent if they are related by action of local invertible (not necessarily unitary) operators. Hence, the question can be asked as follows: Is there any relation between the SLOCC classification and the topological one?

Below a specific way to construct a relation between the topological and SLOCC classifications will be proposed. Although the main part of this work is focused on a specific three-dimensional TQFT, we will start building the intuition from lower dimensional examples, even though the explicit proposals for those cases will be left for a future work.

\paragraph{One-dimensional TQFT.} In one-dimensional TQFT one has a set of points and lines connecting the points. Points are boundaries of lines and for each line one needs a pair of points, so the number of points must be even. Hilbert spaces, or general $\Sigma$, will be associated with groups of points.

It is natural to think of points as particles, so that the lines connect the same kind of particles, although, in general, the particles can be different. For simplicity we will assume that there is only one type of particle. We will also assume particle number conservation modulo two, that is, two particles cannot fuse and produce a single particle, and that the inverse process is also forbidden. With these restrictions the Hilbert spaces can only exist for groups of even number of particles.

It is clear that a single pair of points may represent only a one-dimensional Hilbert space, since there is only one way two points can be connected by a line. The minimal nontrivial case is achieved with two pairs. Larger Hilbert spaces can be obtained through tensor powers of the simplest Hilbert space.

In the case of the simplest four-point Hilbert space there are three distinct topologies, which can be labeled by three vectors
\be
\label{4pointbasis0}
|e_1\rangle \ = \ \begin{array}{c}
\begin{tikzpicture}[thick]
\fill[pink] (-0.1,-0.1) rectangle (0.1,1.0);
\fill[black] (0,0.0) circle (0.05cm);
\fill[black] (0,0.3) circle (0.05cm);
\fill[black] (0,0.6) circle (0.05cm);
\fill[black] (0,0.9) circle (0.05cm);
\draw (0,0) -- (0.4,0) arc (-90:90:0.15cm) -- (0,0.3);
\draw (0,0.6) -- (0.4,0.6) arc (-90:90:0.15cm) -- (0,0.9);
\end{tikzpicture} 
\end{array}\,, \qquad |e_2\rangle \ = \ \begin{array}{c}
\begin{tikzpicture}[thick]
\fill[pink] (-0.1,-0.1) rectangle (0.1,1.0);
\fill[black] (0,0.0) circle (0.05cm);
\fill[black] (0,0.3) circle (0.05cm);
\fill[black] (0,0.6) circle (0.05cm);
\fill[black] (0,0.9) circle (0.05cm);
\draw (0,0) -- (0.1,0) arc (-90:90:0.45cm) -- (0,0.9);
\draw (0,0.6) -- (0.1,0.6) arc (90:-90:0.15cm) -- (0,0.3);
\end{tikzpicture} 
\end{array}
\ee
and
\be
\label{crossedbaselem}
|e_3\rangle \ = \ \begin{array}{c}
\begin{tikzpicture}[thick]
\fill[pink] (-0.1,-0.1) rectangle (0.1,1.0);
\fill[black] (0,0.0) circle (0.05cm);
\fill[black] (0,0.3) circle (0.05cm);
\fill[black] (0,0.6) circle (0.05cm);
\fill[black] (0,0.9) circle (0.05cm);
\draw (0,0) -- (0.4,0) arc (-90:90:0.3cm) -- (0,0.6);
\draw (0,0.3) -- (0.4,0.3) arc (-90:90:0.3cm) -- (0,0.9);
\end{tikzpicture} 
\end{array}\,,
\ee
where the shaded rectangles indicate that the points are joined in the same set $\Sigma$. We can assume that the three vectors above form a basis in the Hilbert space.

According to the axioms of TQFT, scalar products are obtained by gluing pairs of states along the boundary (here shaded rectangles). In order to operate with the above Hilbert space, we need to define the scalar products, for example,
\be
\langle e_1|e_1\rangle \ = \ \begin{array}{c}\scalebox{0.6}{
\begin{tikzpicture}[very thick]
\fill[black] (0,0.0) circle (0.05cm);
\fill[black] (0,0.3) circle (0.05cm);
\fill[black] (0,0.6) circle (0.05cm);
\fill[black] (0,0.9) circle (0.05cm);
\draw (0,0) -- (0.4,0) arc (-90:90:0.15cm) -- (0,0.3);
\draw (0,0.6) -- (0.4,0.6) arc (-90:90:0.15cm) -- (0,0.9);
\draw (0,0) -- (-0.4,0) arc (-90:-270:0.15cm) -- (0,0.3);
\draw (0,0.6) -- (-0.4,0.6) arc (-90:-270:0.15cm) -- (0,0.9);
\end{tikzpicture}} 
\end{array}\,, \qquad \langle e_1|e_2\rangle \ = \ \begin{array}{c}\scalebox{0.6}{
\begin{tikzpicture}[very thick]
\fill[black] (0,0.0) circle (0.05cm);
\fill[black] (0,0.3) circle (0.05cm);
\fill[black] (0,0.6) circle (0.05cm);
\fill[black] (0,0.9) circle (0.05cm);
\draw (0,0) -- (0.1,0) arc (-90:90:0.45cm) -- (0,0.9);
\draw (0,0.6) -- (0.1,0.6) arc (90:-90:0.15cm) -- (0,0.3);
\draw (0,0) -- (-0.4,0) arc (-90:-270:0.15cm) -- (0,0.3);
\draw (0,0.6) -- (-0.4,0.6) arc (-90:-270:0.15cm) -- (0,0.9);
\end{tikzpicture}} 
\end{array}
\,, \qquad \langle e_2|e_2\rangle \ = \ \begin{array}{c}\scalebox{0.6}{
\begin{tikzpicture}[very thick]
\fill[black] (0,0.0) circle (0.05cm);
\fill[black] (0,0.3) circle (0.05cm);
\fill[black] (0,0.6) circle (0.05cm);
\fill[black] (0,0.9) circle (0.05cm);
\draw (0,0) -- (0.1,0) arc (-90:90:0.45cm) -- (0,0.9);
\draw (0,0.6) -- (0.1,0.6) arc (90:-90:0.15cm) -- (0,0.3);
\draw (0,0) -- (-0.1,0) arc (-90:-270:0.45cm) -- (0,0.9);
\draw (0,0.6) -- (-0.1,0.6) arc (90:270:0.15cm) -- (0,0.3);
\end{tikzpicture}} 
\end{array}\,.
\ee
We will use a simple rule to evaluate such diagrams: Each closed circle is replaced by the same number, which for the moment we will denote symbolically as $\begin{array}{c}
\scalebox{0.8}{\begin{tikzpicture}[thick]
\draw[black] (0,0.0) circle (0.2cm);
\end{tikzpicture}} 
\end{array}$. This rule is motivated by some known TQFT examples, where such a circle is similar to a trace of the identity operator in a single-particle Hilbert space, that is, to the dimension of that space. It can also be related to some integrable models and Temperley-Lieb algebra, where such a parameter is called fugacity.

Computing the scalar products, one notices that the above basis is not orthonormal. An orthonormal basis can be constructed through a standard procedure giving, for example,
\be
\label{4pointbasis}
|0\rangle \ = \ \frac{1}{\begin{array}{c}
\begin{tikzpicture}[thick]
\draw[black] (0,0.0) circle (0.2cm);
\end{tikzpicture} 
\end{array}}|e_1\rangle\,, \qquad |1\rangle \ = \ \frac{1}{\sqrt{(\begin{array}{c}
\begin{tikzpicture}[thick]
\draw[black] (0,0.0) circle (0.2cm);
\end{tikzpicture} 
\end{array})^2-1}}\left(|e_2\rangle - \frac{1}{\begin{array}{c}
\begin{tikzpicture}[thick]
\draw[black] (0,0.0) circle (0.2cm);
\end{tikzpicture} 
\end{array}} |e_1\rangle\right) 
\ee
and
\be
\label{basvector3}
|2\rangle \ = \ \sqrt{\frac{\begin{array}{c}
\begin{tikzpicture}[thick]
\draw[black] (0,0.0) circle (0.2cm);
\end{tikzpicture} 
\end{array}+1}{\begin{array}{c}
\begin{tikzpicture}[thick]
\draw[black] (0,0.0) circle (0.2cm);
\end{tikzpicture} 
\end{array}(\begin{array}{c}
\begin{tikzpicture}[thick]
\draw[black] (0,0.0) circle (0.2cm);
\end{tikzpicture} 
\end{array}-1)(\begin{array}{c}
\begin{tikzpicture}[thick]
\draw[black] (0,0.0) circle (0.2cm);
\end{tikzpicture} 
\end{array}+2)}}\left(|e_3\rangle - |0\rangle - \frac{\begin{array}{c}
\begin{tikzpicture}[thick]
\draw[black] (0,0.0) circle (0.2cm);
\end{tikzpicture} 
\end{array}-1}{\sqrt{(\begin{array}{c}
\begin{tikzpicture}[thick]
\draw[black] (0,0.0) circle (0.2cm);
\end{tikzpicture} 
\end{array})^2-1}} |1\rangle\right).
\ee

Let us discuss possible types of entanglement in a system composed of two simplest Hilbert spaces. Starting from the idea that entanglement can be labeled by different topologies connecting $\Sigma_A$ and $\Sigma_B$, as in~(\ref{separable}) and~(\ref{EntState}), one can draw all possible connections between points in the two subsystems. Recall that local invertible operations should yield SLOCC-equivalent states, so we must mod out the topologies by local (within either $\Sigma_A$ or $\Sigma_B$) permutations of points. One is then left with the following three topologies:
\be
\label{2quadruples}
\begin{array}{c}
\begin{tikzpicture}[thick]
\fill[pink] (-0.1,-0.1) rectangle (0.1,1.0);
\fill[cyan] (1.1,-0.1) rectangle (1.3,1.0);
\fill[black] (0,0.0) circle (0.05cm);
\fill[black] (0,0.3) circle (0.05cm);
\fill[black] (0,0.6) circle (0.05cm);
\fill[black] (0,0.9) circle (0.05cm);
\draw (0,0) -- (0.4,0) arc (-90:90:0.15cm) -- (0,0.3);
\draw (0,0.6) -- (0.4,0.6) arc (-90:90:0.15cm) -- (0,0.9);
\fill[black] (1.2,0.0) circle (0.05cm);
\fill[black] (1.2,0.3) circle (0.05cm);
\fill[black] (1.2,0.6) circle (0.05cm);
\fill[black] (1.2,0.9) circle (0.05cm);
\draw (1.2,0) -- (0.8,0) arc (-90:-270:0.15cm) -- (1.2,0.3);
\draw (1.2,0.6) -- (0.8,0.6) arc (-90:-270:0.15cm) -- (1.2,0.9);
\end{tikzpicture} 
\end{array}
\,, \qquad 
\begin{array}{c}
\begin{tikzpicture}[thick]
\fill[pink] (-0.1,-0.1) rectangle (0.1,1.0);
\fill[cyan] (1.1,-0.1) rectangle (1.3,1.0);
\fill[black] (0,0.0) circle (0.05cm);
\fill[black] (0,0.3) circle (0.05cm);
\fill[black] (0,0.6) circle (0.05cm);
\fill[black] (0,0.9) circle (0.05cm);
\draw (0,0) -- (0.4,0) -- (1.2,0.0);
\draw (0,0.6) -- (0.4,0.6) arc (-90:90:0.15cm) -- (0,0.9);
\fill[black] (1.2,0.0) circle (0.05cm);
\fill[black] (1.2,0.3) circle (0.05cm);
\fill[black] (1.2,0.6) circle (0.05cm);
\fill[black] (1.2,0.9) circle (0.05cm);
\draw (1.2,0.3) -- (0,0.3);
\draw (1.2,0.6) -- (0.8,0.6) arc (-90:-270:0.15cm) -- (1.2,0.9);
\end{tikzpicture} 
\end{array}
\qquad \text{and}\qquad  
\begin{array}{c}
\begin{tikzpicture}[thick]
\fill[pink] (-0.1,-0.1) rectangle (0.1,1.0);
\fill[cyan] (1.1,-0.1) rectangle (1.3,1.0);
\fill[black] (0,0.0) circle (0.05cm);
\fill[black] (0,0.3) circle (0.05cm);
\fill[black] (0,0.6) circle (0.05cm);
\fill[black] (0,0.9) circle (0.05cm);
\draw (0,0) -- (1.2,0);
\draw (0,0.6) -- (1.2,0.6);
\fill[black] (1.2,0.0) circle (0.05cm);
\fill[black] (1.2,0.3) circle (0.05cm);
\fill[black] (1.2,0.6) circle (0.05cm);
\fill[black] (1.2,0.9) circle (0.05cm);
\draw (1.2,0.3) -- (0,0.3);
\draw (1.2,0.9) -- (0,0.9);
\end{tikzpicture} 
\end{array}\,,
\ee
where the shaded rectangles indicate the two subsystems.

Since we do not want to distinguish the permutations of the endpoints, the diagrams are labeled by two-node graphs, whose $2\times 2$ adjacency matrices have rows and columns adding up to four (the number of points). We have, consequently, only three such matrices,
\be
\left(
\begin{array}{cc}
   4  &  0 \\
   0  &  4 
\end{array}
\right), \qquad
\left(
\begin{array}{cc}
   2  &  2 \\
   2  &  2 
\end{array}
\right) \qquad
\text{and}
\qquad
\left(
\begin{array}{cc}
   0  &  4 \\
   4  &  0 
\end{array}
\right),
\ee
corresponding to the above diagrams. We will call such diagrams the connectomes, borrowing the terminology from neuroscience. More generally, the connectomes are equivalence classes labeled by the adjacency matrices, of which the above diagrams are the simplest representatives.

It turns out that the first and the second states in~(\ref{2quadruples}) are equivalent. This can be seen explicitly in the basis~(\ref{4pointbasis}) and~(\ref{basvector3}), in which the first two states are annihilated by either projectors $|1\rangle\langle1|$ or $|2\rangle\langle 2|$ applied on the left or on the right state. This is a consequence of the triviality of the Hilbert space of a pair of points and can be explained as follows.

In the second state we can insert the completeness relation,
\be
\begin{array}{c}
\begin{tikzpicture}[thick]
\fill[pink] (-0.1,-0.1) rectangle (0.1,1.0);
\fill[cyan] (1.1,-0.1) rectangle (1.3,1.0);
\fill[black] (0,0.0) circle (0.05cm);
\fill[black] (0,0.3) circle (0.05cm);
\fill[black] (0,0.6) circle (0.05cm);
\fill[black] (0,0.9) circle (0.05cm);
\draw (0,0) -- (0.4,0) -- (1.2,0.0);
\draw (0,0.6) -- (0.4,0.6) arc (-90:90:0.15cm) -- (0,0.9);
\fill[black] (1.2,0.0) circle (0.05cm);
\fill[black] (1.2,0.3) circle (0.05cm);
\fill[black] (1.2,0.6) circle (0.05cm);
\fill[black] (1.2,0.9) circle (0.05cm);
\draw (1.2,0.3) -- (0,0.3);
\draw (1.2,0.6) -- (0.8,0.6) arc (-90:-270:0.15cm) -- (1.2,0.9);
\end{tikzpicture} 
\end{array} \ = \ \sum\limits_n
\begin{array}{c}
\begin{tikzpicture}[thick]
\fill[pink] (-0.1,-0.1) rectangle (0.1,1.0);
\fill[cyan] (2.5,-0.1) rectangle (2.7,1.0);
\fill[olive] (0.65,-0.1) rectangle (0.85,0.4);
\fill[olive] (1.75,-0.1) rectangle (1.95,0.4);
\draw (0,0) -- (0.75,0.0);
\draw (0,0.6) -- (0.4,0.6) arc (-90:90:0.15cm) -- (0,0.9);
\draw (0.75,0.3) -- (0,0.3);
\draw (2.6,0.6) -- (2.2,0.6) arc (-90:-270:0.15cm) -- (2.6,0.9);
\draw (1.85,0) -- (2.6,0.0);
\draw (1.85,0.3) -- (2.6,0.3);
\draw (1.3,0.15) node {$|n\rangle\langle n|$};
\fill[black] (0,0.0) circle (0.05cm);
\fill[black] (0,0.3) circle (0.05cm);
\fill[black] (0,0.6) circle (0.05cm);
\fill[black] (0,0.9) circle (0.05cm);
\fill[black] (0.75,0.0) circle (0.05cm);
\fill[black] (0.75,0.3) circle (0.05cm);
\fill[black] (1.85,0.0) circle (0.05cm);
\fill[black] (1.85,0.3) circle (0.05cm);
\fill[black] (2.6,0.0) circle (0.05cm);
\fill[black] (2.6,0.3) circle (0.05cm);
\fill[black] (2.6,0.6) circle (0.05cm);
\fill[black] (2.6,0.9) circle (0.05cm);
\end{tikzpicture} 
\end{array}
\ = \ \frac{1}{(\!\begin{array}{c}
\scalebox{0.8}{\begin{tikzpicture}[thick]
\draw[black] (0,0.0) circle (0.2cm);
\end{tikzpicture}} 
\end{array}\!)^2}
\begin{array}{c}
\begin{tikzpicture}[thick]
\fill[pink] (-0.1,-0.1) rectangle (0.1,1.0);
\fill[cyan] (1.1,-0.1) rectangle (1.3,1.0);
\fill[black] (0,0.0) circle (0.05cm);
\fill[black] (0,0.3) circle (0.05cm);
\fill[black] (0,0.6) circle (0.05cm);
\fill[black] (0,0.9) circle (0.05cm);
\draw (0,0) -- (0.4,0) arc (-90:90:0.15cm) -- (0,0.3);
\draw (0,0.6) -- (0.4,0.6) arc (-90:90:0.15cm) -- (0,0.9);
\fill[black] (1.2,0.0) circle (0.05cm);
\fill[black] (1.2,0.3) circle (0.05cm);
\fill[black] (1.2,0.6) circle (0.05cm);
\fill[black] (1.2,0.9) circle (0.05cm);
\draw (1.2,0) -- (0.8,0) arc (-90:-270:0.15cm) -- (1.2,0.3);
\draw (1.2,0.6) -- (0.8,0.6) arc (-90:-270:0.15cm) -- (1.2,0.9);
\end{tikzpicture} 
\end{array}\,,
\ee
where in the last step the fact that the Hilbert space of two points is one-dimensional was used.

More generally, if any two parties are connected by only a pair of lines, they are essentially disconnected,
\be
\label{breaking}
\begin{array}{c}
\begin{tikzpicture}[thick]
\fill[black] (0,0.0) circle (0.05cm);
\fill[black] (0,0.3) circle (0.05cm);
\draw (0,0) -- (0.4,0) -- (1.2,0.0);
\fill[black] (1.2,0.0) circle (0.05cm);
\fill[black] (1.2,0.3) circle (0.05cm);
\draw (1.2,0.3) -- (0,0.3);
\fill[gray] (-0.4,0.15) circle (0.5cm);
\fill[gray] (1.6,0.15) circle (0.5cm);
\end{tikzpicture} 
\end{array}
\qquad \sim \qquad
\begin{array}{c}
\begin{tikzpicture}[thick]
\fill[black] (0,0.0) circle (0.05cm);
\fill[black] (0,0.3) circle (0.05cm);
\draw (0,0) -- (0.4,0) arc (-90:90:0.15cm) -- (0,0.3);
\fill[black] (1.2,0.0) circle (0.05cm);
\fill[black] (1.2,0.3) circle (0.05cm);
\draw (1.2,0) -- (0.8,0) arc (-90:-270:0.15cm) -- (1.2,0.3);
\fill[gray] (-0.4,0.15) circle (0.5cm);
\fill[gray] (1.6,0.15) circle (0.5cm);
\end{tikzpicture} 
\end{array}\,.
\ee

With respect to the three-dimensional Hilbert space with the basis~(\ref{4pointbasis}) and~(\ref{basvector3}), the first two states in~(\ref{2quadruples}) correspond to matrices of rank one, while the last state -- to a matrix of rank three. The problem now is that there seems to be no diagrammatic representation for a matrix of rank two. 

The problem can be solved if one projects out the subspace generated by vector $|2\rangle$. In the present case this will help to reduce the Hilbert space to a subspace captured by the diagrams and will allow working with a more basic case of a qubit, rather than a qutrit. Moreover, such a projection will be natural from the point of view of the embedding of the one-dimensional TQFT into the three-dimensional one that we will consider later. In common choices of three-dimensional TQFT the projection occurs naturally, vector $|2\rangle$ is a linear combination of $|0\rangle$ and $|1\rangle$ through the so-called skein relations. The skein relations indicate linear relations between matrices that appear in representations of the braid group. We can derive the analog of the skein relations in the one-dimensional case as well.

Note that state~(\ref{crossedbaselem}) can be obtained from states~(\ref{4pointbasis0}) by action of a permutation operator. Let us denote $B_{k}$ the generator of the permutation of the $k$th and $(k+1)$th points (counted from top to bottom). The permutation generators must be idempotent $B_k^2=\mathbb{I}$ and satisfy the Yang-Baxter equation,
\be
B_kB_{k+1}B_k \ = \ B_{k+1}B_k B_{k+1}\,.
\ee
Let us define new generators $P_k=\mathbb{I}-B_k$. The Yang-Baxter equation and the idempotency implies that
\be
P_kP_{k\pm 1}P_k + P_{k\pm 1}\ = \ P_{k\pm 1}P_{k}P_{k\pm 1} + P_{k}\,,
\ee
which can be solved as
\be
\label{TLrelation2}
P_kP_{k\pm 1}P_k \ = \ P_k\,.
\ee
Next we note that we have the following natural realization of the above algebra in terms of diagrams~(\ref{2quadruples}),
\be
\mathbb{I} \ = \ \begin{array}{c}
\begin{tikzpicture}[thick]
\draw (0,0) -- (1.2,0);
\draw (0,0.6) -- (1.2,0.6);
\draw (1.2,0.3) -- (0,0.3);
\draw (1.2,0.9) -- (0,0.9);
\fill[black] (1.2,0.0) circle (0.05cm);
\fill[black] (1.2,0.3) circle (0.05cm);
\fill[black] (1.2,0.6) circle (0.05cm);
\fill[black] (1.2,0.9) circle (0.05cm);
\fill[black] (0,0.0) circle (0.05cm);
\fill[black] (0,0.3) circle (0.05cm);
\fill[black] (0,0.6) circle (0.05cm);
\fill[black] (0,0.9) circle (0.05cm);
\end{tikzpicture} 
\end{array}, \qquad 
P_1 \ = \ \begin{array}{c}
\begin{tikzpicture}[thick]
\draw (0,0) -- (0.4,0) -- (1.2,0.0);
\draw (0,0.6) -- (0.4,0.6) arc (-90:90:0.15cm) -- (0,0.9);
\draw (1.2,0.3) -- (0,0.3);
\draw (1.2,0.6) -- (0.8,0.6) arc (-90:-270:0.15cm) -- (1.2,0.9);
\fill[black] (1.2,0.0) circle (0.05cm);
\fill[black] (1.2,0.3) circle (0.05cm);
\fill[black] (1.2,0.6) circle (0.05cm);
\fill[black] (1.2,0.9) circle (0.05cm);
\fill[black] (0,0.0) circle (0.05cm);
\fill[black] (0,0.3) circle (0.05cm);
\fill[black] (0,0.6) circle (0.05cm);
\fill[black] (0,0.9) circle (0.05cm);
\end{tikzpicture} 
\end{array},\quad
P_2 \ = \begin{array}{c}
\begin{tikzpicture}[thick]
\draw (0,0) -- (1.2,0.0);
\draw (0,0.3) -- (0.4,0.3) arc (-90:90:0.15cm) -- (0,0.6);
\draw (1.2,0.9) -- (0,0.9);
\draw (1.2,0.3) -- (0.8,0.3) arc (-90:-270:0.15cm) -- (1.2,0.6);
\fill[black] (0,0.0) circle (0.05cm);
\fill[black] (0,0.3) circle (0.05cm);
\fill[black] (0,0.6) circle (0.05cm);
\fill[black] (0,0.9) circle (0.05cm);
\fill[black] (1.2,0.0) circle (0.05cm);
\fill[black] (1.2,0.3) circle (0.05cm);
\fill[black] (1.2,0.6) circle (0.05cm);
\fill[black] (1.2,0.9) circle (0.05cm);
\end{tikzpicture} 
\end{array}, \quad \cdots
\ee
Obviously, $P_k$ are the generators of the Temperley-Lieb algebra, which must satisfy relation~(\ref{TLrelation2}). As such they also satisfy another relation,
\be
P_k^2 \ = \ \begin{array}{c}
\scalebox{0.8}{\begin{tikzpicture}[thick]
\draw[black] (0,0.0) circle (0.2cm);
\end{tikzpicture}} 
\end{array} P_k\,,
\ee
which is apparent from the diagrammatic presentation. Then the permutation condition $B_k^2=\mathbb{I}$ fixes the parameter,
\be
\begin{array}{c}
\scalebox{0.8}{\begin{tikzpicture}[thick]
\draw[black] (0,0.0) circle (0.2cm);
\end{tikzpicture}} 
\end{array} \ = \ -2\,.
\ee
Comparing with~(\ref{basvector3}) one sees indeed that vector $|2\rangle$ is null for this value.

Note that in this realization, $B_k$ are diagrams with crossed lines. The relation between the permutation and Temperey-Lieb generators can be cast diagrammatically as
\be
\begin{tikzpicture}[baseline=0,rounded corners=4]
\draw[thick] (0.5,-0.1) -- (0.3,-0.1) -- (-0.3,0.3) -- (-0.5,0.3);
\draw[thick] (-0.5,-0.1) -- (-0.3,-0.1) -- (0.3,0.3) -- (0.5,0.3);
\end{tikzpicture}  \quad = \quad \begin{tikzpicture}[baseline=0]
\draw[thick] (0.5,-0.1) -- (-0.5,-0.1);
\draw[thick] (0.5,0.3) -- (-0.5,0.3);
\end{tikzpicture} \quad + \quad \begin{tikzpicture}[baseline=0]
\draw[thick] (-0.5,-0.1) -- (-0.3,-0.1) arc (-90:90:0.2) -- (-0.5,0.3);
\draw[thick] (0.5,-0.1) -- (0.3,-0.1) arc (-90:-270:0.2) -- (0.5,0.3);
\end{tikzpicture} \ .
\ee
This is a version of the skein relation for the diagrams related to representations of the permutation group. Hence, we showed that it is quite natural, although not completely generic, to project the basic Hilbert space onto a two-dimensional subspace.

If we focus on the TQFTs with skein relations, that is, linear dependence of the diagrams with crossings of the diagrams without crossings, then the case of a pair of quadruple of points illustrates the situation of a pair of qubits with entanglement types characterized by diagrams 
\be
\label{2qubitcase}
\begin{array}{c}
\begin{tikzpicture}[thick]
\fill[black] (0,0.0) circle (0.05cm);
\fill[black] (0,0.3) circle (0.05cm);
\fill[black] (0,0.6) circle (0.05cm);
\fill[black] (0,0.9) circle (0.05cm);
\draw (0,0) -- (0.4,0) arc (-90:90:0.15cm) -- (0,0.3);
\draw (0,0.6) -- (0.4,0.6) arc (-90:90:0.15cm) -- (0,0.9);
\fill[black] (1.2,0.0) circle (0.05cm);
\fill[black] (1.2,0.3) circle (0.05cm);
\fill[black] (1.2,0.6) circle (0.05cm);
\fill[black] (1.2,0.9) circle (0.05cm);
\draw (1.2,0) -- (0.8,0) arc (-90:-270:0.15cm) -- (1.2,0.3);
\draw (1.2,0.6) -- (0.8,0.6) arc (-90:-270:0.15cm) -- (1.2,0.9);
\end{tikzpicture} 
\end{array}
\,, \qquad 
\text{and}\qquad  
\begin{array}{c}
\begin{tikzpicture}[thick]
\fill[black] (0,0.0) circle (0.05cm);
\fill[black] (0,0.3) circle (0.05cm);
\fill[black] (0,0.6) circle (0.05cm);
\fill[black] (0,0.9) circle (0.05cm);
\draw (0,0) -- (1.2,0);
\draw (0,0.6) -- (1.2,0.6);
\fill[black] (1.2,0.0) circle (0.05cm);
\fill[black] (1.2,0.3) circle (0.05cm);
\fill[black] (1.2,0.6) circle (0.05cm);
\fill[black] (1.2,0.9) circle (0.05cm);
\draw (1.2,0.3) -- (0,0.3);
\draw (1.2,0.9) -- (0,0.9);
\end{tikzpicture} 
\end{array}\,,
\ee
corresponding to the separable and Bell (entangled) classes of SLOCC, respectively. The Schmidt decomposition of the above in the basis~(\ref{4pointbasis}) has one and two terms, respectively. Therefore, with a bit of redundancy due to some additional constraints (here triviality of the Hilbert space associated with two points), the connectivity diagrams characterize the main entanglement patterns of a pair of qubits.

Next we must consider higher dimensional Hilbert spaces in bipartite systems. If we consider the Hilbert space of six points, then one has the following types of connectivity diagrams:
\be
\label{6pointdiags}
\begin{array}{c}
\begin{tikzpicture}[thick]
\fill[black] (0,0.0) circle (0.05cm);
\fill[black] (0,0.3) circle (0.05cm);
\fill[black] (0,0.6) circle (0.05cm);
\fill[black] (0,0.9) circle (0.05cm);
\draw (0,0) -- (0.4,0) arc (-90:90:0.15cm) -- (0,0.3);
\draw (0,0.6) -- (0.4,0.6) arc (-90:90:0.15cm) -- (0,0.9);
\fill[black] (1.2,0.0) circle (0.05cm);
\fill[black] (1.2,0.3) circle (0.05cm);
\fill[black] (1.2,0.6) circle (0.05cm);
\fill[black] (1.2,0.9) circle (0.05cm);
\draw (1.2,0) -- (0.8,0) arc (-90:-270:0.15cm) -- (1.2,0.3);
\draw (1.2,0.6) -- (0.8,0.6) arc (-90:-270:0.15cm) -- (1.2,0.9);
\fill[black] (1.2,1.2) circle (0.05cm);
\fill[black] (1.2,1.5) circle (0.05cm);
\fill[black] (0,1.2) circle (0.05cm);
\fill[black] (0,1.5) circle (0.05cm);
\draw (1.2,1.2) -- (0.8,1.2) arc (-90:-270:0.15cm) -- (1.2,1.5);
\draw (0,1.2) -- (0.4,1.2) arc (-90:90:0.15cm) -- (0,1.5);
\end{tikzpicture} 
\end{array}
\,, \qquad 
\begin{array}{c}
\begin{tikzpicture}[thick]
\fill[black] (0,0.0) circle (0.05cm);
\fill[black] (0,0.3) circle (0.05cm);
\fill[black] (0,0.6) circle (0.05cm);
\fill[black] (0,0.9) circle (0.05cm);
\draw (0,0) -- (0.4,0) -- (1.2,0.0);
\draw (0,0.6) -- (0.4,0.6) arc (-90:90:0.15cm) -- (0,0.9);
\fill[black] (1.2,0.0) circle (0.05cm);
\fill[black] (1.2,0.3) circle (0.05cm);
\fill[black] (1.2,0.6) circle (0.05cm);
\fill[black] (1.2,0.9) circle (0.05cm);
\draw (1.2,0.3) -- (0,0.3);
\draw (1.2,0.6) -- (0.8,0.6) arc (-90:-270:0.15cm) -- (1.2,0.9);
\fill[black] (1.2,1.2) circle (0.05cm);
\fill[black] (1.2,1.5) circle (0.05cm);
\fill[black] (0,1.2) circle (0.05cm);
\fill[black] (0,1.5) circle (0.05cm);
\draw (1.2,1.2) -- (0.8,1.2) arc (-90:-270:0.15cm) -- (1.2,1.5);
\draw (0,1.2) -- (0.4,1.2) arc (-90:90:0.15cm) -- (0,1.5);
\end{tikzpicture} 
\end{array}
\,, \qquad 
\begin{array}{c}
\begin{tikzpicture}[thick]
\fill[black] (0,0.0) circle (0.05cm);
\fill[black] (0,0.3) circle (0.05cm);
\fill[black] (0,0.6) circle (0.05cm);
\fill[black] (0,0.9) circle (0.05cm);
\draw (0,0) -- (0.4,0) -- (1.2,0.0);
\draw (0,1.2) -- (0.4,1.2) arc (-90:90:0.15cm) -- (0,1.5);
\fill[black] (1.2,0.0) circle (0.05cm);
\fill[black] (1.2,0.3) circle (0.05cm);
\fill[black] (1.2,0.6) circle (0.05cm);
\fill[black] (1.2,0.9) circle (0.05cm);
\draw (1.2,0.3) -- (0,0.3);
\draw (1.2,1.2) -- (0.8,1.2) arc (-90:-270:0.15cm) -- (1.2,1.5);
\fill[black] (1.2,1.2) circle (0.05cm);
\fill[black] (1.2,1.5) circle (0.05cm);
\fill[black] (0,1.2) circle (0.05cm);
\fill[black] (0,1.5) circle (0.05cm);
\draw (0,0.6) -- (1.2,0.6);
\draw (1.2,0.9) -- (0,0.9);
\end{tikzpicture} 
\end{array}
\,,\qquad \text{and}\qquad  
\begin{array}{c}
\begin{tikzpicture}[thick]
\fill[black] (0,0.0) circle (0.05cm);
\fill[black] (0,0.3) circle (0.05cm);
\fill[black] (0,0.6) circle (0.05cm);
\fill[black] (0,0.9) circle (0.05cm);
\draw (0,0) -- (1.2,0);
\draw (0,0.6) -- (1.2,0.6);
\fill[black] (1.2,0.0) circle (0.05cm);
\fill[black] (1.2,0.3) circle (0.05cm);
\fill[black] (1.2,0.6) circle (0.05cm);
\fill[black] (1.2,0.9) circle (0.05cm);
\draw (1.2,0.3) -- (0,0.3);
\draw (1.2,0.9) -- (0,0.9);
\fill[black] (1.2,1.2) circle (0.05cm);
\fill[black] (1.2,1.5) circle (0.05cm);
\fill[black] (0,1.2) circle (0.05cm);
\fill[black] (0,1.5) circle (0.05cm);
\draw (0,1.2) -- (1.2,1.2);
\draw (1.2,1.5) -- (0,1.5);
\end{tikzpicture} 
\end{array}\,.
\ee
Again, the first two diagrams are equivalent, so there are three independent classes of states up to local operations. The first three diagrams then correspond to matrices of rank one and two, and the last one -- to a matrix of the maximal rank. If the dimension of the associated Hilbert space (maximal rank) were three, then we would be dealing with the case of a pair of qutrits. However, this is not the case if we define the Hilbert space as before. There are five crossing-free diagrams to make the basis:
\be
\label{5basis}
|e_1\rangle \ = \ \begin{array}{c}
\begin{tikzpicture}[thick]
 \fill[black] (0,0.0) circle (0.05cm);
\fill[black] (0,0.3) circle (0.05cm);
\fill[black] (0,0.6) circle (0.05cm);
\fill[black] (0,0.9) circle (0.05cm);
\draw (0,0) -- (0.4,0) arc (-90:90:0.15cm) -- (0,0.3);
\draw (0,0.6) -- (0.4,0.6) arc (-90:90:0.15cm) -- (0,0.9);
\fill[black] (0,1.2) circle (0.05cm);
\fill[black] (0,1.5) circle (0.05cm);
\draw (0,1.2) -- (0.4,1.2) arc (-90:90:0.15cm) -- (0,1.5);
\end{tikzpicture} 
\end{array} , \quad 
|e_2\rangle \ = \ \begin{array}{c}
\begin{tikzpicture}[thick]
\fill[black] (0,0.0) circle (0.05cm);
\fill[black] (0,0.3) circle (0.05cm);
\fill[black] (0,0.6) circle (0.05cm);
\fill[black] (0,0.9) circle (0.05cm);
\draw (0,0) -- (0.1,0) arc (-90:90:0.45cm) -- (0,0.9);
\draw (0,0.6) -- (0.1,0.6) arc (90:-90:0.15cm) -- (0,0.3);
\fill[black] (0,1.2) circle (0.05cm);
\fill[black] (0,1.5) circle (0.05cm);
\draw (0,1.2) -- (0.4,1.2) arc (-90:90:0.15cm) -- (0,1.5);
\end{tikzpicture} 
\end{array}
, \quad
|e_3\rangle \ = \ \begin{array}{c}
\begin{tikzpicture}[thick]
\fill[black] (0,0.0) circle (0.05cm);
\fill[black] (0,0.3) circle (0.05cm);
\fill[black] (0,0.6) circle (0.05cm);
\fill[black] (0,0.9) circle (0.05cm);
\draw (0,0.6) -- (0.1,0.6) arc (-90:90:0.45cm) -- (0,1.5);
\draw (0,1.2) -- (0.1,1.2) arc (90:-90:0.15cm) -- (0,0.9);
\fill[black] (0,1.2) circle (0.05cm);
\fill[black] (0,1.5) circle (0.05cm);
\draw (0,0) -- (0.4,0) arc (-90:90:0.15cm) -- (0,0.3);
\end{tikzpicture} 
\end{array}
, \quad 
|e_4\rangle \ = \  \begin{array}{c}
\begin{tikzpicture}[thick]
\fill[black] (0,0.0) circle (0.05cm);
\fill[black] (0,0.3) circle (0.05cm);
\fill[black] (0,0.6) circle (0.05cm);
\fill[black] (0,0.9) circle (0.05cm);
\draw (0,0.0) -- (0.1,0.0) arc (-90:0:0.45cm) -- (0.55,1.05) arc (0:90:0.45) -- (0,1.5);
\draw (0,1.2) -- (0.1,1.2) arc (90:-90:0.15cm) -- (0,0.9);
\fill[black] (0,1.2) circle (0.05cm);
\fill[black] (0,1.5) circle (0.05cm);
\draw (0,0.3) -- (0.1,0.3) arc (-90:90:0.15cm) -- (0,0.6);
\end{tikzpicture} 
\end{array}
, \quad 
|e_5\rangle \ = \ \begin{array}{c}
\begin{tikzpicture}[thick]
\fill[black] (0,0.0) circle (0.05cm);
\fill[black] (0,0.3) circle (0.05cm);
\fill[black] (0,0.6) circle (0.05cm);
\fill[black] (0,0.9) circle (0.05cm);
\draw (0,0.0) -- (0.1,0.0) arc (-90:0:0.45cm) -- (0.55,1.05) arc (0:90:0.45) -- (0,1.5);
\draw (0,1.2) -- (0.1,1.2) arc (90:0:0.3cm) -- (0.4,0.6) arc (0:-90:0.3) -- (0,0.3);
\fill[black] (0,1.2) circle (0.05cm);
\fill[black] (0,1.5) circle (0.05cm);
\draw (0,0.6) -- (0.1,0.6) arc (-90:90:0.15cm) -- (0,0.9);
\end{tikzpicture} 
\end{array}\,.
\ee
In general $2n$-point Hilbert space has a basis labeled by the elements of Temperley-Lieb algebra $TL_n$, which has $n-1$ generators and the number of elements given by Catalan numbers $C_n=1,2,5,14,\ldots$ for $n=1,2,3,4,\ldots$

For the five-dimensional Hilbert space of~(\ref{5basis}) the diagrams of~(\ref{6pointdiags}) are not enough to capture all the types of the SLOCC entanglement. In particular, they miss the rank three and rank four cases. It is obvious that the situation is even worse for higher dimensional Hilbert spaces, since the number of connectivity diagrams grows linearly with the number of points, while the dimension of the Hilbert space -- exponentially.

As before, to solve the mismatch problem, we will need to project the Hilbert space on a subspace with appropriate dimension. We do this in section~\ref{sec:Jones-Wenzl} by building a more systematic generalization of the qubit example. Instead of the obvious generalization of the diagrams, we will focus on constructing appropriate Hilbert spaces with a general dimension, not only those with dimensions given by the Catalan numbers. The appropriate projectors were originally defined for three-dimensional TQFT, but the results can be reduced to one dimension in the limit where the 3D braiding reduces to ordinary permutation because the 3D construction will essentially be an embedding of the 1D story.

\paragraph{Two-dimensional TQFT.} Closed boundaries $\Sigma$ of two-dimensional manifolds are topological circles $S^1$. The cobordisms are realized by smooth two-dimensional surfaces interpolating between boundary circles. If such a surface is oriented it is called a Seifert surface. Every compact oriented surface with nonempty boundary in three dimensions is a Seifert surface of some link (a collection of circles that might be linked), and there are many distinct surfaces for a given link. Consequently, the topology of the surface determines an equivalence class of quantum states. 

Local operations correspond to gluing surfaces with only two circular boundaries to a chosen boundary circle. If the glued surface has a nontrivial genus then the topology of the original state will change. In general, such operations are expected to reduce entanglement since there is no smooth topological inverse operation, which reduces the genus. Here, we will only give some heuristic arguments supporting this statement, leaving the proof for a future work. Some explicit calculations confirming this property in a three-dimensional TQFT have been recently shown in~\cite{Melnikov:2023nzn}.

We have seen that the state with the simply connected topology is expected to have a density matrix of maximal rank, that is why we expect that adding handles should, in general, reduce the amount of entanglement. A surface of infinite genus is essentially a disconnected topology, corresponding to a separable state. Finite genus surfaces should correspond to local invertible, but nonunitary operations. 

Further details of the correspondence in the two-dimensional case will be studied elsewhere. Here we only comment that the discussion of the one-dimensional case can be embedded in the two-dimensional one if one considers open boundaries. For example, the following diagrams may be considered as the embedding of basis~(\ref{4pointbasis0}) into a two-dimensional TQFT:
\be
|e_1\rangle \ = \ \begin{array}{c}
\begin{tikzpicture}[thick]
\fill[gray] (0,0) -- (0.4,0) arc (-90:90:0.15cm) -- (0,0.3);
\fill[gray] (0,0.6) -- (0.4,0.6) arc (-90:90:0.15cm) -- (0,0.9);
\fill[black] (0,0.0) circle (0.05cm);
\fill[black] (0,0.3) circle (0.05cm);
\fill[black] (0,0.6) circle (0.05cm);
\fill[black] (0,0.9) circle (0.05cm);
\draw (0,0) -- (0.4,0) arc (-90:90:0.15cm) -- (0,0.3);
\draw (0,0.6) -- (0.4,0.6) arc (-90:90:0.15cm) -- (0,0.9);
\end{tikzpicture} 
\end{array}\,, \qquad |e_2\rangle \ = \ \begin{array}{c}
\begin{tikzpicture}[thick]
\fill[gray] (0,0) -- (0.1,0) arc (-90:90:0.45cm) -- (0,0.9);
\fill[white] (0,0.6) -- (0.1,0.6) arc (90:-90:0.15cm) -- (0,0.3);
\fill[black] (0,0.0) circle (0.05cm);
\fill[black] (0,0.3) circle (0.05cm);
\fill[black] (0,0.6) circle (0.05cm);
\fill[black] (0,0.9) circle (0.05cm);
\draw (0,0) -- (0.1,0) arc (-90:90:0.45cm) -- (0,0.9);
\draw (0,0.6) -- (0.1,0.6) arc (90:-90:0.15cm) -- (0,0.3);
\end{tikzpicture} 
\end{array}\,.
\ee
In this case permutation of lines yields twists of ribbons, so one may want to relax the permutation constraint $B_k^2=\mathbb{I}$.

\paragraph{Three-dimensional TQFT.} TQFT in three dimensions are perhaps the most known ones. Chern-Simons theory is a canonical example realizing the TQFT axioms.

In the three-dimensional case we can take $\Sigma$ to be any closed Riemann surface. The simplest examples are sphere $\Sigma=S^2$ and torus $\Sigma=T^2$. The states are three-dimensional manifolds filling the space between spheres, tori, or more general $\Sigma$. Distinct 3D topologies can be classified by the Seifert manifolds -- closed 3D manifolds that can be obtained from the states by gluing 3-balls to all open $S^2$ boundaries, solid tori to all $T^2$, etc. Seifert manifolds can be obtained from a three-sphere $S^3$ by cutting out an arbitrary number of solid tori, twisting the tori by modular transformations, and gluing the twisted tori back -- the surgery operation.

As in two dimensions, one can also consider $\Sigma$ with its own boundaries. The latter can be introduced by removing points from the Riemann surface (introducing punctures). Boundaries of $\Sigma$ (punctures) extend in three-dimensional bulk as curves, connecting punctures at the boundaries. Another possibility is that such curves do not extend to the boundaries and close in the bulk, producing loops. 

In Chern-Simons theory realization such curves are called Wilson lines. Each such line is characterized by an integrable representations $R$ of the appropriate Kac-Moody algebra. This is to say that the lines have their own internal attributes. Moreover, representation theory of Kac-Moody algebra induces the property of fusion in lines or TQFT states, which tells that a pair of lines with associated representations $R_1$ and $R_2$ are equivalent to a linear combination of lines with representations appearing in the expansion of $R_1\otimes R_2$ in irreducible representations. 

Wilson lines realize embedding of 1D TQFT in 3D ones, adding importance to the relative position of the lines, thus relaxing the constraint $B_k^2=\mathbb{I}$ so that $B_k$ became the generators of the braid group, and endowing some of aspects of one-dimensional theories with a specific mathematical meaning. We can consider the example of level $k$ $SU(2)$ Chern-Simons theory and $\Sigma$ being 2-spheres $S^2$. Then the fact that the Hilbert space associated with two points is trivial (one-dimensional) can be related to the properties of holomorphic sections of line bundles on the sphere with two punctures~\cite{Witten:1988hf}. Basis~(\ref{4pointbasis}) obtained from general considerations in one-dimensional TQFT is related to the basis of conformal blocks in the Wess-Zumino conformal field theory with $su(2)_k$ Kac-Moody algebra. State~(\ref{crossedbaselem}) is  a linear combination of the basis states~(\ref{4pointbasis}) through the skein relations
\be
\label{skein}
\begin{array}{c}
\begin{tikzpicture}[baseline=0]
\draw[thick] (0.5,-0.1) -- (0.3,-0.1) -- (0.15,0);
\draw[thick] (-0.15,0.2) -- (-0.3,0.3) -- (-0.5,0.3);
\draw[thick] (-0.5,-0.1) -- (-0.3,-0.1) -- (0.3,0.3) -- (0.5,0.3);
\end{tikzpicture}\end{array}  \quad = \ A\quad \begin{array}{c}\begin{tikzpicture}[baseline=0]
\draw[thick] (-0.5,-0.1) -- (-0.3,-0.1) arc (-90:90:0.2) -- (-0.5,0.3);
\draw[thick] (0.5,-0.1) -- (0.3,-0.1) arc (-90:-270:0.2) -- (0.5,0.3);
\end{tikzpicture}\end{array} \quad +\ A^{-1}\quad\begin{array}{c} \begin{tikzpicture}[baseline=0]
\draw[thick] (0.5,-0.1) -- (-0.5,-0.1);
\draw[thick] (0.5,0.3) -- (-0.5,0.3);
\end{tikzpicture}\end{array}\,.
\ee
As in the skein relations derived in the one-dimensional case, (\ref{skein}) expresses linear relations between operators in the Hilbert space, which means that any operator or state represented by a diagram containing an intersection can be replaced by a linear combination of operators or states with two alternative wirings. For $SU(2)_k$ Chern-Simons they were derived in~\cite{Witten:1988hf}, but they are a direct consequence of the fact that lines in three dimensions should form a representation of the braid group.

Version~(\ref{skein}) of skein relations is due to Kauffman~\cite{Kauffman:1987sta}. Parameter $A$ is expressed in terms of the $SU(2)$ Chern-Simons data as follows:
\be
A^{-4} \equiv q \ = \ \exp\left(\frac{2\pi i}{k+2}\right).
\ee
Finally, if a 3D manifold containing a contractible Wilson loop (homeomorphic to a trivial circle) is mapped to a tensor $T_\circ$ (scalar, vector, operator, etc.) under the TQFT category map, then $T_\circ=\begin{array}{c}
\begin{tikzpicture}[thick]
\draw[black] (0,0.0) circle (0.2cm);
\end{tikzpicture} 
\end{array} T$, where $T$ is the image of a manifold without the loop. In other words, unlinked circles can be replaced by numerical factors. In the $SU(2)$ Chern-Simons TQFT this number is
\be
\label{d}
\begin{array}{c}
\begin{tikzpicture}[thick]
\draw[black] (0,0.0) circle (0.2cm);
\end{tikzpicture} 
\end{array} \equiv d \ = \ - A^2 - A^{-2}\,.
\ee
This can be derived as a consistency relation, applying~(\ref{skein}) to a twist followed by an inverse twist to a pair of lines, which should be equivalent to an identity operation. Number $-d$ is also referred to as ``quantum dimension'' of the fundamental representation of $su(2)_k$ Kac-Moody algebra, or ``fugacity'' in statistical physics models based on Temperley-Lieb algebra representations.

\begin{figure}
    $$\begin{array}{c}\includegraphics[width=0.2\linewidth]{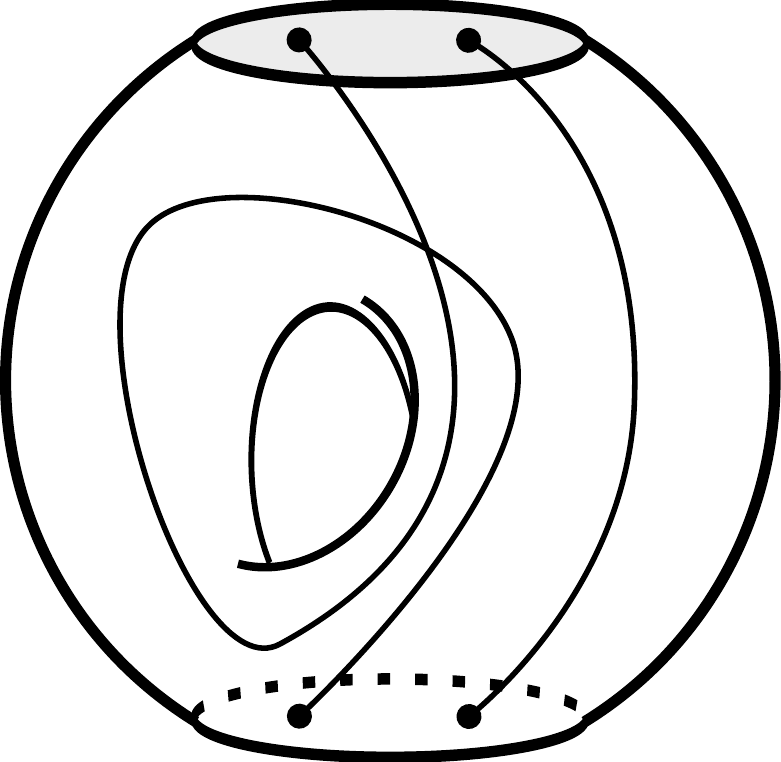}\end{array}
    {\qquad \Huge = \quad \sum\limits_R S^{R0}} \quad \begin{array}{c}\includegraphics[width=0.2\linewidth]{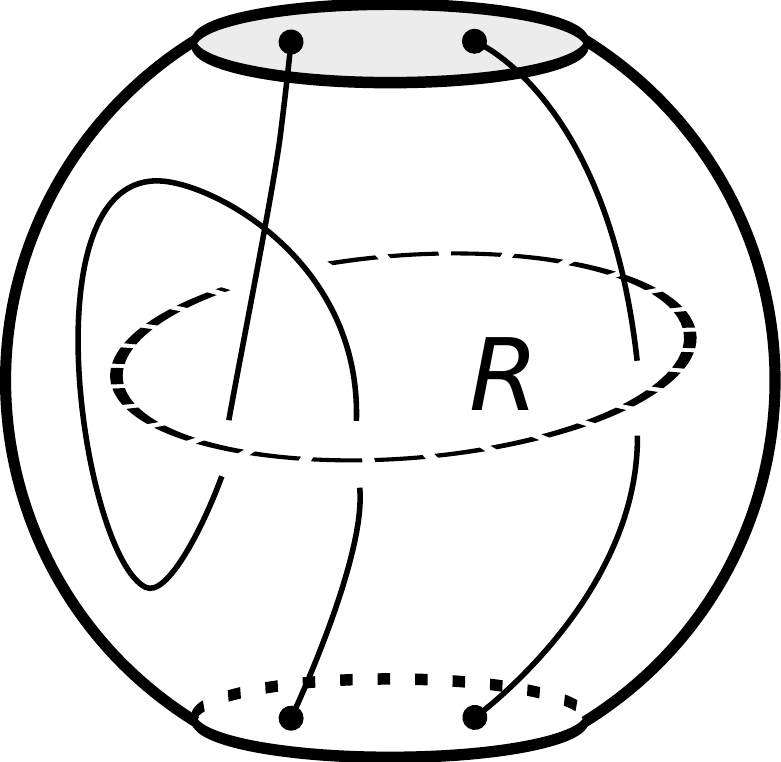}\end{array}$$
    \caption{Surgery expresses nontrivial 3D topologies (with ``holes'') as linear combinations of simpler ones (with holes filled) but with extra Wilson lines. The diagrams are heuristic depictions of 3-dimensional spaces interpolating between two 2-sphere boundaries. $S^{R_1R_2}$ denotes the modular $S$ transformation in the basis labeled by irreducible representations~\cite{Witten:1988hf,Verlinde:1988sn}.}
    \label{fig:surgery}
\end{figure}

Wilson lines and surgeries provide a plethora of realizations of different three-dimensional topologies. Skein relations, together with the circle factorization rule, allow reducing any knotted topology to a linear combination of topologies with unknotted and unlinked Wilson lines. Similarly, surgeries allow reducing nontrivial three-dimensional topologies of Seifert manifolds to that of $R^3$ or $S^3$. Figure~\ref{fig:surgery} heuristically illustrates how surgery replaces a hole in a 3D topology by a linear combination of spaces without holes but with additional Wilson lines. Therefore, as long as the basis in the Hilbert space is concerned, we can restrict ourselves to a discussion of unknotted and unlinked Wilson lines embedded in a simply-connected 3D topology.

Furthermore, it is sufficient to consider connected topologies due to a property analogous to~(\ref{breaking}). If a 3-manifold can be cut into two pieces along an $S^2$ that does not cut any Wilson lines, then such a manifold is equivalent to a pair of disconnected 3-manifolds obtained from the original one contracting the $S^2$ cut to a point. The reason, again, is that $S^2$ with no punctures corresponds to a trivial Hilbert space. This can be illustrated heuristically by the equivalence of states~(\ref{separable}) and~(\ref{EntState}) up to normalization, if the corresponding diagrams correspond to cobordisms of a pair of $S^2$. In other words, disconnected three-dimensional topologies are equivalent to fully connected ones. More generally, the manifolds are truly connected if they cannot be cut into two pieces along an $S^2$ in such a way that at most two Wilson lines are cut, cf.~(\ref{breaking}). This has an important consequence for entanglement: one needs Wilson lines to support it if the states are constructed as cobordisms between 2-spheres.

The main purpose of the discussion here, apart from introducing an ample set of options in 3D TQFT as compared to the 1D ones, is to show that if we restrict ourselves to simply-connected 3D manifolds with $S^2$ boundaries and add Wilson lines in the fundamental representation of $su(2)_k$ to them, then the discussion essentially returns to the 1D case with some additional features, such as a more general linear dependence of states with crossings, fusion of lines, and the value of $\!\begin{array}{c}\scalebox{0.8}{
\begin{tikzpicture}[thick]
\draw[black] (0,0.0) circle (0.2cm);
\end{tikzpicture}} 
\end{array}\!$. This is the specific example of three-dimensional theories that will be considered here. Consequently, the purpose is to extend the connectome classification, introduced in the 1D case, to the 3D theories.

Since in 3D theories there are much more options of topology, the question is whether the classification by the simplest connectome diagrams is a reasonable coarse graining of different forms of entanglement. What we are going to show is that for the bipartite systems the classification is as good as the SLOCC one, although the situation is more complicated for the multipartite entanglement.

\paragraph{Qubits.} To realize a qubit in the three-dimensional theory we will use a sphere with four punctures as $\Sigma$. For Wilson lines in fundamental representation of $su(2)_k$ Kac-Moody, this is the minimal example, similarly to the one-dimensional case. In other words, the dimension of the Hilbert space of $SU(2)_k$ Chern-Simons in a 3-ball with two Wilson lines in the fundamental representation, with the endpoints on the boundary sphere is two. Diagrams~(\ref{4pointbasis0}) can serve as a basis in this Hilbert space, if we understand the shaded rectangles as the boundary $S^2$. For simplicity, we will not draw the boundary explicitly in most of the discussion below, neither we will illustrate the ambient space, which in most cases is $\mathbb{R}^3$. Multiple $S^2$ boundaries will be distinguished by grouping the endpoints of the lines.

So, states~(\ref{4pointbasis}) then represent the standard qubit basis $|0\rangle$ and $|1\rangle$. All the scalar products in this space can be computed as explained in the one-dimensional example, with topologically trivial circles replaced by $d$, which is now given by~(\ref{d}) and with skein relation~(\ref{skein}).

In the case of a pair of qubits we will have two $S^2$ boundaries. We already know how to construct a separable state, say $|00\rangle$, and a maximally entangled state. They can be represented by diagrams~(\ref{entanglement}), or by the connectome diagrams~(\ref{2qubitcase}), by gluing basis states~(\ref{4pointbasis}) to the maximally entangled state one can find its explicit form in that basis,
\be
\label{maxent}
\begin{array}{c}
\begin{tikzpicture}[thick]
\fill[black] (0,0.0) circle (0.05cm);
\fill[black] (0,0.3) circle (0.05cm);
\fill[black] (0,0.6) circle (0.05cm);
\fill[black] (0,0.9) circle (0.05cm);
\draw (0,0) -- (1.2,0);
\draw (0,0.6) -- (1.2,0.6);
\fill[black] (1.2,0.0) circle (0.05cm);
\fill[black] (1.2,0.3) circle (0.05cm);
\fill[black] (1.2,0.6) circle (0.05cm);
\fill[black] (1.2,0.9) circle (0.05cm);
\draw (1.2,0.3) -- (0,0.3);
\draw (1.2,0.9) -- (0,0.9);
\end{tikzpicture} 
\end{array} \ = \ |00\rangle + |11\rangle\,.
\ee
Since the Hilbert space of Chern-Simons on $S^2$ with two punctures has at most dimension one, one can repeat the discussion in the one-dimensional case and conclude that there are only two inequivalent connectome diagrams in the case of two qubits. These two diagrams are equivalent to the two classes of SLOCC entanglement of qubits.

Let us consider an example of a nontrivial topologically entangled state and show that it has less entanglement than~(\ref{maxent}). Consider
\be
\label{chained}
\begin{array}{c}\includegraphics[scale=0.25]{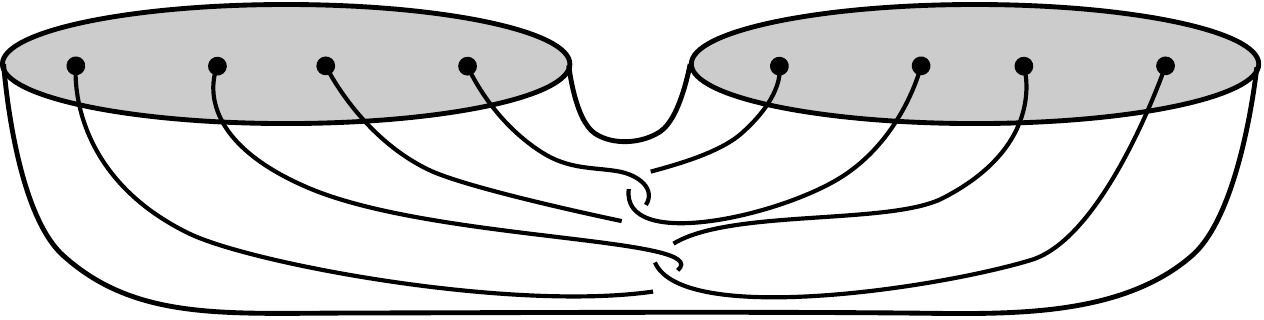}\end{array} 
\ee
This state can be expanded via skein relations as follows:
\be
\label{chainedstate0}
\begin{array}{c}
\begin{tikzpicture}[thick]
\fill[black] (0,0.0) circle (0.05cm);
\fill[black] (0,0.3) circle (0.05cm);
\fill[black] (0,0.6) circle (0.05cm);
\fill[black] (0,0.9) circle (0.05cm);
\draw (0,0) -- (0.5,0) arc (-90:30:0.15cm);
\draw (0,0.3) -- (0.5,0.3) arc (90:60:0.15cm);
\draw (0,0.6) -- (0.5,0.6) arc (-90:30:0.15cm);
\draw (0,0.9) -- (0.5,0.9) arc (90:60:0.15cm);
\fill[black] (1.2,0.0) circle (0.05cm);
\fill[black] (1.2,0.3) circle (0.05cm);
\fill[black] (1.2,0.6) circle (0.05cm);
\fill[black] (1.2,0.9) circle (0.05cm);
\draw (1.2,0) -- (0.7,0) arc (-90:-120:0.15cm);
\draw (1.2,0.3) -- (0.7,0.3) arc (90:210:0.15cm);
\draw (1.2,0.6) -- (0.7,0.6) arc (-90:-120:0.15cm);
\draw (1.2,0.9) -- (0.7,0.9) arc (90:210:0.15cm);
\end{tikzpicture} 
\end{array}
=  {A^{4}} 
\begin{array}{c}
\begin{tikzpicture}[thick]
\fill[black] (0,0.0) circle (0.05cm);
\fill[black] (0,0.3) circle (0.05cm);
\fill[black] (0,0.6) circle (0.05cm);
\fill[black] (0,0.9) circle (0.05cm);
\draw (0,0) -- (0.4,0) arc (-90:90:0.15cm) -- (0,0.3);
\draw (0,0.6) -- (0.4,0.6) arc (-90:90:0.15cm) -- (0,0.9);
\fill[black] (1.2,0.0) circle (0.05cm);
\fill[black] (1.2,0.3) circle (0.05cm);
\fill[black] (1.2,0.6) circle (0.05cm);
\fill[black] (1.2,0.9) circle (0.05cm);
\draw (1.2,0) -- (0.8,0) arc (-90:-270:0.15cm) -- (1.2,0.3);
\draw (1.2,0.6) -- (0.8,0.6) arc (-90:-270:0.15cm) -- (1.2,0.9);
\end{tikzpicture} 
\end{array}
 + (A^2-A^{-2}) \left(
\begin{array}{c}
\begin{tikzpicture}[thick]
\fill[black] (0,0.0) circle (0.05cm);
\fill[black] (0,0.3) circle (0.05cm);
\fill[black] (0,0.6) circle (0.05cm);
\fill[black] (0,0.9) circle (0.05cm);
\draw (0,0) -- (0.4,0) -- (1.2,0.0);
\draw (0,0.6) -- (0.4,0.6) arc (-90:90:0.15cm) -- (0,0.9);
\fill[black] (1.2,0.0) circle (0.05cm);
\fill[black] (1.2,0.3) circle (0.05cm);
\fill[black] (1.2,0.6) circle (0.05cm);
\fill[black] (1.2,0.9) circle (0.05cm);
\draw (1.2,0.3) -- (0,0.3);
\draw (1.2,0.6) -- (0.8,0.6) arc (-90:-270:0.15cm) -- (1.2,0.9);
\end{tikzpicture} 
\end{array}
 +  
\begin{array}{c}
\begin{tikzpicture}[thick]
\fill[black] (0,0.0) circle (0.05cm);
\fill[black] (0,0.3) circle (0.05cm);
\fill[black] (0,0.6) circle (0.05cm);
\fill[black] (0,0.9) circle (0.05cm);
\draw (0,0) -- (0.4,0) arc (-90:90:0.15cm) -- (0,0.3);
\draw (0,0.6) -- (1.2,0.6);
\fill[black] (1.2,0.0) circle (0.05cm);
\fill[black] (1.2,0.3) circle (0.05cm);
\fill[black] (1.2,0.6) circle (0.05cm);
\fill[black] (1.2,0.9) circle (0.05cm);
\draw (1.2,0) -- (0.8,0) arc (-90:-270:0.15cm) -- (1.2,0.3);
\draw (1.2,0.9) -- (0,0.9);
\end{tikzpicture} 
\end{array} \right)
 +  (1-A^{-4})^2  
\begin{array}{c}
\begin{tikzpicture}[thick]
\fill[black] (0,0.0) circle (0.05cm);
\fill[black] (0,0.3) circle (0.05cm);
\fill[black] (0,0.6) circle (0.05cm);
\fill[black] (0,0.9) circle (0.05cm);
\draw (0,0) -- (1.2,0);
\draw (0,0.6) -- (1.2,0.6);
\fill[black] (1.2,0.0) circle (0.05cm);
\fill[black] (1.2,0.3) circle (0.05cm);
\fill[black] (1.2,0.6) circle (0.05cm);
\fill[black] (1.2,0.9) circle (0.05cm);
\draw (1.2,0.3) -- (0,0.3);
\draw (1.2,0.9) -- (0,0.9);
\end{tikzpicture} 
\end{array}.
\ee
In basis~(\ref{4pointbasis}) this state can be written as
\begin{multline}
\begin{array}{c}\includegraphics[scale=0.25]{lines3}\end{array} \ = \ (A^4+A^{-4})^2|00\rangle + {(1-A^{-4})^2}|11\rangle\,. \\ = \  4\cos^2\left(\frac{2\pi}{k+2}\right)|00\rangle -  4e^{-4i\theta}\sin^2\left(\frac{\pi}{k+2}\right)|11\rangle\,.
\end{multline}
This is an entangled state of the Bell SLOCC type, but it is not maximally entangled for $k\neq 4$ and can only be converted to the maximally entangled one~\cite{Vidal:1999vh} with probability given by
\be
p \ = \ 2\min\left\{\frac{\cos^4\left(\frac{2\pi}{k+2}\right)}{\cos^4\left(\frac{2\pi}{k+2}\right)+\sin^4\left(\frac{\pi}{k+2}\right)},\frac{\sin^4\left(\frac{\pi}{k+2}\right)}{\cos^4\left(\frac{2\pi}{k+2}\right)+\sin^4\left(\frac{\pi}{k+2}\right)}\right\}.
\ee
Conversely, the maximally entangled state can be converted to any nonmaximally entangled state with probability one. In the topological realization this asymmetry of the conversion is expressed in the state itself. One can view state~(\ref{chained}) as a result of an application of a local operator on the maximally entangled state~(\ref{maxent}). This local operator, which can act on either the first or the second qubit, is given by the same diagram as state~(\ref{chained}) itself. However, there is no inverse local topological operation that can undo such an action.

So far the connectome classification worked in the case of a pair of qubits. However, if we try to generalize it to the case of general bipartite entanglement, we will run into the same problem as in the one-dimensional case. As we have seen, in the 1D TQFT the number of naive diagrams is smaller than the number of the SLOCC classes if we grow the Hilbert spaces just by adding extra punctures to the spheres. In the next section we will solve the classification mismatch using a connection to representation theory and introducing appropriate projectors, which will allow us to extend the connectome classification to arbitrary pairs of qudits.


\section{Jones-Wenzl projectors and bipartite entanglement}
\label{sec:Jones-Wenzl}

As specified above, we understand the ordinary lines in the diagrams as labeled by fundamental representations of $su(2)_k$. Diagrams composed of open lines are, in general, operators acting on the linear spaces of punctures with the basis provided by diagrams like~(\ref{4pointbasis0}) or their generalizations. In the previous section we gave all the necessary tools to convert the diagrammatic presentation into linear-algebraic form. See also~\cite{Kauffman:2013bh,Kauffman:1994tem}. 

Jones-Wenzl projectors~\cite{Jones:1985dw,Wenzl:1985seq} are useful in the discussion of the diagrammatic presentation of the irreducible representations. For $su(2)_k$, operator
\be
\label{JWprojector}
\begin{tikzpicture}[baseline=2]
\draw[thick] (0,0) rectangle (1,0.5);
\draw[ultra thick] (0.2,1) node[anchor=east] {$n$} -- (0.2,0.5);
\draw[ultra thick] (0.2,0) -- (0.2,-0.5);
\draw[thick] (0.5,-0.5) -- (0.5,0);
\draw[thick] (0.5,1) -- (0.5,0.5);
\draw[thick] (0.8,-0.5) -- (0.8,0);
\draw[thick] (0.8,1) -- (0.8,0.5);
\end{tikzpicture}\quad \ = \ \quad 
\begin{tikzpicture}[baseline=2]
\draw[thick] (0,0) rectangle (1,0.5);
\draw[ultra thick] (0.3,-0.5) -- (0.3,0);
\draw[ultra thick] (0.3,1) node[anchor=east] {$n$} -- (0.3,0.5);
\draw[thick] (0.7,-0.5) -- (0.7,0);
\draw[thick] (0.7,1) -- (0.7,0.5);
\draw[thick] (1.2,-0.5) -- (1.2,1);
\end{tikzpicture}\quad \ - \ \frac{\Delta_n}{\Delta_{n+1}} \quad \begin{tikzpicture}[baseline=2]
\draw[thick] (0,0.5) rectangle (1,0.8);
\draw[ultra thick] (0.3,-0.) -- (0.3,0.5);
\draw[ultra thick] (0.3,-0.5) -- (0.3,-0.3);
\draw[ultra thick] (0.3,1) node[anchor=east] {$n$} -- (0.3,0.8);
\draw[thick] (0.7,-0.)  arc (180:0:0.2) -- (1.1,-0.5);
\draw[thick] (0.7,1) -- (0.7,0.8);
\draw[thick] (0.7,-0.5) -- (0.7,-0.3);
\draw[thick] (1.1,1) -- (1.1,0.5) arc (0:-180:0.2);
\draw[thick] (0,-0.3) rectangle (1,0.);
\end{tikzpicture}\quad
\ee
acts as a projector on the maximal spin (symmetric) representation in the tensor product of $n+2$ fundamental irreps. Here, the thick line with label $n$ substitutes $n$ ordinary lines. In~(\ref{JWprojector}) the operator is defined recursively and so is $\Delta_n$:
\be
\Delta_{-1}\ = \ 0\,,\qquad \Delta_{0} \ = \ 1\,, \qquad \Delta_{n+1} \ = \ d\Delta_{n} - \Delta_{n-1}\,.
\ee

Let us consider a few examples of Jones-Wenzl projectors explicitly. The simplest projector, is the one of a single line. It is the line itself, as follows from~(\ref{JWprojector}),
\be
\begin{array}{c}
     \begin{tikzpicture}
         \draw[thick] (0.25,0) -- (0.25,1.3);
         \fill[white] (-0.2,0.4) rectangle (0.7,0.9);
         \draw[thick] (-0.2,0.4) rectangle (0.7,0.9);
     \end{tikzpicture} 
\end{array}
\ = \ \begin{array}{c}
     \begin{tikzpicture}
         \draw[thick] (0.25,0) -- (0.25,1.3);
     \end{tikzpicture} 
\end{array}\,.
\ee

A pair of lines corresponds to a tensor product of two fundamental representations. These have the fusion rule ${\bf 2}\otimes {\bf 2}={\bf 3}\oplus{\bf 1}$. Then the Jones-Wenzl projector on ${\bf 3}$ is given by
\be
\label{P3}
\begin{tikzpicture}[baseline=2]
\draw[thick] (0,0) rectangle (1,0.5);
\draw[thick] (0.3,-0.5) -- (0.3,0);
\draw[thick] (0.3,1) -- (0.3,0.5);
\draw[thick] (0.7,-0.5) -- (0.7,0);
\draw[thick] (0.7,1) -- (0.7,0.5);
\end{tikzpicture}
\quad \ = \ \quad 
\begin{tikzpicture}[baseline=2]
\draw[thick] (0.3,-0.5) -- (0.3,1);
\draw[thick] (0.7,-0.5) -- (0.7,1);
\end{tikzpicture} \ - \ \frac{1}{d} \quad \begin{tikzpicture}[baseline=2]
\draw[thick] (0.7,-0.)  arc (180:0:0.2) -- (1.1,-0.5);
\draw[thick] (0.7,1) -- (0.7,0.5);
\draw[thick] (0.7,-0.5) -- (0.7,0);
\draw[thick] (1.1,1) -- (1.1,0.5) arc (0:-180:0.2);
\end{tikzpicture}\quad\,.
\ee
The first term in this projector is simply the identity operator, while the second one is the projector on the subspace orthogonal to that of representation ${\bf 3}$. Indeed, in the second term the two lines fuse to nothing, which correspond to the production of a singlet. Using~(\ref{P3}) it is easy to check the following properties of the projector:
\be
\begin{array}{c}
     \begin{tikzpicture}
         \draw[thick] (0,0) -- (1.4,0);
         \draw[thick] (0,0.5) -- (1.4,0.5);
         \fill[white] (0.5,-0.2) rectangle (1.0,0.7);
         \draw[thick] (0.5,-0.2) rectangle (1.0,0.7);
         \draw[thick] (1.5,0) -- (1.75,0) arc (-90:90:0.25) -- (1.5,0.5);
         \draw[thick] (2.7,0) -- (2.45,0) arc (270:90:0.25) -- (2.7,0.5);
     \end{tikzpicture} 
\end{array} \ = \ \begin{array}{c}
     \begin{tikzpicture}
         \draw[thick] (1.0,0) -- (1.75,0) arc (-90:90:0.25) -- (1.0,0.5);
         \draw[thick] (2.7,0) -- (2.45,0) arc (270:90:0.25) -- (2.7,0.5);
     \end{tikzpicture} 
\end{array} \ - \ 
\frac{1}{d}\begin{array}{c}
     \begin{tikzpicture}
         \draw[thick] (0,0) -- (0.65,0) arc (-90:90:0.25) -- (0,0.5);
         \draw[thick] (1.45,0) -- (1.75,0) arc (-90:90:0.25) -- (1.45,0.5) arc (90:270:0.25);
         \draw[thick] (2.7,0) -- (2.45,0) arc (270:90:0.25) -- (2.7,0.5);
     \end{tikzpicture} 
\end{array} \ = \ 0\,,  
\ee
where the closed line was substituted by $d$, and
\be
\begin{array}{c}
     \begin{tikzpicture}
         \draw[thick] (0,0) -- (1.4,0);
         \draw[thick] (0,0.5) -- (1.4,0.5);
         \fill[white] (0.5,-0.2) rectangle (1.0,0.7);
         \draw[thick] (0.5,-0.2) rectangle (1.0,0.7);
         \draw[thick] (1.5,0) -- (2.9,0);
         \draw[thick] (1.5,0.5) -- (2.9,0.5);
         \fill[white] (2.0,-0.2) rectangle (2.5,0.7);
         \draw[thick] (2.0,-0.2) rectangle (2.5,0.7);
     \end{tikzpicture} 
\end{array}
\ = \ \begin{array}{c}
     \begin{tikzpicture}
         \draw[thick] (0,0) -- (1.4,0);
         \draw[thick] (0,0.5) -- (1.4,0.5);
         \fill[white] (0.5,-0.2) rectangle (1.0,0.7);
         \draw[thick] (0.5,-0.2) rectangle (1.0,0.7);
     \end{tikzpicture} 
\end{array} \,.
\ee
In terms of these projectors the basis states~(\ref{4pointbasis}) can be cast as 
\be
|0\rangle \ = \ \frac{1}{d}\begin{array}{c}
\begin{tikzpicture}[thick]
\fill[black] (0,0.0) circle (0.05cm);
\fill[black] (0,0.3) circle (0.05cm);
\fill[black] (0,0.6) circle (0.05cm);
\fill[black] (0,0.9) circle (0.05cm);
\draw (0,0) -- (0.4,0) arc (-90:90:0.15cm) -- (0,0.3);
\draw (0,0.6) -- (0.4,0.6) arc (-90:90:0.15cm) -- (0,0.9);
\end{tikzpicture} 
\end{array}\,, \qquad |1\rangle \ = \ \frac{1}{\sqrt{\Delta_2}}\begin{array}{c}
\begin{tikzpicture}[thick]
\fill[black] (0,0.0) circle (0.05cm);
\fill[black] (0,0.3) circle (0.05cm);
\fill[black] (0,0.6) circle (0.05cm);
\fill[black] (0,0.9) circle (0.05cm);
\draw (0,0) -- (0.3,0) arc (-90:90:0.45cm) -- (0,0.9);
\draw (0,0.6) -- (0.3,0.6) arc (90:-90:0.15cm) -- (0,0.3);
\fill[white] (0.1,-0.1) rectangle (0.3,0.4);
\draw (0.1,-0.1) rectangle (0.3,0.4);
\fill[white] (0.1,0.5) rectangle (0.3,1);
\draw (0.1,0.5) rectangle (0.3,1);
\end{tikzpicture} 
\end{array}\,.
\ee
One can see explicitly that the basis states correspond to two fusion channels.

Note that the $n$-line projectors are constructed from the elements of the Temperley-Lieb algebra $TL_n$. Projector~(\ref{P3}) and the orthogonal projector are equivalent to the basis elements~(\ref{4pointbasis}).

In another example, $TL_3$ consists of five elements and the corresponding Jones-Wenzl projector,
\be
\begin{tikzpicture}[baseline=2]
\draw[thick] (0,0) rectangle (1,0.5);
\draw[thick] (0.19,-0.5) -- (0.19,0);
\draw[thick] (0.19,1) -- (0.19,0.5);
\draw[thick] (0.5,-0.5) -- (0.5,0);
\draw[thick] (0.5,1) -- (0.5,0.5);
\draw[thick] (0.8,-0.5) -- (0.8,0);
\draw[thick] (0.8,1) -- (0.8,0.5);
\end{tikzpicture}\quad \ = \ \quad 
\begin{tikzpicture}[baseline=2]
\draw[thick] (0.19,-0.5) -- (0.19,1);
\draw[thick] (0.5,-0.5) -- (0.5,1);
\draw[thick] (0.8,-0.5) -- (0.8,1);
\end{tikzpicture}\quad \ - \ \frac{d}{d^2-1}\left(
\begin{array}{c}
     \begin{tikzpicture}[baseline=2]
\draw[thick] (0.3,-0.5) -- (0.3,1);
\draw[thick] (0.7,-0.5) -- (0.7,-0.)  arc (180:0:0.2) -- (1.1,-0.5);
\draw[thick] (1.1,1) -- (1.1,0.5) arc (0:-180:0.2) -- (0.7,1);
\end{tikzpicture} 
\end{array}
+
\begin{array}{c}
     \begin{tikzpicture}[baseline=2]
\draw[thick] (1.1,-0.5) -- (1.1,1);
\draw[thick] (0.3,-0.5) -- (0.3,-0.)  arc (180:0:0.2) -- (0.7,-0.5);
\draw[thick] (0.7,1) -- (0.7,0.5) arc (0:-180:0.2) -- (0.3,1);
\end{tikzpicture} 
\end{array}
\right)
\quad +
\quad \frac{1}{d^2-1} \left(
\begin{array}{c}
     \begin{tikzpicture}[baseline=2]
\draw[thick] (0.3,-0.5) -- (0.3,-0.2) arc (180:90:0.4) arc (-90:0:0.4) -- (1.1,1);
\draw[thick] (0.7,-0.5) -- (0.7,-0.2)  arc (180:0:0.2) -- (1.1,-0.5);
\draw[thick] (0.7,1) -- (0.7,0.7) arc (0:-180:0.2) -- (0.3,1);
\end{tikzpicture} 
\end{array}
+
\begin{array}{c}
     \begin{tikzpicture}[baseline=2]
\draw[thick] (1.1,-0.5) -- (1.1,-0.2) arc (0:90:0.4) arc (270:180:0.4) -- (0.3,1);
\draw[thick] (0.3,-0.5) -- (0.3,-0.2)  arc (180:0:0.2) -- (0.7,-0.5);
\draw[thick] (1.1,1) -- (1.1,0.7) arc (0:-180:0.2) -- (0.7,1);
\end{tikzpicture} 
\end{array}
\right)
\label{P4}
\ee
projects onto representation ${\bf 4}$ in the tensor product ${\bf 2}\otimes {\bf 2}\otimes {\bf 2}$. Elements appearing in~(\ref{P4}) are the same as in the basis~(\ref{5basis}). One can notice that the diagrams in the bracket correspond to fusion of ${\bf 2}$ with a singlet, so they are projecting on the subspace orthogonal to ${\bf 4}$.

The Jones-Wenzl projectors have a few useful properties. Their tracelike closure computes the quantum dimension $(-1)^n\Delta_n$ of the corresponding irrep:
\be
\begin{tikzpicture}[baseline=2]
\draw[thick] (0,0) rectangle (1,0.5);
\draw[ultra thick] (0.5,0.7) node[anchor=east] {$n$} -- (0.5,0.5);
\draw[ultra thick] (0.5,0) -- (0.5,-0.2) arc (-180:-90:0.2) -- (1,-0.4) arc (-90:90:0.65) -- (0.7,0.9) arc (90:180:0.2);
\end{tikzpicture}
\ = \ \Delta_n\,, \qquad \text{in particular,} \qquad \begin{tikzpicture}[baseline=2]
\draw[thick] (0,0) rectangle (1,0.5);
\draw[thick] (0.5,0.7) -- (0.5,0.5);
\draw[thick] (0.5,0) -- (0.5,-0.2) arc (-180:-90:0.2) -- (1,-0.4) arc (-90:90:0.65) -- (0.7,0.9) arc (90:180:0.2);
\end{tikzpicture} \ = \ \begin{array}{c}
\begin{tikzpicture}[thick]
\draw[black] (0,0.0) circle (0.2cm);
\end{tikzpicture} 
\end{array} \ \equiv\  d=\Delta_1\,. 
\ee
Platlike closure of the projector (pairwise closure of any neighboring pair of input or output lines) nullifies it,
\be
\label{platclosure}
\begin{tikzpicture}[baseline=2]
\draw[thick] (0,0) rectangle (1.5,0.5);
\draw[thick] (0.2,0.5) -- (0.2,0.7) arc (180:0:0.15) -- (0.5,0.5);
\draw[thick] (0.8,0.5) -- (0.8,0.85);
\draw[thick] (0.5,0) -- (0.5,-0.35);
\draw[thick] (0.2,0) -- (0.2,-0.35);
\draw[thick] (0.8,0) -- (0.8,-0.35);
\draw (1.2,0.7) node {\bf $\cdots$};
\draw (1.2,-0.2) node {\bf $\cdots$};
\end{tikzpicture}
\ = \  \begin{tikzpicture}[baseline=2]
\draw[thick] (0,0) rectangle (1.5,0.5);
\draw[thick] (0.5,0.5) -- (0.5,0.7) arc (180:0:0.15) -- (0.8,0.5);
\draw[thick] (0.2,0.5) -- (0.2,0.85);
\draw[thick] (0.5,0) -- (0.5,-0.35);
\draw[thick] (0.2,0) -- (0.2,-0.35);
\draw[thick] (0.8,0) -- (0.8,-0.35);
\draw (1.2,0.7) node {\bf $\cdots$};
\draw (1.2,-0.2) node {\bf $\cdots$};
\end{tikzpicture} \ = \ \begin{tikzpicture}[baseline=2]
\draw[thick] (0,0) rectangle (1.5,0.5);
\draw[thick] (0.2,0.5) -- (0.2,0.85);
\draw[thick] (0.8,0.5) -- (0.8,0.85);
\draw[thick] (0.5,0.5) -- (0.5,0.85);
\draw[thick] (0.2,0) -- (0.2,-0.2) arc (-180:0:0.15) -- (0.5,0);
\draw[thick] (0.8,0) -- (0.8,-0.35);
\draw (1.2,0.7) node {\bf $\cdots$};
\draw (1.2,-0.2) node {\bf $\cdots$};
\end{tikzpicture} \ = \ 0\,.
\ee

We will now use the Jones-Wenzl projectors to construct the Hilbert spaces of qudits. The qutrit space can be built from four {\bf 3}-projectors~(\ref{P3}). Such space has a basis
\be
|e_1\rangle \ = \ \begin{array}{c}
\begin{tikzpicture}[thick]
\fill[black] (0,0.0) circle (0.05cm);
\fill[black] (0,0.3) circle (0.05cm);
\fill[black] (0,0.6) circle (0.05cm);
\fill[black] (0,0.9) circle (0.05cm);
\draw (0,0) -- (0.3,0) arc (-90:90:0.45cm) -- (0,0.9);
\draw (0,0.6) -- (0.3,0.6) arc (90:-90:0.15cm) -- (0,0.3);
\fill[white] (0.1,-0.1) rectangle (0.3,0.4);
\draw (0.1,-0.1) rectangle (0.3,0.4);
\fill[white] (0.1,0.5) rectangle (0.3,1);
\draw (0.1,0.5) rectangle (0.3,1);
\newcommand\y{1.2};
\fill[black] (0,0.0+\y) circle (0.05cm);
\fill[black] (0,0.3+\y) circle (0.05cm);
\fill[black] (0,0.6+\y) circle (0.05cm);
\fill[black] (0,0.9+\y) circle (0.05cm);
\draw (0,0+\y) -- (0.3,0+\y) arc (-90:90:0.45cm) -- (0,0.9+\y);
\draw (0,0.6+\y) -- (0.3,0.6+\y) arc (90:-90:0.15cm) -- (0,0.3+\y);
\fill[white] (0.1,-0.1+\y) rectangle (0.3,0.4+\y);
\draw (0.1,-0.1+\y) rectangle (0.3,0.4+\y);
\fill[white] (0.1,0.5+\y) rectangle (0.3,1+\y);
\draw (0.1,0.5+\y) rectangle (0.3,1+\y);
\end{tikzpicture} 
\end{array}\,, \qquad
|e_2\rangle \ = \ \begin{array}{c}
\begin{tikzpicture}[thick]
\newcommand\y{1.2};
\fill[black] (0,0.0) circle (0.05cm);
\fill[black] (0,0.3) circle (0.05cm);
\fill[black] (0,0.6) circle (0.05cm);
\fill[black] (0,0.9) circle (0.05cm);
\draw (0,0) -- (0.3,0) arc (-90:90:1.05cm) -- (0,0.9+\y);
\draw (0,0.6) -- (0.3,0.6) arc (90:-90:0.15cm) -- (0,0.3);
\fill[black] (0,0.0+\y) circle (0.05cm);
\fill[black] (0,0.3+\y) circle (0.05cm);
\fill[black] (0,0.6+\y) circle (0.05cm);
\fill[black] (0,0.9+\y) circle (0.05cm);
\draw (0,0+\y) -- (0.9,0+\y) arc (90:-90:0.15cm) -- (0,0.9);
\draw (0,0.6+\y) -- (0.3,0.6+\y) arc (90:-90:0.15cm) -- (0,0.3+\y);
\fill[white] (0.1,-0.1) rectangle (0.3,0.4);
\draw (0.1,-0.1) rectangle (0.3,0.4);
\fill[white] (0.1,0.5) rectangle (0.3,1);
\draw (0.1,0.5) rectangle (0.3,1);
\fill[white] (0.1,-0.1+\y) rectangle (0.3,0.4+\y);
\draw (0.1,-0.1+\y) rectangle (0.3,0.4+\y);
\fill[white] (0.1,0.5+\y) rectangle (0.3,1+\y);
\draw (0.1,0.5+\y) rectangle (0.3,1+\y);
\end{tikzpicture} 
\end{array}\,, \qquad
|e_3\rangle \ = \ \begin{array}{c}
\begin{tikzpicture}[thick]
\newcommand\y{1.2};
\fill[black] (0,0.0) circle (0.05cm);
\fill[black] (0,0.3) circle (0.05cm);
\fill[black] (0,0.6) circle (0.05cm);
\fill[black] (0,0.9) circle (0.05cm);
\draw (0,0) -- (0.3,0) arc (-90:90:1.05cm) -- (0,0.9+\y);
\draw (0,0.6) -- (0.3,0.6) arc (-90:90:0.45cm) -- (0,0.3+\y);
\fill[black] (0,0.0+\y) circle (0.05cm);
\fill[black] (0,0.3+\y) circle (0.05cm);
\fill[black] (0,0.6+\y) circle (0.05cm);
\fill[black] (0,0.9+\y) circle (0.05cm);
\draw (0,0+\y) -- (0.3,0+\y) arc (90:-90:0.15cm) -- (0,0.9);
\draw (0,0.6+\y) -- (0.3,0.6+\y) arc (90:-90:0.75cm) -- (0,0.3);
\fill[white] (0.1,-0.1) rectangle (0.3,0.4);
\draw (0.1,-0.1) rectangle (0.3,0.4);
\fill[white] (0.1,0.5) rectangle (0.3,1);
\draw (0.1,0.5) rectangle (0.3,1);
\fill[white] (0.1,-0.1+\y) rectangle (0.3,0.4+\y);
\draw (0.1,-0.1+\y) rectangle (0.3,0.4+\y);
\fill[white] (0.1,0.5+\y) rectangle (0.3,1+\y);
\draw (0.1,0.5+\y) rectangle (0.3,1+\y);
\end{tikzpicture} 
\end{array}
\ee
It is not hard to see that these are the only non-null crossing-free options of connecting the projectors allowed by~(\ref{platclosure}), and their number matches the number of irreps in the expansion of ${\bf 3}\otimes {\bf 3}={\bf 5}\oplus{\bf 3}\oplus {\bf 1}$. States $|e_1\rangle$ and $|e_3\rangle$ are similar to states~(\ref{4pointbasis0}), especially if in the qutrit case one fuses a pair of ordinary lines in one gross line and notices that a single line in~(\ref{4pointbasis0}) is itself a projector on {\bf 2}. The fine structure of the gross lines, on the other hand, allows for more connection options.

Again, we can construct an orthonormal basis. For example,
\be
\label{8pointbasis}
|0\rangle \ = \ \frac{1}{\Delta_2}\,|e_1\rangle\,, \qquad |1\rangle \ = \ \frac{d}{(\Delta_2-1)\sqrt{\Delta_2}}|\hat{e}_2\rangle\,, \qquad |2\rangle \ = \ \frac{1}{\sqrt{\Delta_2^2-\Delta_2-1}}\left(|e_3\rangle - |0\rangle-\sqrt{\Delta_2}|1\rangle\right)
\ee
Here, for convenience, we redefined the state $|e_2\rangle$:
\be
|\hat{e}_2\rangle \ = \ |e_2\rangle -\frac{1}{d}|e_1\rangle \ = \ \begin{array}{c}
\begin{tikzpicture}[thick]
\newcommand\y{1.2};
\fill[black] (0,0.0) circle (0.05cm);
\fill[black] (0,0.3) circle (0.05cm);
\fill[black] (0,0.6) circle (0.05cm);
\fill[black] (0,0.9) circle (0.05cm);
\draw (0,0) -- (0.3,0) arc (-90:90:1.05cm) -- (0,0.9+\y);
\draw (0,0.6) -- (0.3,0.6) arc (90:-90:0.15cm) -- (0,0.3);
\fill[black] (0,0.0+\y) circle (0.05cm);
\fill[black] (0,0.3+\y) circle (0.05cm);
\fill[black] (0,0.6+\y) circle (0.05cm);
\fill[black] (0,0.9+\y) circle (0.05cm);
\draw (0,0+\y) -- (0.9,0+\y) arc (90:-90:0.15cm) -- (0,0.9);
\draw (0,0.6+\y) -- (0.3,0.6+\y) arc (90:-90:0.15cm) -- (0,0.3+\y);
\fill[white] (0.1,-0.1) rectangle (0.3,0.4);
\draw (0.1,-0.1) rectangle (0.3,0.4);
\fill[white] (0.1,0.5) rectangle (0.3,1);
\draw (0.1,0.5) rectangle (0.3,1);
\fill[white] (0.1,-0.1+\y) rectangle (0.3,0.4+\y);
\draw (0.1,-0.1+\y) rectangle (0.3,0.4+\y);
\fill[white] (0.1,0.5+\y) rectangle (0.3,1+\y);
\draw (0.1,0.5+\y) rectangle (0.3,1+\y);
\fill[white] (0.9,1.0) rectangle (1.5,1.1);
\draw (0.9,1.0) rectangle (1.5,1.1);
\end{tikzpicture} 
\end{array}\,,
\ee

One can now study possible two-qutrit topologies, that is, diagrams connecting eight pairs of points, taking into account local permutations and relations~(\ref{platclosure}). Since there is a larger Hilbert space associated with eight lines, some permutations may act as projectors away from the subspace spanned by basis~(\ref{8pointbasis}). The valid permutations are only those that do not break the projectors: swapping a pair of projectors or permutation of lines within the same projector. Similarly, from the point of view of this subspace, there will be several linearly dependent diagrams.  A natural choice of linearly independent ones is the following:
\be
\label{8pointdiags}
\begin{array}{c}
\begin{tikzpicture}[thick]
\fill[black] (0,0.0) circle (0.05cm);
\fill[black] (0,0.3) circle (0.05cm);
\fill[black] (0,0.6) circle (0.05cm);
\fill[black] (0,0.9) circle (0.05cm);
\draw (0,0) -- (0.4,0) -- (1.2,0.0);
\draw (0,1.2) -- (0.1,1.2) arc (-90:90:0.45cm) -- (0,2.1);
\fill[black] (1.2,0.0) circle (0.05cm);
\fill[black] (1.2,0.3) circle (0.05cm);
\fill[black] (1.2,0.6) circle (0.05cm);
\fill[black] (1.2,0.9) circle (0.05cm);
\draw (1.2,0.3) -- (0,0.3);
\draw (1.2,1.2) -- (1.1,1.2) arc (-90:-270:0.45cm) -- (1.2,2.1);
\fill[black] (1.2,1.2) circle (0.05cm);
\fill[black] (1.2,1.5) circle (0.05cm);
\fill[black] (0,1.2) circle (0.05cm);
\fill[black] (0,1.5) circle (0.05cm);
\draw (0,0.6) -- (1.2,0.6);
\draw (1.2,0.9) -- (0,0.9);
\fill[black] (1.2,1.8) circle (0.05cm);
\fill[black] (1.2,2.1) circle (0.05cm);
\fill[black] (0,1.8) circle (0.05cm);
\fill[black] (0,2.1) circle (0.05cm);
\draw (0,1.8) -- (0.1,1.8) arc (90:-90:0.15) -- (0,1.5);
\draw (1.2,1.8) -- (1.1,1.8) arc (90:270:0.15) -- (1.2,1.5);
\end{tikzpicture} 
\end{array}
\,,\qquad \begin{array}{c}
\begin{tikzpicture}[thick]
\fill[black] (0,0.0) circle (0.05cm);
\fill[black] (0,0.3) circle (0.05cm);
\fill[black] (0,0.6) circle (0.05cm);
\fill[black] (0,0.9) circle (0.05cm);
\draw (0,0) -- (0.4,0) -- (1.2,0.0);
\draw (0,2.1) -- (1.2,2.1);
\fill[black] (1.2,0.0) circle (0.05cm);
\fill[black] (1.2,0.3) circle (0.05cm);
\fill[black] (1.2,0.6) circle (0.05cm);
\fill[black] (1.2,0.9) circle (0.05cm);
\draw (1.2,0.3) -- (0,0.3);
\draw (1.2,0.6) -- (0,0.6);
\fill[black] (1.2,1.2) circle (0.05cm);
\fill[black] (1.2,1.5) circle (0.05cm);
\fill[black] (0,1.2) circle (0.05cm);
\fill[black] (0,1.5) circle (0.05cm);
\draw (0,0.9) -- (1.2,0.9);
\draw (0,1.2) -- (1.2,1.2);
\fill[black] (1.2,1.8) circle (0.05cm);
\fill[black] (1.2,2.1) circle (0.05cm);
\fill[black] (0,1.8) circle (0.05cm);
\fill[black] (0,2.1) circle (0.05cm);
\draw (0,1.8) -- (0.1,1.8) arc (90:-90:0.15) -- (0,1.5);
\draw (1.2,1.8) -- (1.1,1.8) arc (90:270:0.15) -- (1.2,1.5);
\end{tikzpicture} 
\end{array}
\,, \qquad \text{and}\qquad  
\begin{array}{c}
\begin{tikzpicture}[thick]
\fill[black] (0,0.0) circle (0.05cm);
\fill[black] (0,0.3) circle (0.05cm);
\fill[black] (0,0.6) circle (0.05cm);
\fill[black] (0,0.9) circle (0.05cm);
\draw (0,0) -- (1.2,0);
\draw (0,0.6) -- (1.2,0.6);
\fill[black] (1.2,0.0) circle (0.05cm);
\fill[black] (1.2,0.3) circle (0.05cm);
\fill[black] (1.2,0.6) circle (0.05cm);
\fill[black] (1.2,0.9) circle (0.05cm);
\draw (1.2,0.3) -- (0,0.3);
\draw (1.2,0.9) -- (0,0.9);
\fill[black] (1.2,1.2) circle (0.05cm);
\fill[black] (1.2,1.5) circle (0.05cm);
\fill[black] (0,1.2) circle (0.05cm);
\fill[black] (0,1.5) circle (0.05cm);
\draw (0,1.2) -- (1.2,1.2);
\draw (1.2,1.5) -- (0,1.5);
\fill[black] (1.2,1.8) circle (0.05cm);
\fill[black] (1.2,2.1) circle (0.05cm);
\fill[black] (0,1.8) circle (0.05cm);
\fill[black] (0,2.1) circle (0.05cm);
\draw (0,1.8) -- (1.2,1.8);
\draw (1.2,2.1) -- (0,2.1);
\end{tikzpicture} 
\end{array}\,.
\ee
Using basis~(\ref{8pointbasis}) one can check that the three independent diagrams correspond to matrices of rank one, two, and three, respectively.

Note that in the above list of diagrams at most four connections between the left and the right subsystems are broken. It is easy to see that breaking more connections would produce operators equivalent to the first diagram, that is a rank one matrix. In other words, one can always consider half of the lines, say the lower four, as unbroken and count the options existing for the remaining half. In the case of qutrits one ends up with the three options appearing above.

We can now review the case of two qubits. By the same argument the only possible choices are the diagrams
\be
\begin{array}{c}
\begin{tikzpicture}[thick]
\fill[black] (0,0.0) circle (0.05cm);
\fill[black] (0,0.3) circle (0.05cm);
\fill[black] (0,0.6) circle (0.05cm);
\fill[black] (0,0.9) circle (0.05cm);
\draw (0,0) -- (0.4,0) -- (1.2,0.0);
\draw (0,0.6) -- (0.4,0.6) arc (-90:90:0.15cm) -- (0,0.9);
\fill[black] (1.2,0.0) circle (0.05cm);
\fill[black] (1.2,0.3) circle (0.05cm);
\fill[black] (1.2,0.6) circle (0.05cm);
\fill[black] (1.2,0.9) circle (0.05cm);
\draw (1.2,0.3) -- (0,0.3);
\draw (1.2,0.6) -- (0.8,0.6) arc (-90:-270:0.15cm) -- (1.2,0.9);
\end{tikzpicture} 
\end{array}
\,,\qquad \text{and}\qquad  
\begin{array}{c}
\begin{tikzpicture}[thick]
\fill[black] (0,0.0) circle (0.05cm);
\fill[black] (0,0.3) circle (0.05cm);
\fill[black] (0,0.6) circle (0.05cm);
\fill[black] (0,0.9) circle (0.05cm);
\draw (0,0) -- (1.2,0);
\draw (0,0.6) -- (1.2,0.6);
\fill[black] (1.2,0.0) circle (0.05cm);
\fill[black] (1.2,0.3) circle (0.05cm);
\fill[black] (1.2,0.6) circle (0.05cm);
\fill[black] (1.2,0.9) circle (0.05cm);
\draw (1.2,0.3) -- (0,0.3);
\draw (1.2,0.9) -- (0,0.9);
\end{tikzpicture} 
\end{array}\,.
\ee

Now one can generalize this classification to arbitrary qudits, assuming each to be a state in a Hilbert space of dimension $n$. We can describe all the SLOCC classes of entangled pairs of qudits $A$ and $B$ by diagrams connecting $4(n-1)$ pairs of points. One choice of linearly independent diagrams corresponds to those with at least $2n-2$ connections between $A$ and $B$, so we can simply classify all the types of entanglement between the qudits by the remaining $2n-2$ lines. From these $2n-2$ points in either $A$ or $B$ only those can be connected by a line that belongs to different projectors. Up to permutations within each projector, this approach gives $d$ independent diagrams, matching the number of SLOCC classes. For example, $n=2,3,4$, one obtains the following \emph{reduced} diagrams:
\begin{eqnarray}
n=2\,: & & \begin{array}{c}
\begin{tikzpicture}[thick]
\fill[black] (0,0.6) circle (0.05cm);
\fill[black] (0,0.9) circle (0.05cm);
\draw (0,0.6) -- (0.4,0.6) arc (-90:90:0.15cm) -- (0,0.9);
\fill[black] (1.2,0.6) circle (0.05cm);
\fill[black] (1.2,0.9) circle (0.05cm);
\draw (1.2,0.6) -- (0.8,0.6) arc (-90:-270:0.15cm) -- (1.2,0.9);
\end{tikzpicture} 
\end{array}
\,,\qquad 
\begin{array}{c}
\begin{tikzpicture}[thick]
\fill[black] (0,0.6) circle (0.05cm);
\fill[black] (0,0.9) circle (0.05cm);
\draw (0,0.6) -- (1.2,0.6);
\fill[black] (1.2,0.6) circle (0.05cm);
\fill[black] (1.2,0.9) circle (0.05cm);
\draw (1.2,0.9) -- (0,0.9);
\end{tikzpicture} 
\end{array}\,, \\
n=3\,: & & \begin{array}{c}
\begin{tikzpicture}[thick]
\fill[black] (0,0.0) circle (0.05cm);
\fill[black] (0,0.3) circle (0.05cm);
\fill[black] (0,0.6) circle (0.05cm);
\fill[black] (0,0.9) circle (0.05cm);
\draw (0,0) -- (0.1,0) arc (-90:90:0.45) -- (0.,0.9);
\draw (0,0.6) -- (0.1,0.6) arc (90:-90:0.15cm) -- (0,0.3);
\fill[black] (1.2,0.0) circle (0.05cm);
\fill[black] (1.2,0.3) circle (0.05cm);
\fill[black] (1.2,0.6) circle (0.05cm);
\fill[black] (1.2,0.9) circle (0.05cm);
\draw (1.2,0.9) -- (1.1,0.9) arc (90:270:0.45) -- (1.2,0);
\draw (1.2,0.6) -- (1.1,0.6) arc (90:270:0.15cm) -- (1.2,0.3);
\end{tikzpicture} 
\end{array}
\,,\qquad  
\begin{array}{c}
\begin{tikzpicture}[thick]
\fill[black] (0,0.0) circle (0.05cm);
\fill[black] (0,0.3) circle (0.05cm);
\fill[black] (0,0.6) circle (0.05cm);
\fill[black] (0,0.9) circle (0.05cm);
\draw (0,0) -- (0.4,0) -- (1.2,0.0);
\draw (0,0.6) -- (0.1,0.6) arc (90:-90:0.15cm) -- (0,0.3);
\fill[black] (1.2,0.0) circle (0.05cm);
\fill[black] (1.2,0.3) circle (0.05cm);
\fill[black] (1.2,0.6) circle (0.05cm);
\fill[black] (1.2,0.9) circle (0.05cm);
\draw (1.2,0.9) -- (0,0.9);
\draw (1.2,0.6) -- (1.1,0.6) arc (90:270:0.15cm) -- (1.2,0.3);
\end{tikzpicture} 
\end{array}
\,,\qquad  
\begin{array}{c}
\begin{tikzpicture}[thick]
\fill[black] (0,0.0) circle (0.05cm);
\fill[black] (0,0.3) circle (0.05cm);
\fill[black] (0,0.6) circle (0.05cm);
\fill[black] (0,0.9) circle (0.05cm);
\draw (0,0) -- (1.2,0);
\draw (0,0.6) -- (1.2,0.6);
\fill[black] (1.2,0.0) circle (0.05cm);
\fill[black] (1.2,0.3) circle (0.05cm);
\fill[black] (1.2,0.6) circle (0.05cm);
\fill[black] (1.2,0.9) circle (0.05cm);
\draw (1.2,0.3) -- (0,0.3);
\draw (1.2,0.9) -- (0,0.9);
\end{tikzpicture} 
\end{array}\,,\\
n=4\,: & & \begin{array}{c}
\begin{tikzpicture}[thick]
\fill[black] (0,0.0) circle (0.05cm);
\fill[black] (0,0.3) circle (0.05cm);
\fill[black] (0,0.6) circle (0.05cm);
\fill[black] (0,0.9) circle (0.05cm);
\draw (0,0.0) -- (0.1,0.0) arc (-90:0:0.45cm) -- (0.55,1.05) arc (0:90:0.45) -- (0,1.5);
\draw (0,1.2) -- (0.1,1.2) arc (90:0:0.3cm) -- (0.4,0.6) arc (0:-90:0.3) -- (0,0.3);
\fill[black] (0,1.2) circle (0.05cm);
\fill[black] (0,1.5) circle (0.05cm);
\draw (0,0.6) -- (0.1,0.6) arc (-90:90:0.15cm) -- (0,0.9);
\fill[black] (1.2,0.0) circle (0.05cm);
\fill[black] (1.2,0.3) circle (0.05cm);
\fill[black] (1.2,0.6) circle (0.05cm);
\fill[black] (1.2,0.9) circle (0.05cm);
\fill[black] (1.2,1.2) circle (0.05cm);
\fill[black] (1.2,1.5) circle (0.05cm);
\draw (1.2,0.0) -- (1.1,0.0) arc (270:180:0.45cm) -- (0.65,1.05) arc (180:90:0.45) -- (1.2,1.5);
\draw (1.2,1.2) -- (1.1,1.2) arc (90:180:0.3cm) -- (0.8,0.6) arc (180:270:0.3) -- (1.2,0.3);
\draw (1.2,0.6) -- (1.1,0.6) arc (270:90:0.15cm) -- (1.2,0.9);
\end{tikzpicture} 
\end{array}
\,,\qquad \begin{array}{c}
\begin{tikzpicture}[thick]
\fill[black] (0,0.0) circle (0.05cm);
\fill[black] (0,0.3) circle (0.05cm);
\fill[black] (0,0.6) circle (0.05cm);
\fill[black] (0,0.9) circle (0.05cm);
\draw (0,0) -- (0.1,0) arc (-90:90:0.45) -- (0.,0.9);
\draw (0,0.6) -- (0.1,0.6) arc (90:-90:0.15cm) -- (0,0.3);
\fill[black] (1.2,0.0) circle (0.05cm);
\fill[black] (1.2,0.3) circle (0.05cm);
\fill[black] (1.2,0.6) circle (0.05cm);
\fill[black] (1.2,0.9) circle (0.05cm);
\draw (1.2,0.9) -- (1.1,0.9) arc (90:270:0.45) -- (1.2,0);
\draw (1.2,0.6) -- (1.1,0.6) arc (90:270:0.15cm) -- (1.2,0.3);
\fill[black] (0,-0.3) circle (0.05cm);
\fill[black] (0,1.2) circle (0.05cm);
\fill[black] (1.2,-0.3) circle (0.05cm);
\fill[black] (1.2,1.2) circle (0.05cm);
\draw (1.2,-0.3) -- (0,-0.3);
\draw (1.2,1.2) -- (0,1.2);
\end{tikzpicture} 
\end{array}
\,,\qquad  
\begin{array}{c}
\begin{tikzpicture}[thick]
\fill[black] (0,0.0) circle (0.05cm);
\fill[black] (0,0.3) circle (0.05cm);
\fill[black] (0,0.6) circle (0.05cm);
\fill[black] (0,0.9) circle (0.05cm);
\draw (0,0) -- (0.4,0) -- (1.2,0.0);
\draw (0,0.6) -- (0.1,0.6) arc (90:-90:0.15cm) -- (0,0.3);
\fill[black] (1.2,0.0) circle (0.05cm);
\fill[black] (1.2,0.3) circle (0.05cm);
\fill[black] (1.2,0.6) circle (0.05cm);
\fill[black] (1.2,0.9) circle (0.05cm);
\draw (1.2,0.9) -- (0,0.9);
\draw (1.2,0.6) -- (1.1,0.6) arc (90:270:0.15cm) -- (1.2,0.3);
\fill[black] (0,-0.3) circle (0.05cm);
\fill[black] (0,1.2) circle (0.05cm);
\fill[black] (1.2,-0.3) circle (0.05cm);
\fill[black] (1.2,1.2) circle (0.05cm);
\draw (1.2,-0.3) -- (0,-0.3);
\draw (1.2,1.2) -- (0,1.2);
\end{tikzpicture} 
\end{array}
\,,\qquad  
\begin{array}{c}
\begin{tikzpicture}[thick]
\fill[black] (0,0.0) circle (0.05cm);
\fill[black] (0,0.3) circle (0.05cm);
\fill[black] (0,0.6) circle (0.05cm);
\fill[black] (0,0.9) circle (0.05cm);
\draw (0,0) -- (1.2,0);
\draw (0,0.6) -- (1.2,0.6);
\fill[black] (1.2,0.0) circle (0.05cm);
\fill[black] (1.2,0.3) circle (0.05cm);
\fill[black] (1.2,0.6) circle (0.05cm);
\fill[black] (1.2,0.9) circle (0.05cm);
\draw (1.2,0.3) -- (0,0.3);
\draw (1.2,0.9) -- (0,0.9);
\fill[black] (0,-0.3) circle (0.05cm);
\fill[black] (0,1.2) circle (0.05cm);
\fill[black] (1.2,-0.3) circle (0.05cm);
\fill[black] (1.2,1.2) circle (0.05cm);
\draw (1.2,-0.3) -- (0,-0.3);
\draw (1.2,1.2) -- (0,1.2);
\end{tikzpicture} 
\end{array}\,.
\end{eqnarray}

This classification has a straightforward generalization to the case of a pair of inequivalent qudits.


\section{Tripartite entanglement}
\label{sec:3partite}

In this section we extend the topological classification method to the case of tripartite entanglement. We will focus on the situation with three qubits, for which the classification reduces to drawing all possible connections, with no need of projectors.

In the case of tripartite entanglement between parties $A$, $B$ and $C$, SLOCC classification yields four distinct classes of entanglement~\cite{Dur:2000zz}. Among the four classes there are the fully separable cases (in terms of reduced density matrices of either of subsystems $A$, $B$, or $C$ corresponding to ${\rm rk}\rho_{A}={\rm rk}\rho_{B}={\rm rk}\rho_{C}=1$), the partially separable Bell case (only one of $\rho_A$, $\rho_B$, or $\rho_C$ having rank one) and two genuinely tripartite classes named GHZ and W. In a general TQFT these characteristic classes should be captured by the following cobordism diagrams:
\begin{eqnarray}
\label{3party0}
\begin{array}{c}\includegraphics[scale=0.2]{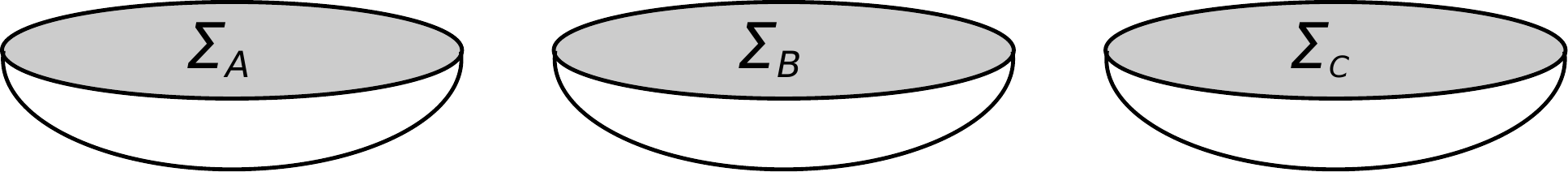}\end{array}, & \begin{array}{c}\includegraphics[scale=0.2]{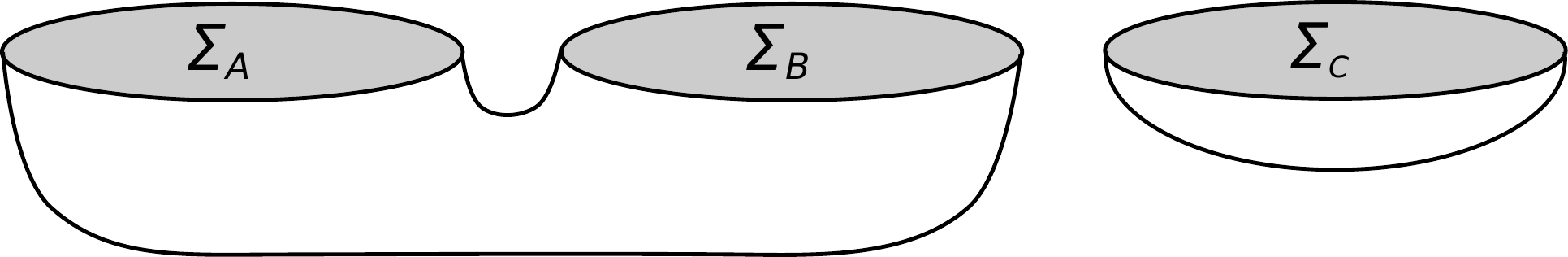}\end{array}, & \begin{array}{c}\includegraphics[scale=0.2]{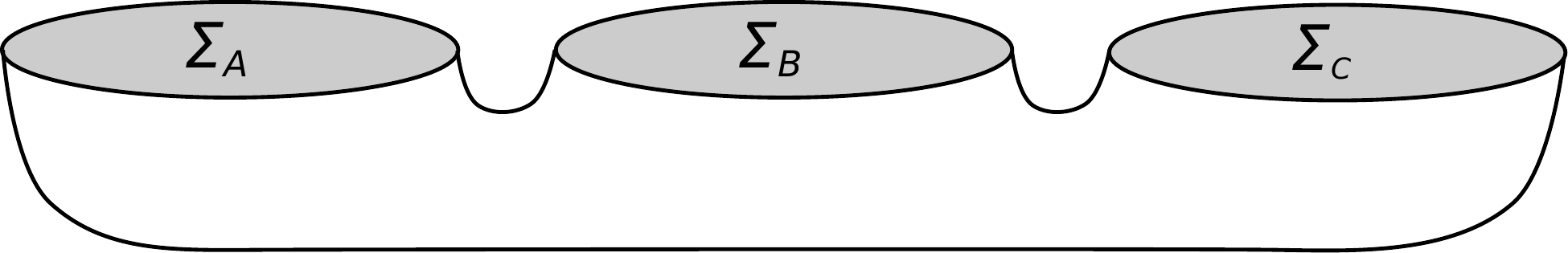}\end{array}.
\end{eqnarray}
The first two diagrams clearly represent the separable and the Bell case. The last one typically means the GHZ entanglement, as we will discuss. W class is more intricate. We will only be able to obtain it as a special case of the fully-connected diagram above.

We will focus on 1D TQFT or on 3D TQFT with $S^2$ boundaries, as explained before. This means that the entanglement will be supported by a skeleton of Wilson lines. The Hilbert space of a qubit is represented by four points (punctures). We will consider states of three qubits up to local permutations of the points. 

There are seven nonequivalent ways of connecting the punctures, up to local permutation of points and permutation of parties $A$, $B$ and $C$:
\be
\newcommand{\x}{0.75}
\begin{array}{c}
\scalebox{\x}{\begin{tikzpicture}[thick]
\fill[black] (0,0.0) circle (0.05cm);
\fill[black] (0,0.2) circle (0.05cm);
\fill[black] (0,0.4) circle (0.05cm);
\fill[black] (0,0.6) circle (0.05cm);
\draw (0,0) -- (0.2,0) arc (-90:90:0.1) -- (0,0.2);
\draw (0,0.4) -- (0.2,0.4) arc (-90:90:0.1) -- (0,0.6);
\fill[black] (1.8,0.0) circle (0.05cm);
\fill[black] (1.8,0.2) circle (0.05cm);
\fill[black] (1.8,0.4) circle (0.05cm);
\fill[black] (1.8,0.6) circle (0.05cm);
\draw (1.8,0.4) -- (1.6,0.4) arc (270:90:0.1) -- (1.8,0.6);
\draw (1.8,0.) -- (1.6,0.) arc (270:90:0.1) -- (1.8,0.2);
\fill[black] (0.6,1.2) circle (0.05cm);
\fill[black] (0.8,1.2) circle (0.05cm);
\fill[black] (1.,1.2) circle (0.05cm);
\fill[black] (1.2,1.2) circle (0.05cm);
\draw (0.6,1.2) -- (0.6,1.0) arc (-180:0:0.1) -- (0.8,1.2);
\draw (1.0,1.2) -- (1.0,1.0) arc (-180:0:0.1) -- (1.2,1.2);
\end{tikzpicture}} 
\end{array}
\quad
\begin{array}{c}
\scalebox{\x}{\begin{tikzpicture}[thick]
\fill[black] (0,0.0) circle (0.05cm);
\fill[black] (0,0.2) circle (0.05cm);
\fill[black] (0,0.4) circle (0.05cm);
\fill[black] (0,0.6) circle (0.05cm);
\draw (0,0.4) -- (0.2,0.4) arc (-90:90:0.1) -- (0,0.6);
\draw (0,0) -- (1.8,0.0);
\fill[black] (1.8,0.0) circle (0.05cm);
\fill[black] (1.8,0.2) circle (0.05cm);
\fill[black] (1.8,0.4) circle (0.05cm);
\fill[black] (1.8,0.6) circle (0.05cm);
\draw (1.8,0.2) -- (0,0.2);
\draw (1.8,0.4) -- (1.6,0.4) arc (270:90:0.1) -- (1.8,0.6);
\fill[black] (0.6,1.2) circle (0.05cm);
\fill[black] (0.8,1.2) circle (0.05cm);
\fill[black] (1.,1.2) circle (0.05cm);
\fill[black] (1.2,1.2) circle (0.05cm);
\draw (0.6,1.2) -- (0.6,1.0) arc (-180:0:0.1) -- (0.8,1.2);
\draw (1.0,1.2) -- (1.0,1.0) arc (-180:0:0.1) -- (1.2,1.2);
\end{tikzpicture}} 
\end{array}
\quad
\begin{array}{c}
\scalebox{\x}{\begin{tikzpicture}[thick]
\fill[black] (0,0.0) circle (0.05cm);
\fill[black] (0,0.2) circle (0.05cm);
\fill[black] (0,0.4) circle (0.05cm);
\fill[black] (0,0.6) circle (0.05cm);
\fill[black] (1.8,0.0) circle (0.05cm);
\fill[black] (1.8,0.2) circle (0.05cm);
\fill[black] (1.8,0.4) circle (0.05cm);
\fill[black] (1.8,0.6) circle (0.05cm);
\fill[black] (0.6,1.2) circle (0.05cm);
\fill[black] (0.8,1.2) circle (0.05cm);
\fill[black] (1.,1.2) circle (0.05cm);
\fill[black] (1.2,1.2) circle (0.05cm);
\draw (0.6,1.2) -- (0.6,1.0) arc (0:-90:0.4) -- (0.,0.6);
\draw (0.8,1.2) -- (0.8,1.0) arc (0:-90:0.6) -- (0.,0.4);
\draw (1.0,1.2) -- (1.0,1.0) arc (180:270:0.6) -- (1.8,0.4);
\draw (1.2,1.2) -- (1.2,1.0) arc (180:270:0.4) -- (1.8,0.6);
\draw (0,0) -- (0.2,0) arc (-90:90:0.1) -- (0,0.2);
\draw (1.8,0.) -- (1.6,0.) arc (270:90:0.1) -- (1.8,0.2);
\end{tikzpicture}}
\end{array}
\quad
\begin{array}{c}
\scalebox{\x}{\begin{tikzpicture}[thick]
\fill[black] (0,0.0) circle (0.05cm);
\fill[black] (0,0.2) circle (0.05cm);
\fill[black] (0,0.4) circle (0.05cm);
\fill[black] (0,0.6) circle (0.05cm);
\draw (0,0.4) -- (0.2,0.4) arc (90:-90:0.1) -- (0,0.2);
\draw (0,0) -- (1.8,0.0);
\fill[black] (1.8,0.0) circle (0.05cm);
\fill[black] (1.8,0.2) circle (0.05cm);
\fill[black] (1.8,0.4) circle (0.05cm);
\fill[black] (1.8,0.6) circle (0.05cm);
\draw (1.8,0.4) -- (1.6,0.4) arc (90:270:0.1) -- (1.8,0.2);
\fill[black] (0.6,1.2) circle (0.05cm);
\fill[black] (0.8,1.2) circle (0.05cm);
\fill[black] (1.,1.2) circle (0.05cm);
\fill[black] (1.2,1.2) circle (0.05cm);
\draw (0.6,1.2) -- (0.6,1.0) arc (0:-90:0.4) -- (0.,0.6);
\draw (0.8,1.2) -- (0.8,1.0) arc (-180:0:0.1) -- (1.0,1.2);
\draw (1.2,1.2) -- (1.2,1.0) arc (180:270:0.4) -- (1.8,0.6);
\end{tikzpicture}}
\end{array}
\quad
\begin{array}{c}
\scalebox{\x}{\begin{tikzpicture}[thick]
\fill[black] (0,0.0) circle (0.05cm);
\fill[black] (0,0.2) circle (0.05cm);
\fill[black] (0,0.4) circle (0.05cm);
\fill[black] (0,0.6) circle (0.05cm);
\fill[black] (1.8,0.0) circle (0.05cm);
\fill[black] (1.8,0.2) circle (0.05cm);
\fill[black] (1.8,0.4) circle (0.05cm);
\fill[black] (1.8,0.6) circle (0.05cm);
\draw (1.8,0.2) -- (0,0.2);
\draw (1.8,0.) -- (0,0.);
\draw (1.8,0.6) -- (0,0.6);
\draw (1.8,0.4) -- (0,0.4);
\fill[black] (0.6,1.2) circle (0.05cm);
\fill[black] (0.8,1.2) circle (0.05cm);
\fill[black] (1.,1.2) circle (0.05cm);
\fill[black] (1.2,1.2) circle (0.05cm);
\draw (0.6,1.2) -- (0.6,1.0) arc (-180:0:0.1) -- (0.8,1.2);
\draw (1.0,1.2) -- (1.0,1.0) arc (-180:0:0.1) -- (1.2,1.2);
\end{tikzpicture}}
\end{array}
\quad
\begin{array}{c}
\scalebox{\x}{\begin{tikzpicture}[thick]
\fill[black] (0,0.0) circle (0.05cm);
\fill[black] (0,0.2) circle (0.05cm);
\fill[black] (0,0.4) circle (0.05cm);
\fill[black] (0,0.6) circle (0.05cm);
\draw (0,0.4) -- (1.8,0.4);
\draw (0,0) -- (1.8,0.0);
\fill[black] (1.8,0.0) circle (0.05cm);
\fill[black] (1.8,0.2) circle (0.05cm);
\fill[black] (1.8,0.4) circle (0.05cm);
\fill[black] (1.8,0.6) circle (0.05cm);
\draw (0,0.2) -- (1.8,0.2);
\fill[black] (0.6,1.2) circle (0.05cm);
\fill[black] (0.8,1.2) circle (0.05cm);
\fill[black] (1.,1.2) circle (0.05cm);
\fill[black] (1.2,1.2) circle (0.05cm);
\draw (0.6,1.2) -- (0.6,1.0) arc (0:-90:0.4) -- (0.,0.6);
\draw (0.8,1.2) -- (0.8,1.0) arc (-180:0:0.1) -- (1.0,1.2);
\draw (1.2,1.2) -- (1.2,1.0) arc (180:270:0.4) -- (1.8,0.6);
\end{tikzpicture}}
\end{array}
\quad
\begin{array}{c}
\scalebox{\x}{\begin{tikzpicture}[thick]
\fill[black] (0,0.0) circle (0.05cm);
\fill[black] (0,0.2) circle (0.05cm);
\fill[black] (0,0.4) circle (0.05cm);
\fill[black] (0,0.6) circle (0.05cm);
\draw (0,0) -- (1.8,0.0);
\fill[black] (1.8,0.0) circle (0.05cm);
\fill[black] (1.8,0.2) circle (0.05cm);
\fill[black] (1.8,0.4) circle (0.05cm);
\fill[black] (1.8,0.6) circle (0.05cm);
\draw (0,0.2) -- (1.8,0.2);
\fill[black] (0.6,1.2) circle (0.05cm);
\fill[black] (0.8,1.2) circle (0.05cm);
\fill[black] (1.,1.2) circle (0.05cm);
\fill[black] (1.2,1.2) circle (0.05cm);
\draw (0.6,1.2) -- (0.6,1.0) arc (0:-90:0.4) -- (0.,0.6);
\draw (0.8,1.2) -- (0.8,1.0) arc (0:-90:0.6) -- (0.,0.4);
\draw (1.0,1.2) -- (1.0,1.0) arc (180:270:0.6) -- (1.8,0.4);
\draw (1.2,1.2) -- (1.2,1.0) arc (180:270:0.4) -- (1.8,0.6);
\end{tikzpicture}} 
\end{array}
\label{3partydiags}
\ee
Since a party connected to the rest of the system by only two lines is not entangled with the rest as~(\ref{breaking}) suggests (the lines can be cut), we conclude that the diagrams only encode three nonequivalent classes, as the heuristic classification in equation~(\ref{3party0}). Namely, the first four diagrams describe separable states, the next pair -- the Bell type of entanglement, and the last one -- a truly tripartite entanglement. The class of the last state can be deduced from explicit evaluation in basis~(\ref{4pointbasis}). One thus obtains
\begin{equation}
        \begin{array}{c}
\begin{tikzpicture}[thick]
\fill[black] (0,0.0) circle (0.05cm);
\fill[black] (0,0.2) circle (0.05cm);
\fill[black] (0,0.4) circle (0.05cm);
\fill[black] (0,0.6) circle (0.05cm);
\draw (0,0) -- (1.8,0.0);
\fill[black] (1.8,0.0) circle (0.05cm);
\fill[black] (1.8,0.2) circle (0.05cm);
\fill[black] (1.8,0.4) circle (0.05cm);
\fill[black] (1.8,0.6) circle (0.05cm);
\draw (0,0.2) -- (1.8,0.2);
\fill[black] (0.6,1.2) circle (0.05cm);
\fill[black] (0.8,1.2) circle (0.05cm);
\fill[black] (1.,1.2) circle (0.05cm);
\fill[black] (1.2,1.2) circle (0.05cm);
\draw (0.6,1.2) -- (0.6,1.0) arc (0:-90:0.4) -- (0.,0.6);
\draw (0.8,1.2) -- (0.8,1.0) arc (0:-90:0.6) -- (0.,0.4);
\draw (1.0,1.2) -- (1.0,1.0) arc (180:270:0.6) -- (1.8,0.4);
\draw (1.2,1.2) -- (1.2,1.0) arc (180:270:0.4) -- (1.8,0.6);
\end{tikzpicture} 
\end{array} \ =\  |000\rangle + \frac{1}{\sqrt{d^2 - 1}} |111\rangle \,,
    \end{equation}
which is a GHZ state.

For completeness we also give the expressions for the remaining diagrams in this basis:
\begin{eqnarray}
\begin{array}{c}
\begin{tikzpicture}[thick]
\fill[black] (0,0.0) circle (0.05cm);
\fill[black] (0,0.2) circle (0.05cm);
\fill[black] (0,0.4) circle (0.05cm);
\fill[black] (0,0.6) circle (0.05cm);
\draw (0,0) -- (0.2,0) arc (-90:90:0.1) -- (0,0.2);
\draw (0,0.4) -- (0.2,0.4) arc (-90:90:0.1) -- (0,0.6);
\fill[black] (1.8,0.0) circle (0.05cm);
\fill[black] (1.8,0.2) circle (0.05cm);
\fill[black] (1.8,0.4) circle (0.05cm);
\fill[black] (1.8,0.6) circle (0.05cm);
\draw (1.8,0.4) -- (1.6,0.4) arc (270:90:0.1) -- (1.8,0.6);
\draw (1.8,0.) -- (1.6,0.) arc (270:90:0.1) -- (1.8,0.2);
\fill[black] (0.6,1.2) circle (0.05cm);
\fill[black] (0.8,1.2) circle (0.05cm);
\fill[black] (1.,1.2) circle (0.05cm);
\fill[black] (1.2,1.2) circle (0.05cm);
\draw (0.6,1.2) -- (0.6,1.0) arc (-180:0:0.1) -- (0.8,1.2);
\draw (1.0,1.2) -- (1.0,1.0) arc (-180:0:0.1) -- (1.2,1.2);
\end{tikzpicture} 
\end{array}
& = &  d^3|000\rangle \,, \\
\begin{array}{c}
\begin{tikzpicture}[thick]
\fill[black] (0,0.0) circle (0.05cm);
\fill[black] (0,0.2) circle (0.05cm);
\fill[black] (0,0.4) circle (0.05cm);
\fill[black] (0,0.6) circle (0.05cm);
\draw (0,0.4) -- (0.2,0.4) arc (-90:90:0.1) -- (0,0.6);
\draw (0,0) -- (1.8,0.0);
\fill[black] (1.8,0.0) circle (0.05cm);
\fill[black] (1.8,0.2) circle (0.05cm);
\fill[black] (1.8,0.4) circle (0.05cm);
\fill[black] (1.8,0.6) circle (0.05cm);
\draw (1.8,0.2) -- (0,0.2);
\draw (1.8,0.4) -- (1.6,0.4) arc (270:90:0.1) -- (1.8,0.6);
\fill[black] (0.6,1.2) circle (0.05cm);
\fill[black] (0.8,1.2) circle (0.05cm);
\fill[black] (1.,1.2) circle (0.05cm);
\fill[black] (1.2,1.2) circle (0.05cm);
\draw (0.6,1.2) -- (0.6,1.0) arc (-180:0:0.1) -- (0.8,1.2);
\draw (1.0,1.2) -- (1.0,1.0) arc (-180:0:0.1) -- (1.2,1.2);
\end{tikzpicture} 
\end{array}
& = &   d^2|000\rangle\,,\\
\begin{array}{c}
\begin{tikzpicture}[thick]
\fill[black] (0,0.0) circle (0.05cm);
\fill[black] (0,0.2) circle (0.05cm);
\fill[black] (0,0.4) circle (0.05cm);
\fill[black] (0,0.6) circle (0.05cm);
\fill[black] (1.8,0.0) circle (0.05cm);
\fill[black] (1.8,0.2) circle (0.05cm);
\fill[black] (1.8,0.4) circle (0.05cm);
\fill[black] (1.8,0.6) circle (0.05cm);
\fill[black] (0.6,1.2) circle (0.05cm);
\fill[black] (0.8,1.2) circle (0.05cm);
\fill[black] (1.,1.2) circle (0.05cm);
\fill[black] (1.2,1.2) circle (0.05cm);
\draw (0.6,1.2) -- (0.6,1.0) arc (0:-90:0.4) -- (0.,0.6);
\draw (0.8,1.2) -- (0.8,1.0) arc (0:-90:0.6) -- (0.,0.4);
\draw (1.0,1.2) -- (1.0,1.0) arc (180:270:0.6) -- (1.8,0.4);
\draw (1.2,1.2) -- (1.2,1.0) arc (180:270:0.4) -- (1.8,0.6);
\draw (0,0) -- (0.2,0) arc (-90:90:0.1) -- (0,0.2);
\draw (1.8,0.) -- (1.6,0.) arc (270:90:0.1) -- (1.8,0.2);
\end{tikzpicture}
\end{array} & = & d|000\rangle\,, \\
\begin{array}{c}
\begin{tikzpicture}[thick]
\fill[black] (0,0.0) circle (0.05cm);
\fill[black] (0,0.2) circle (0.05cm);
\fill[black] (0,0.4) circle (0.05cm);
\fill[black] (0,0.6) circle (0.05cm);
\draw (0,0.4) -- (0.2,0.4) arc (90:-90:0.1) -- (0,0.2);
\draw (0,0) -- (1.8,0.0);
\fill[black] (1.8,0.0) circle (0.05cm);
\fill[black] (1.8,0.2) circle (0.05cm);
\fill[black] (1.8,0.4) circle (0.05cm);
\fill[black] (1.8,0.6) circle (0.05cm);
\draw (1.8,0.4) -- (1.6,0.4) arc (90:270:0.1) -- (1.8,0.2);
\fill[black] (0.6,1.2) circle (0.05cm);
\fill[black] (0.8,1.2) circle (0.05cm);
\fill[black] (1.,1.2) circle (0.05cm);
\fill[black] (1.2,1.2) circle (0.05cm);
\draw (0.6,1.2) -- (0.6,1.0) arc (0:-90:0.4) -- (0.,0.6);
\draw (0.8,1.2) -- (0.8,1.0) arc (-180:0:0.1) -- (1.0,1.2);
\draw (1.2,1.2) -- (1.2,1.0) arc (180:270:0.4) -- (1.8,0.6);
\end{tikzpicture} 
\end{array}
& = &   \frac{1}{d^2} \bigg\{ |000\rangle  + \big( d^2 -1\big)^{1/2} 
        \big( |001\rangle + |010\rangle + |100\rangle \big) \nn \\ 
        & & + \big( d^2 -1\big) \big( |011\rangle + |101\rangle + |110\rangle \big)+ \big( d^2 -1\big)^{3/2} |111\rangle \bigg\} \,,\\
\begin{array}{c}
\begin{tikzpicture}[thick]
\fill[black] (0,0.0) circle (0.05cm);
\fill[black] (0,0.2) circle (0.05cm);
\fill[black] (0,0.4) circle (0.05cm);
\fill[black] (0,0.6) circle (0.05cm);
\fill[black] (1.8,0.0) circle (0.05cm);
\fill[black] (1.8,0.2) circle (0.05cm);
\fill[black] (1.8,0.4) circle (0.05cm);
\fill[black] (1.8,0.6) circle (0.05cm);
\draw (1.8,0.2) -- (0,0.2);
\draw (1.8,0.) -- (0,0.);
\draw (1.8,0.6) -- (0,0.6);
\draw (1.8,0.4) -- (0,0.4);
\fill[black] (0.6,1.2) circle (0.05cm);
\fill[black] (0.8,1.2) circle (0.05cm);
\fill[black] (1.,1.2) circle (0.05cm);
\fill[black] (1.2,1.2) circle (0.05cm);
\draw (0.6,1.2) -- (0.6,1.0) arc (-180:0:0.1) -- (0.8,1.2);
\draw (1.0,1.2) -- (1.0,1.0) arc (-180:0:0.1) -- (1.2,1.2);
\end{tikzpicture} 
\end{array}
& = &  d\left(|000\rangle + |011\rangle\right)\,,\\
\begin{array}{c}
\begin{tikzpicture}[thick]
\fill[black] (0,0.0) circle (0.05cm);
\fill[black] (0,0.2) circle (0.05cm);
\fill[black] (0,0.4) circle (0.05cm);
\fill[black] (0,0.6) circle (0.05cm);
\draw (0,0.4) -- (1.8,0.4);
\draw (0,0) -- (1.8,0.0);
\fill[black] (1.8,0.0) circle (0.05cm);
\fill[black] (1.8,0.2) circle (0.05cm);
\fill[black] (1.8,0.4) circle (0.05cm);
\fill[black] (1.8,0.6) circle (0.05cm);
\draw (0,0.2) -- (1.8,0.2);
\fill[black] (0.6,1.2) circle (0.05cm);
\fill[black] (0.8,1.2) circle (0.05cm);
\fill[black] (1.,1.2) circle (0.05cm);
\fill[black] (1.2,1.2) circle (0.05cm);
\draw (0.6,1.2) -- (0.6,1.0) arc (0:-90:0.4) -- (0.,0.6);
\draw (0.8,1.2) -- (0.8,1.0) arc (-180:0:0.1) -- (1.0,1.2);
\draw (1.2,1.2) -- (1.2,1.0) arc (180:270:0.4) -- (1.8,0.6);
\end{tikzpicture} 
\end{array}
& = & \frac{1}{d}  \bigg[|000\rangle + \big(d^2 - 1\big)^{1/2}|010\rangle +|101\rangle + \big(d^2 - 1\big)^{1/2}|111\rangle \bigg]  \,.
\end{eqnarray}

The diagrams are equivalent to unordered 3-vertex graphs with adjacency matrices, whose entries in the lines and columns add up to four. For example,
\be
\begin{array}{c}
\scalebox{0.8}{\begin{tikzpicture}[thick]
\fill[black] (0,0.0) circle (0.05cm);
\fill[black] (0,0.2) circle (0.05cm);
\fill[black] (0,0.4) circle (0.05cm);
\fill[black] (0,0.6) circle (0.05cm);
\draw (0,0) -- (0.2,0) arc (-90:90:0.1) -- (0,0.2);
\draw (0,0.4) -- (0.2,0.4) arc (-90:90:0.1) -- (0,0.6);
\fill[black] (1.8,0.0) circle (0.05cm);
\fill[black] (1.8,0.2) circle (0.05cm);
\fill[black] (1.8,0.4) circle (0.05cm);
\fill[black] (1.8,0.6) circle (0.05cm);
\draw (1.8,0.4) -- (1.6,0.4) arc (270:90:0.1) -- (1.8,0.6);
\draw (1.8,0.) -- (1.6,0.) arc (270:90:0.1) -- (1.8,0.2);
\fill[black] (0.6,1.2) circle (0.05cm);
\fill[black] (0.8,1.2) circle (0.05cm);
\fill[black] (1.,1.2) circle (0.05cm);
\fill[black] (1.2,1.2) circle (0.05cm);
\draw (0.6,1.2) -- (0.6,1.0) arc (-180:0:0.1) -- (0.8,1.2);
\draw (1.0,1.2) -- (1.0,1.0) arc (-180:0:0.1) -- (1.2,1.2);
\end{tikzpicture}} 
\end{array} \ \leftrightarrow \ \left(\begin{array}{ccc}
     4 & 0 & 0 \\
     0 & 4 & 0\\
     0 & 0 & 4
\end{array}\right), \quad 
\begin{array}{c}
\scalebox{0.8}{\begin{tikzpicture}[thick]
\fill[black] (0,0.0) circle (0.05cm);
\fill[black] (0,0.2) circle (0.05cm);
\fill[black] (0,0.4) circle (0.05cm);
\fill[black] (0,0.6) circle (0.05cm);
\fill[black] (1.8,0.0) circle (0.05cm);
\fill[black] (1.8,0.2) circle (0.05cm);
\fill[black] (1.8,0.4) circle (0.05cm);
\fill[black] (1.8,0.6) circle (0.05cm);
\draw (1.8,0.2) -- (0,0.2);
\draw (1.8,0.) -- (0,0.);
\draw (1.8,0.6) -- (0,0.6);
\draw (1.8,0.4) -- (0,0.4);
\fill[black] (0.6,1.2) circle (0.05cm);
\fill[black] (0.8,1.2) circle (0.05cm);
\fill[black] (1.,1.2) circle (0.05cm);
\fill[black] (1.2,1.2) circle (0.05cm);
\draw (0.6,1.2) -- (0.6,1.0) arc (-180:0:0.1) -- (0.8,1.2);
\draw (1.0,1.2) -- (1.0,1.0) arc (-180:0:0.1) -- (1.2,1.2);
\end{tikzpicture}} 
\end{array}
\ \leftrightarrow \ \left(\begin{array}{ccc}
     0 & 0 & 4 \\
     0 & 4 & 0\\
     4 & 0 & 0
\end{array}\right),\quad
\begin{array}{c}
\scalebox{0.8}{\begin{tikzpicture}[thick]
\fill[black] (0,0.0) circle (0.05cm);
\fill[black] (0,0.2) circle (0.05cm);
\fill[black] (0,0.4) circle (0.05cm);
\fill[black] (0,0.6) circle (0.05cm);
\draw (0,0) -- (1.8,0.0);
\fill[black] (1.8,0.0) circle (0.05cm);
\fill[black] (1.8,0.2) circle (0.05cm);
\fill[black] (1.8,0.4) circle (0.05cm);
\fill[black] (1.8,0.6) circle (0.05cm);
\draw (0,0.2) -- (1.8,0.2);
\fill[black] (0.6,1.2) circle (0.05cm);
\fill[black] (0.8,1.2) circle (0.05cm);
\fill[black] (1.,1.2) circle (0.05cm);
\fill[black] (1.2,1.2) circle (0.05cm);
\draw (0.6,1.2) -- (0.6,1.0) arc (0:-90:0.4) -- (0.,0.6);
\draw (0.8,1.2) -- (0.8,1.0) arc (0:-90:0.6) -- (0.,0.4);
\draw (1.0,1.2) -- (1.0,1.0) arc (180:270:0.6) -- (1.8,0.4);
\draw (1.2,1.2) -- (1.2,1.0) arc (180:270:0.4) -- (1.8,0.6);
\end{tikzpicture}}
\end{array} 
\ \leftrightarrow \ \left(\begin{array}{ccc}
     0 & 2 & 2 \\
     2 & 0 & 2\\
     2 & 2 & 0
\end{array}\right).
\ee

This classification misses the W class of SLOCC. Local invertible operations can only generate a W state from a GHZ state asymptotically, that is with a set accuracy, so the above set of diagrams is a more coarse classification of entanglement classes than the SLOCC one.

One can try to make a finer classification by allowing nonlocal braiding of the lines. As was mentioned, braiding offers an infinite number of options, so the question that remains is how one can systematically restrict these options to a finite number of classes, or whether there is a minimal diagram for the W state. In what follows we will present a candidate state, which is at least an approximation to a W state, though no finer but finite classification will be attempted in this work.


The W-type entanglement can be detected with the help of 3-tangle $\tau_{ABC}$ introduced in~\cite{Coffman:1999jd}. It is an entanglement monotone, which means it cannot increase under the action of LOCC. For the GHZ type of entanglement $\tau_{ABC}>0$, with $\tau_{ABC}=1$ for the GHZ state. For the W (as well as for the Bell and for the separable) entanglement $\tau_{ABC}=0$. Let us try to design a tripartite entangled state that is not related to GHZ states by a local operator.

Let us consider the state: 
\begin{multline}
\label{quasiW}
\begin{array}{c}
\begin{tikzpicture}[thick]
\draw[draw=white,double=black] (0.6,0.65) arc (180:135:0.3) arc (-45:0:0.4);
\draw[draw=white,double=black] (0.9,0.35) arc (270:360:0.3);
\draw[draw=white,double=black] (0,0.2) -- (0.8,0.2) arc (-90:180:0.3) arc (0:-90:0.1) -- (0,0.4);
\draw[draw=white,double=black] (1.8,0.2) -- (1.0,0.2) arc (270:0:0.3) arc (180:270:0.1) -- (1.8,0.4);
\draw[draw=white,double=black] (0.6,0.65) arc (180:270:0.3);
\draw[draw=white,double=black] (1.2,0.65) arc (360:405:0.3) arc (225:180:0.4);
\fill[black] (0,0.0) circle (0.05cm);
\fill[black] (0,0.2) circle (0.05cm);
\fill[black] (0,0.4) circle (0.05cm);
\fill[black] (0,0.6) circle (0.05cm);
\draw (0,0) -- (1.8,0.0);
\fill[black] (1.8,0.0) circle (0.05cm);
\fill[black] (1.8,0.2) circle (0.05cm);
\fill[black] (1.8,0.4) circle (0.05cm);
\fill[black] (1.8,0.6) circle (0.05cm);
\fill[black] (0.6,1.2) circle (0.05cm);
\fill[black] (0.8,1.2) circle (0.05cm);
\fill[black] (1.,1.2) circle (0.05cm);
\fill[black] (1.2,1.2) circle (0.05cm);
\draw (0.6,1.2) -- (0.6,1.0) arc (0:-90:0.4) -- (0.,0.6);
\draw (1.2,1.2) -- (1.2,1.0) arc (180:270:0.4) -- (1.8,0.6);
\end{tikzpicture} 
\end{array} \quad = \ \frac{A^{12}+A^4-1}{A^{12} \left(A^4+1\right)^2}|000\rangle
- \frac{1+ A^4\left(A^8+1\right) \left(A^{20}-3 A^{16}+A^8-3 A^4-1\right)}{A^{12} \left(A^4+1\right)^2 \sqrt{d^2-1}}|111\rangle
\\ + \frac{\sqrt{d^2-1} \left(A^8-A^4+1\right)}{\left(A^4+1\right)^2}\big( |001\rangle + |010\rangle + |100\rangle \big)
+ \frac{-A^{16}+2 A^{12}+A^4+1}{\left(A^4+1\right)^2}\big( |011\rangle + |101\rangle + |110\rangle \big)\,.
\end{multline}
The lines in the middle are linked in the Borromean ring fashion. In order to prepare such a state one needs to permute endpoints belonging to different parties. 

It turns out that state~(\ref{quasiW}) is, in general, a GHZ class state. This can be understood from computing the 3-tangle, which is nonzero for generic values of $\theta = -i\log A$. Yet the 3-tangle has a number of discrete zeroes. For some of the zeroes, e.g. $\theta=0$ and $\theta=\pi/8$, the state is actually separable, but there are also zeroes with all local density matrices entangled, for example, for $\theta\simeq 0.0945868\pi$, which does not correspond to a simple rational multiple of $\pi$. For this approximate zero the state is at least very close to a W-type state.

A number of other examples with different ways of braiding of the lines was considered, but none of them gave a state that would have zero 3-tangle, apart from discrete points corresponding to biseparable states. This indicates that the naively nonlocal operations are, in fact, equivalent to some local ones. The question that remains is the following: What are the truly nonlocal operations and what is the natural presentation for the W state?

Let us also note that in a physical realization of this, or a similar TQFT, the parameters like $k$ or $\theta$ would take a discrete set of values, e.g. commensurate with the discrete filling fractions in the quantum Hall effect. In such a setup it would be difficult to tune the system to a point where~(\ref{quasiW}) is a W state. One should rather aim at a specific local protocol that can generate the desired W states with necessary accuracy.

Finally, if one uses the same method for four qubits then one will find six nonequivalent classes, of which only two are nonbiseparable. This is to be contrasted with nine nonbiseparable, albeit infinite, families of SLOCC in the classification found by~\cite{Verstraete:2002four}, or seven such classes found in the coarse-grained geometric classification in~\cite{Sawicki:2012cri,Walter:2013ent,Sawicki:2014con,Maciazek:2018asy}. The number of classes in the connectivity topological classification instead equals the number of highest weights of irreducible representations in the fourth tensor power of ${\bf 2}$.

\section{Conclusions}
\label{sec:conclusions}

In this paper the possibility of classification of different types of quantum entanglement via states of TQFT with special topology was investigated. In the specific case of 3D TQFT with $S^2$ boundaries, where each boundary represents a party in an entangled system, the topology was realized by different arrangements of (Wilson) lines connecting the parties. It was shown that for the Hilbert spaces of qudits constructed through the Jones-Wenzl projectors, the given topological classification is equivalent to the SLOCC classification in the case of bipartite entanglement. In particular, the classes of bipartite entanglement for a pair of qudits, each considered as a state in $\mathbb{C}^m$ and $\mathbb{C}^n$ ($m\geq n$), can be classified by the following $n$ diagrams:
\be
2n-2\left\{\quad \begin{array}{c}
\begin{tikzpicture}[thick]
\newcommand{\x}{-0.3}
\fill[black] (0+\x,-0.3) circle (0.05cm);
\fill[black] (0+\x,0.3) circle (0.05cm);
\fill[black] (0+\x,0.6) circle (0.05cm);
\fill[black] (0+\x,0.9) circle (0.05cm);
\draw (0+\x,-0.3) -- (0.4+\x,-0.3) arc (-90:0:0.45cm) -- (0.85+\x,1.35) arc (0:90:0.45) -- (0+\x,1.8);
\draw (0+\x,1.2) -- (0.1+\x,1.2) arc (90:0:0.3cm) -- (0.4+\x,0.6) arc (0:-90:0.3) -- (0+\x,0.3);
\fill[black] (0+\x,1.2) circle (0.05cm);
\fill[black] (0+\x,1.8) circle (0.05cm);
\draw (0+\x,0.6) -- (0.1+\x,0.6) arc (-90:90:0.15cm) -- (0+\x,0.9);
\draw (0.33,0.75) node {$\cdots$};
\draw (\x,1.6) node {$\vdots$};
\draw (\x,0.1) node {$\vdots$};
\fill[black] (1.2-\x,-0.3) circle (0.05cm);
\fill[black] (1.2-\x,0.3) circle (0.05cm);
\fill[black] (1.2-\x,0.6) circle (0.05cm);
\fill[black] (1.2-\x,0.9) circle (0.05cm);
\fill[black] (1.2-\x,1.2) circle (0.05cm);
\fill[black] (1.2-\x,1.8) circle (0.05cm);
\draw (1.2-\x,-0.3) -- (0.8-\x,-0.3) arc (270:180:0.45cm) -- (0.35-\x,1.35) arc (180:90:0.45) -- (1.2-\x,1.8);
\draw (1.2-\x,1.2) -- (1.1-\x,1.2) arc (90:180:0.3cm) -- (0.8-\x,0.6) arc (180:270:0.3) -- (1.2-\x,0.3);
\draw (1.2-\x,0.6) -- (1.1-\x,0.6) arc (270:90:0.15cm) -- (1.2-\x,0.9);
\draw (1.2-0.32,0.75) node {$\cdots$};
\draw (1.2-\x,1.6) node {$\vdots$};
\draw (1.2-\x,0.1) node {$\vdots$};
\end{tikzpicture} 
\end{array}\right.
\,,\qquad \begin{array}{c}
\begin{tikzpicture}[thick]
\newcommand{\x}{-0.3}
\fill[black] (0+\x,-0.3) circle (0.05cm);
\fill[black] (0+\x,0.3) circle (0.05cm);
\fill[black] (0+\x,0.6) circle (0.05cm);
\fill[black] (0+\x,0.9) circle (0.05cm);
\draw (0+\x,-0.3) -- (1.2-\x,-0.3);
\draw (0+\x,1.2) -- (0.1+\x,1.2) arc (90:0:0.3cm) -- (0.4+\x,0.6) arc (0:-90:0.3) -- (0+\x,0.3);
\fill[black] (0+\x,1.2) circle (0.05cm);
\fill[black] (0+\x,1.8) circle (0.05cm);
\draw (0+\x,0.6) -- (0.1+\x,0.6) arc (-90:90:0.15cm) -- (0+\x,0.9);
\draw (0.33,0.75) node {$\cdots$};
\draw (\x,1.6) node {$\vdots$};
\draw (\x,0.1) node {$\vdots$};
\fill[black] (1.2-\x,-0.3) circle (0.05cm);
\fill[black] (1.2-\x,0.3) circle (0.05cm);
\fill[black] (1.2-\x,0.6) circle (0.05cm);
\fill[black] (1.2-\x,0.9) circle (0.05cm);
\fill[black] (1.2-\x,1.2) circle (0.05cm);
\fill[black] (1.2-\x,1.8) circle (0.05cm);
\draw (1.2-\x,1.8) -- (0+\x,1.8);
\draw (1.2-\x,1.2) -- (1.1-\x,1.2) arc (90:180:0.3cm) -- (0.8-\x,0.6) arc (180:270:0.3) -- (1.2-\x,0.3);
\draw (1.2-\x,0.6) -- (1.1-\x,0.6) arc (270:90:0.15cm) -- (1.2-\x,0.9);
\draw (1.2-0.32,0.75) node {$\cdots$};
\draw (1.2-\x,1.6) node {$\vdots$};
\draw (1.2-\x,0.1) node {$\vdots$};
\end{tikzpicture} 
\end{array}
\,,\qquad  
\begin{array}{c}
\begin{tikzpicture}[thick]
\newcommand{\x}{-0.3}
\fill[black] (0+\x,-0.3) circle (0.05cm);
\fill[black] (0+\x,0.3) circle (0.05cm);
\fill[black] (0+\x,0.6) circle (0.05cm);
\fill[black] (0+\x,0.9) circle (0.05cm);
\draw (0+\x,-0.3) -- (1.2-\x,-0.3);
\draw (0+\x,0.3) -- (1.2-\x,0.3);
\fill[black] (0+\x,1.2) circle (0.05cm);
\fill[black] (0+\x,1.8) circle (0.05cm);
\draw (0+\x,0.6) -- (0.1+\x,0.6) arc (-90:90:0.15cm) -- (0+\x,0.9);
\draw (\x,1.6) node {$\vdots$};
\draw (\x,0.1) node {$\vdots$};
\fill[black] (1.2-\x,-0.3) circle (0.05cm);
\fill[black] (1.2-\x,0.3) circle (0.05cm);
\fill[black] (1.2-\x,0.6) circle (0.05cm);
\fill[black] (1.2-\x,0.9) circle (0.05cm);
\fill[black] (1.2-\x,1.2) circle (0.05cm);
\fill[black] (1.2-\x,1.8) circle (0.05cm);
\draw (1.2-\x,1.8) -- (0+\x,1.8);
\draw (1.2-\x,1.2) -- (0+\x,1.2);
\draw (1.2-\x,0.6) -- (1.1-\x,0.6) arc (270:90:0.15cm) -- (1.2-\x,0.9);
\draw (1.2-\x,1.6) node {$\vdots$};
\draw (1.2-\x,0.1) node {$\vdots$};
\end{tikzpicture} 
\end{array}
\,,\qquad  
\begin{array}{c}
\begin{tikzpicture}[thick]
\newcommand{\x}{-0.3}
\fill[black] (0+\x,-0.3) circle (0.05cm);
\fill[black] (0+\x,0.3) circle (0.05cm);
\fill[black] (0+\x,0.6) circle (0.05cm);
\fill[black] (0+\x,0.9) circle (0.05cm);
\draw (0+\x,-0.3) -- (1.2-\x,-0.3);
\draw (0+\x,0.3) -- (1.2-\x,0.3);
\fill[black] (0+\x,1.2) circle (0.05cm);
\fill[black] (0+\x,1.8) circle (0.05cm);
\draw (0+\x,0.6) -- (1.2-\x,0.6);
\draw (\x,1.6) node {$\vdots$};
\draw (\x,0.1) node {$\vdots$};
\fill[black] (1.2-\x,-0.3) circle (0.05cm);
\fill[black] (1.2-\x,0.3) circle (0.05cm);
\fill[black] (1.2-\x,0.6) circle (0.05cm);
\fill[black] (1.2-\x,0.9) circle (0.05cm);
\fill[black] (1.2-\x,1.2) circle (0.05cm);
\fill[black] (1.2-\x,1.8) circle (0.05cm);
\draw (1.2-\x,1.8) -- (0+\x,1.8);
\draw (1.2-\x,1.2) -- (0+\x,1.2);
\draw (1.2-\x,0.9) -- (0+\x,0.9);
\draw (1.2-\x,1.6) node {$\vdots$};
\draw (1.2-\x,0.1) node {$\vdots$};
\end{tikzpicture} 
\end{array}\,,
\ee
which show how the correlations are distributed between the two parties. These diagrams can be viewed as a heuristic depiction of (reduced density) matrices or general tensors of rank varying from one (leftmost) to $n$ (rightmost). In the case of generic matrices and tensors such association with lines is less evident since the dimension of the Hilbert space grows exponentially rather than linearly with the number of lines. In TQFT, on the other side, there are natural projectors and subspaces that make the dimension of the Hilbert space grow as needed. Connection with representation theory then ensures that there is a proper match between the classes of diagrams and the classes of entanglement.

The discussion can be generalized to other TQFT as well. The 3D TQFTs with sphere boundaries are also examples of 1D TQFT, as explained in the text. One can also consider 3D theories with other type of boundaries. For the case of $T^2$, for example, one can adopt the construction of~\cite{Balasubramanian:2016sro} and of the works that followed, where related questions were investigated.

The considered topological classification based on the connectivity turns out to be more coarse than the SLOCC one for higher number of parties. In the tripartite case it fails to capture the W state of three qubits. As far as nonbiseparable tripartite classes are concerned, the topological classification only captures the GHZ class. The GHZ states are dense in the space of all tripartite qubit states, which means that they can also approximate the W states with a desired accuracy, and the topological classification is complete in this sense, but more coarse, than the SLOCC one. The situation repeats in the case of four qubits, in which the connectivity classification only detects two nonbiseparable classes. Unlike the SLOCC classification, it remains finite in the four-qubit case.

The possibility that W states have a simple topological presentation has not been excluded by this work, but it looks more likely that their TQFT version is not simple. They are of coarse present as linear combinations of simple states, but quantum protocols that generate them from simple states are likely complex. After all, the discreteness of topology makes it inevitable that some states can only be approximated by the available discrete dense set. It would be interesting to understand how different entanglement classes can be systematically obtained from the dense GHZ class, for example, by specific braiding operations.

The connectivity classification and its possible extensions have a number of advantages. First, it is straightforward to generalize it to an arbitrary number of parties and to arbitrary party dimensions. Most importantly, however, the classification has a very intuitive nature, which tells that to be more entangled, one needs more cables between the parties, while the entanglement distribution depends on how the cables are wired. Even with the current restrictions it may help in designing new quantum states for specific quantum tasks, new measures of entanglements, and quantum protocols. For example, an indicator of genuine tripartite entanglement between parties $A$, $B$, and $C$ can be constructed from the diagrams representing the state as follows:
\be
\tau_3(A) \ = - \Tr_{\Hc_A\times\Hc_A} L\log L\,, \qquad L  = \ \frac{\hat L}{\Tr_{\Hc_A\times\Hc_A}\hat{L}}\,, \qquad \hat L \ = \!\!
\begin{array}{c}
\begin{tikzpicture}[thick]
\newcommand{\x}{1.4}
\newcommand{\y}{0.8}
\fill[black] (0,0.0) circle (0.05cm);
\fill[black] (0,0.2) circle (0.05cm);
\fill[black] (0,0.8) circle (0.05cm);
\fill[black] (0,1) circle (0.05cm);
\draw (0,0.2) -- (0.4,0.2) arc (-90:0:0.1) -- (0.5,0.7) arc (0:90:0.1) -- (0,0.8);
\draw (0.8,0.2) -- (1.2,0.2) arc (-90:0:0.1) -- (1.3,0.7) arc (0:90:0.1) -- (0.8,0.8) arc (90:180:0.1) -- (0.7,0.3) arc (180:270:0.1);
\fill[black] (0.5,0.5) circle (0.05cm);
\fill[black] (0.7,0.5) circle (0.05cm);
\fill[black] (1,0.8) circle (0.05cm);
\fill[black] (1,1) circle (0.05cm);
\fill[black] (1,0.2) circle (0.05cm);
\fill[black] (1,0) circle (0.05cm);
\draw (0.8+\y,0.2) -- (1.2+\y,0.2) arc (-90:0:0.1) -- (1.3+\y,0.7) arc (0:90:0.1) -- (0.8+\y,0.8) arc (90:180:0.1) -- (0.7+\y,0.3) arc (180:270:0.1);
\fill[black] (0.5+\y,0.5) circle (0.05cm);
\fill[black] (0.7+\y,0.5) circle (0.05cm);
\fill[black] (1+\y,0.8) circle (0.05cm);
\fill[black] (1+\y,1) circle (0.05cm);
\fill[black] (1+\y,0.2) circle (0.05cm);
\fill[black] (1+\y,0) circle (0.05cm);
\draw (0,0) -- (1.4+\x,0.0);
\fill[black] (1.3+\y,0.5) circle (0.05cm);
\fill[black] (1.5+\y,0.5) circle (0.05cm);
\fill[black] (1.4+\x,0.0) circle (0.05cm);
\fill[black] (1.4+\x,0.2) circle (0.05cm);
\fill[black] (1.4+\x,0.8) circle (0.05cm);
\fill[black] (1.4+\x,1) circle (0.05cm);
\draw (1.4+\x,1) -- (0,1);
\draw (1.4+\x,0.2) -- (1.+\x,0.2) arc (270:180:0.1) -- (0.9+\x,0.7) arc (180:90:0.1) -- (1.4+\x,0.8);
\draw (-0.3,0.1) node {\small $A$};
\draw (-0.3,0.9) node {\small $A$};
\draw (3.1,0.1) node {\small $A$};
\draw (3.1,0.9) node {\small $A$};
\draw (0.3,0.5) node {\small $B$};
\draw (1.1,0.5) node {\small $A$};
\draw (1.9,0.5) node {\small $C$};
\draw (1.,-0.25) node {\small $C$};
\draw (1.,1.25) node {\small $C$};
\draw (1.8,-0.25) node {\small $B$};
\draw (1.8,1.25) node {\small $B$};
\end{tikzpicture} 
\end{array}
\!,
\ee
where $\Hc_A$ is the Hilbert space associated with party $A$ and the double-line 3-vertex denotes a generic tripartite state of a pure system. This ladder, which is made of six such vertices (wave functions), can be similarly built for $B$ and $C$, and the result will be nonzero, for either $\tau_3(A)$, $\tau_3(B)$, or $\tau_3(C)$, only if there is three-way entanglement between $A$, $B$ and $C$. The idea here is that the ladder will factorize if any of the chain elements $A$, $B$, or $C$ is broken, as in the case of the first two diagrams in~(\ref{3party0}). Further details and generalizations will be discussed elsewhere~\cite{Melnikov:2023nzn}.

\paragraph{Acknowledgments.} I would like to thank João Aires for collaboration at early stages of this project and Rafael Chaves for useful comments. I would also like to thank Gabriel Cardoso, Suresh Nampuri, and Roger Picken for the opportunity to present my results at IST Lisbon. This work was supported by RSF grant No.~18-71-10073.

\bibliographystyle{JHEP}
\bibliography{refs}

\providecommand{\href}[2]{#2}\begingroup\raggedright\begin{thebibliography}{10}

\bibitem{Aravind:1997}
P.~Aravind, \emph{Borromean entanglement of the ghz state},  in
  \emph{Potentiality, entanglement and passion-at-a-distance}, pp.~53--59.
\newblock Springer, 1997.

\bibitem{Atiyah:1989vu}
M.~Atiyah, \emph{{Topological quantum field theories}},
  \href{https://doi.org/10.1007/BF02698547}{\emph{Inst. Hautes Etudes Sci.
  Publ. Math.} {\bfseries 68} (1989) 175}.

\bibitem{Witten:1988ze}
E.~Witten, \emph{{Topological Quantum Field Theory}},
  \href{https://doi.org/10.1007/BF01223371}{\emph{Commun. Math. Phys.}
  {\bfseries 117} (1988) 353}.

\bibitem{Witten:1988hf}
E.~Witten, \emph{{Quantum Field Theory and the Jones Polynomial}},
  \href{https://doi.org/10.1007/BF01217730}{\emph{Commun. Math. Phys.}
  {\bfseries 121} (1989) 351}.

\bibitem{Kitaev:1997wr}
A.~Y. Kitaev, \emph{{Fault tolerant quantum computation by anyons}},
  \href{https://doi.org/10.1016/S0003-4916(02)00018-0}{\emph{Annals Phys.}
  {\bfseries 303} (2003) 2}
  [\href{https://arxiv.org/abs/quant-ph/9707021}{{\ttfamily
  quant-ph/9707021}}].

\bibitem{Freedman:2000rc}
M.~H. Freedman, A.~Kitaev and Z.~Wang, \emph{{Simulation of topological field
  theories by quantum computers}},
  \href{https://doi.org/10.1007/s002200200635}{\emph{Commun. Math. Phys.}
  {\bfseries 227} (2002) 587}
  [\href{https://arxiv.org/abs/quant-ph/0001071}{{\ttfamily
  quant-ph/0001071}}].

\bibitem{Kitaev:2000nmw}
A.~Y. Kitaev, \emph{{Unpaired Majorana fermions in quantum wires}},
  \href{https://doi.org/10.1070/1063-7869/44/10S/S29}{\emph{Phys. Usp.}
  {\bfseries 44} (2001) 131}
  [\href{https://arxiv.org/abs/cond-mat/0010440}{{\ttfamily
  cond-mat/0010440}}].

\bibitem{Nayak:2008zza}
C.~Nayak, S.~H. Simon, A.~Stern, M.~Freedman and S.~Das~Sarma,
  \emph{{Non-Abelian anyons and topological quantum computation}},
  \href{https://doi.org/10.1103/RevModPhys.80.1083}{\emph{Rev. Mod. Phys.}
  {\bfseries 80} (2008) 1083}
  [\href{https://arxiv.org/abs/0707.1889}{{\ttfamily 0707.1889}}].

\bibitem{Kauffman:2002qua}
L.~H. Kauffman, \emph{Quantum computing and the jones polynomial},
  {\emph{CONTEMPORARY MATHEMATICS} {\bfseries 305} (2002) 101}.

\bibitem{Kauffman:2003cri}
L.~H. Kauffman and S.~J. Lomonaco~Jr, \emph{Entanglement criteria: quantum and
  topological},  in \emph{Quantum Information and Computation}, vol.~5105,
  pp.~51--58, International Society for Optics and Photonics, 2003.

\bibitem{Kauffman:2004bra}
L.~H. Kauffman and S.~J. Lomonaco~Jr, \emph{Braiding operators are universal
  quantum gates}, {\emph{New Journal of Physics} {\bfseries 6} (2004) 134}.

\bibitem{Kauffman:2004qua}
L.~H. Kauffman and S.~J. Lomonaco~Jr, \emph{Quantum knots},  in \emph{Quantum
  Information and Computation II}, vol.~5436, pp.~268--284, International
  Society for Optics and Photonics, 2004.

\bibitem{Lomonaco:2008qua}
S.~J. Lomonaco and L.~H. Kauffman, \emph{Quantum knots and mosaics},
  {\emph{Quantum Information Processing} {\bfseries 7} (2008) 85}.

\bibitem{Melnikov:2017bjb}
D.~Melnikov, A.~Mironov, S.~Mironov, A.~Morozov and A.~Morozov, \emph{{Towards
  topological quantum computer}},
  \href{https://doi.org/10.1016/j.nuclphysb.2017.11.016}{\emph{Nucl. Phys. B}
  {\bfseries 926} (2018) 491}
  [\href{https://arxiv.org/abs/1703.00431}{{\ttfamily 1703.00431}}].

\bibitem{Kauffman:2019top}
L.~H. Kauffman and E.~Mehrotra, \emph{Topological aspects of quantum
  entanglement}, {\emph{Quantum Information Processing} {\bfseries 18} (2019)
  1} [\href{https://arxiv.org/abs/1611.08047}{{\ttfamily 1611.08047}}].

\bibitem{Padmanabhan:2019qed}
P.~Padmanabhan, F.~Sugino and D.~Trancanelli, \emph{{Quantum entanglement,
  supersymmetry, and the generalized Yang-Baxter equation}},
  \href{https://arxiv.org/abs/1911.02577}{{\ttfamily 1911.02577}}.

\bibitem{Kauffman:2013bh}
L.~H. Kauffman, \emph{{Knot Logic and Topological Quantum Computing with
  Majorana Fermions}},  \href{https://arxiv.org/abs/1301.6214}{{\ttfamily
  1301.6214}}.

\bibitem{Dong:2008ft}
S.~Dong, E.~Fradkin, R.~G. Leigh and S.~Nowling, \emph{{Topological
  Entanglement Entropy in Chern-Simons Theories and Quantum Hall Fluids}},
  \href{https://doi.org/10.1088/1126-6708/2008/05/016}{\emph{JHEP} {\bfseries
  05} (2008) 016} [\href{https://arxiv.org/abs/0802.3231}{{\ttfamily
  0802.3231}}].

\bibitem{Salton:2016qpp}
G.~Salton, B.~Swingle and M.~Walter, \emph{{Entanglement from Topology in
  Chern-Simons Theory}},
  \href{https://doi.org/10.1103/PhysRevD.95.105007}{\emph{Phys. Rev. D}
  {\bfseries 95} (2017) 105007}
  [\href{https://arxiv.org/abs/1611.01516}{{\ttfamily 1611.01516}}].

\bibitem{Balasubramanian:2016sro}
V.~Balasubramanian, J.~R. Fliss, R.~G. Leigh and O.~Parrikar,
  \emph{{Multi-Boundary Entanglement in Chern-Simons Theory and Link
  Invariants}}, \href{https://doi.org/10.1007/JHEP04(2017)061}{\emph{JHEP}
  {\bfseries 04} (2017) 061}
  [\href{https://arxiv.org/abs/1611.05460}{{\ttfamily 1611.05460}}].

\bibitem{Chun:2017hja}
S.~Chun and N.~Bao, \emph{{Entanglement entropy from SU(2) Chern-Simons theory
  and symmetric webs}},  \href{https://arxiv.org/abs/1707.03525}{{\ttfamily
  1707.03525}}.

\bibitem{Dwivedi:2017rnj}
S.~Dwivedi, V.~K. Singh, S.~Dhara, P.~Ramadevi, Y.~Zhou and L.~K. Joshi,
  \emph{{Entanglement on linked boundaries in Chern-Simons theory with generic
  gauge groups}}, \href{https://doi.org/10.1007/JHEP02(2018)163}{\emph{JHEP}
  {\bfseries 02} (2018) 163}
  [\href{https://arxiv.org/abs/1711.06474}{{\ttfamily 1711.06474}}].

\bibitem{Balasubramanian:2018por}
V.~Balasubramanian, M.~DeCross, J.~Fliss, A.~Kar, R.~G. Leigh and O.~Parrikar,
  \emph{{Entanglement Entropy and the Colored Jones Polynomial}},
  \href{https://doi.org/10.1007/JHEP05(2018)038}{\emph{JHEP} {\bfseries 05}
  (2018) 038} [\href{https://arxiv.org/abs/1801.01131}{{\ttfamily
  1801.01131}}].

\bibitem{Melnikov:2018zfn}
D.~Melnikov, A.~Mironov, S.~Mironov, A.~Morozov and A.~Morozov, \emph{{From
  Topological to Quantum Entanglement}},
  \href{https://doi.org/10.1007/JHEP05(2019)116}{\emph{JHEP} {\bfseries 05}
  (2019) 116} [\href{https://arxiv.org/abs/1809.04574}{{\ttfamily
  1809.04574}}].

\bibitem{Quinta:2018scm}
G.~M. Quinta and R.~Andr\'e, \emph{{Classifying quantum entanglement through
  topological links}},
  \href{https://doi.org/10.1103/PhysRevA.97.042307}{\emph{Phys. Rev. A}
  {\bfseries 97} (2018) 042307}
  [\href{https://arxiv.org/abs/1803.08935}{{\ttfamily 1803.08935}}].

\bibitem{Dwivedi:2019bzh}
S.~Dwivedi, V.~K. Singh, P.~Ramadevi, Y.~Zhou and S.~Dhara, \emph{{Entanglement
  on multiple $S^2$ boundaries in Chern-Simons theory}},
  \href{https://doi.org/10.1007/JHEP08(2019)034}{\emph{JHEP} {\bfseries 08}
  (2019) 034} [\href{https://arxiv.org/abs/1906.11489}{{\ttfamily
  1906.11489}}].

\bibitem{Dwivedi:2020jyx}
S.~Dwivedi, A.~Addazi, Y.~Zhou and P.~Sharma, \emph{{Multi-boundary
  entanglement in Chern-Simons theory with finite gauge groups}},
  \href{https://doi.org/10.1007/JHEP04(2020)158}{\emph{JHEP} {\bfseries 04}
  (2020) 158} [\href{https://arxiv.org/abs/2003.01404}{{\ttfamily
  2003.01404}}].

\bibitem{Bao:2021gzu}
N.~Bao, N.~Cheng, S.~Hern\'andez-Cuenca and V.~P. Su, \emph{{Topological Link
  Models of Multipartite Entanglement}},
  \href{https://doi.org/10.22331/q-2022-06-20-741}{\emph{Quantum} {\bfseries 6}
  (2022) 741} [\href{https://arxiv.org/abs/2109.01150}{{\ttfamily
  2109.01150}}].

\bibitem{Buican:2021axn}
M.~Buican and R.~Radhakrishnan, \emph{{Galois orbits of TQFTs: symmetries and
  unitarity}}, \href{https://doi.org/10.1007/JHEP01(2022)004}{\emph{JHEP}
  {\bfseries 01} (2022) 004}
  [\href{https://arxiv.org/abs/2109.02766}{{\ttfamily 2109.02766}}].

\bibitem{Jones:1985dw}
V.~F.~R. Jones, \emph{{A polynomial invariant for knots via von Neumann
  algebras}},
  \href{https://doi.org/10.1090/S0273-0979-1985-15304-2}{\emph{Bull. Am. Math.
  Soc.} {\bfseries 12} (1985) 103}.

\bibitem{Wenzl:1985seq}
H.~Wenzl, \emph{On sequences of projections}, {\emph{CR Math. Rep. Acad. Sci.
  Canada} {\bfseries 9} (1987) 5}.

\bibitem{Dur:2000zz}
W.~Dur, G.~Vidal and J.~I. Cirac, \emph{{Three qubits can be entangled in two
  inequivalent ways}},
  \href{https://doi.org/10.1103/PhysRevA.62.062314}{\emph{Phys. Rev. A}
  {\bfseries 62} (2000) 062314}
  [\href{https://arxiv.org/abs/quant-ph/0005115}{{\ttfamily
  quant-ph/0005115}}].

\bibitem{Verstraete:2002four}
F.~Verstraete, J.~Dehaene, B.~De~Moor and H.~Verschelde, \emph{Four qubits can
  be entangled in nine different ways}, {\emph{Physical Review A} {\bfseries
  65} (2002) 052112}.

\bibitem{Horodecki:2009zz}
R.~Horodecki, P.~Horodecki, M.~Horodecki and K.~Horodecki, \emph{{Quantum
  entanglement}}, \href{https://doi.org/10.1103/RevModPhys.81.865}{\emph{Rev.
  Mod. Phys.} {\bfseries 81} (2009) 865}
  [\href{https://arxiv.org/abs/quant-ph/0702225}{{\ttfamily
  quant-ph/0702225}}].

\bibitem{Sawicki:2012cri}
A.~Sawicki, M.~Oszmaniec and M.~Ku\ifmmode~\acute{s}\else \'{s}\fi{},
  \emph{Critical sets of the total variance can detect all stochastic local
  operations and classical communication classes of multiparticle
  entanglement}, \href{https://doi.org/10.1103/PhysRevA.86.040304}{\emph{Phys.
  Rev. A} {\bfseries 86} (2012) 040304}.

\bibitem{Walter:2013ent}
M.~Walter, B.~Doran, D.~Gross and M.~Christandl, \emph{Entanglement polytopes:
  multiparticle entanglement from single-particle information}, {\emph{Science}
  {\bfseries 340} (2013) 1205}.

\bibitem{Sawicki:2014con}
A.~Sawicki, M.~Oszmaniec and M.~Ku{\'s}, \emph{Convexity of momentum map, morse
  index, and quantum entanglement}, {\emph{Reviews in Mathematical Physics}
  {\bfseries 26} (2014) 1450004}.

\bibitem{Maciazek:2018asy}
T.~Maciazek and A.~Sawicki, \emph{Asymptotic properties of entanglement
  polytopes for large number of qubits}, {\emph{Journal of Physics A:
  Mathematical and Theoretical} {\bfseries 51} (2018) 07LT01}.

\bibitem{Melnikov:2020mno}
D.~Melnikov, \emph{{Topological View on Entanglement and Complexity}},
  \href{https://doi.org/10.1007/978-3-030-35473-2_11}{\emph{Springer Proc.
  Phys.} {\bfseries 239} (2020) 271}.

\bibitem{Melnikov:2023nzn}
D.~Melnikov, \emph{{Connectomes and Properties of Quantum Entanglement}},
  \href{https://arxiv.org/abs/2302.08548}{{\ttfamily 2302.08548}}.

\bibitem{Kauffman:1987sta}
L.~H. Kauffman, \emph{State models and the jones polynomial}, {\emph{Topology}
  {\bfseries 26} (1987) 395}.

\bibitem{Verlinde:1988sn}
E.~P. Verlinde, \emph{{Fusion Rules and Modular Transformations in 2D Conformal
  Field Theory}},
  \href{https://doi.org/10.1016/0550-3213(88)90603-7}{\emph{Nucl. Phys. B}
  {\bfseries 300} (1988) 360}.

\bibitem{Vidal:1999vh}
G.~Vidal, \emph{{Entanglement of pure states for a single copy}},
  \href{https://doi.org/10.1103/PhysRevLett.83.1046}{\emph{Phys. Rev. Lett.}
  {\bfseries 83} (1999) 1046}
  [\href{https://arxiv.org/abs/quant-ph/9902033}{{\ttfamily
  quant-ph/9902033}}].

\bibitem{Kauffman:1994tem}
L.~H. Kauffman, S.~L. Lins and S.~Lins, \emph{Temperley-Lieb recoupling theory
  and invariants of 3-manifolds}, no.~134. Princeton University Press, 1994.

\bibitem{Coffman:1999jd}
V.~Coffman, J.~Kundu and W.~K. Wootters, \emph{{Distributed entanglement}},
  \href{https://doi.org/10.1103/PhysRevA.61.052306}{\emph{Phys. Rev. A}
  {\bfseries 61} (2000) 052306}
  [\href{https://arxiv.org/abs/quant-ph/9907047}{{\ttfamily
  quant-ph/9907047}}].

\end{thebibliography}\endgroup



\providecommand{\href}[2]{#2}\begingroup\raggedright\endgroup

\end{document}